\documentclass[prx, reprint, 10pt, superscriptaddress,
notitlepage, nofootinbib, 
longbibliography,
floatfix]{revtex4-1}
\usepackage{gpkmac}

\newcommand{\TLI}{\mathbf{)\,(}}
\newcommand{\TLE}{\rotatebox[origin=c]{-90}{$\mathbf{)\,(}$}}

\usepackage{xcolor}

\begin{document}

\title{Entanglement Negativity at Measurement-Induced Criticality}
\author{Shengqi Sang}
\thanks{SS and YL contributed equally to this work.}
\affiliation{Perimeter Institute for Theoretical Physics, Waterloo, Ontario N2L 2Y5, Canada}
\author{Yaodong Li}
\thanks{SS and YL contributed equally to this work.}
\affiliation{Department of Physics, University of California, Santa Barbara, CA 93106, USA}
\author{Tianci Zhou}
\affiliation{Kavli Institute for Theoretical Physics, University of California, Santa Barbara, CA 93106, USA}
\author{Xiao Chen}
\affiliation{Department of Physics, Boston College, Chestnut Hill, MA 02467, USA}
\author{Timothy H. Hsieh}
\affiliation{Perimeter Institute for Theoretical Physics, Waterloo, Ontario N2L 2Y5, Canada}
\author{Matthew P. A. Fisher}
\affiliation{Department of Physics, University of California, Santa Barbara, CA 93106, USA}

\date{November 30, 2020}

\begin{abstract}




We propose entanglement negativity as a
fine-grained probe of 
measurement-induced criticality. 
We motivate this proposal in stabilizer states,
where for two disjoint subregions, comparing their ``mutual negativity'' and their mutual information leads to a precise distinction between bipartite and multipartite entanglement.
In a measurement-only stabilizer circuit that maps exactly to two-dimensional critical percolation, we show that the mutual information and the mutual negativity are governed by boundary conformal fields of different scaling dimensions at long distances.
We then consider a class of ``hybrid'' circuit models obtained by perturbing the measurement-only circuit with
unitary gates of progressive levels of complexity.
While other critical exponents vary appreciably for different choices of unitary gate ensembles at their respective critical points, the mutual negativity has scaling dimension $3$ across remarkably many of the hybrid circuits, which is notably different from that in percolation.
We contrast our results with
limiting cases where a geometrical minimal-cut picture is available.


\end{abstract}

\maketitle


\tableofcontents

\section{Introduction\label{sec:intro}}

Quantum dynamics involving unitary evolution interspersed with measurements~\cite{nahum2018hybrid, li1808hybrid, nandkishore2018hybrid} have provided a wealth of new phenomena being actively explored.  When the measurement rate is low, the dominant unitaries generically lead to steady states which have volume-law entanglement but are nonetheless sharply distinct from random Page states and exhibit novel instances of quantum error-correcting codes~\cite{choi2019qec, gullans1905purification, fan2020selforganized, li2007capillary_qecc, fidkowski2008forget, gullans2010lowdepth}.  When the measurement rate is high, the steady state generically has area-law entanglement but can harbor nontrivial long-range entanglement~\cite{nahum1911majorana, hsieh_sang_2004_protected, barkeshli2004symmetric, buechler2006projectiveTFIM}.  

Remarkably, the transition between the volume and area law phases exhibits conformal invariance and has been actively studied with both numerical and analytic approaches~\cite{nahum2018hybrid, li1901hybrid, andreas2019hybrid, choi2019spin, huse1911tripartite, li2003cft}.
On one hand, Haar and Clifford random circuit simulations provide abundant data on the scaling of entanglement and mutual information among other quantities, {revealing many new universality classes of measurement-induced criticality (MIC) remaining to be  classified~\cite{ippoliti2004measurementonly, hsieh_sang_2004_protected, barkeshli2004symmetric}.}
On the other hand, analytical progress {for the random Haar circuit} requires taking the limit of large local Hilbert space dimension $d$, in which the entanglement transition maps to critical percolation in two dimensions~\cite{andreas2019hybrid}.  The distinction between the percolation universality class in the large $d$ limit and the critical data of (finite $d$) models accessed in numerics is still somewhat unclear. 
{These considerations} motivate a finer characterization of the latter.

In this work, we employ entanglement negativity as a fruitful probe of measurement-induced criticality.  Negativity was originally introduced as a measure of mixed state entanglement~\cite{neg1, neg2, neg3, vidal_werner_2001}: while von Neumann entanglement entropy is sensitive to both quantum and classical correlations in a bipartite mixed state, negativity only detects quantum correlations.
This property also makes it a useful probe for distinguishing bipartite and multipartite quantum correlations in a pure state: given a {tripartite} pure state $\Psi_{ABC}$ (see for example Fig.~\ref{fig:struct_thm}), the negativity of the reduced density matrix $\rho_{AB}$ -- which we refer to as “mutual negativity” -- detects quantum correlations between $A$ and $B$.   In contrast, mutual information $I_{A,B}$ is sensitive to both quantum and ``classical'' correlation, which in this case arises from tripartite correlation between $A$, $B$, and $C$.  For general states, it is difficult to classify multipartite entanglement.  However, as we will review, for stabilizer states there is a precisely-defined separation between bipartite and tripartite entanglement, and this is completely captured by the difference between mutual negativity and mutual information.

We thus use negativity as a finer diagnostic of bipartite versus multipartite entanglement in finite $d$ measurement-induced critical points.  We find that mutual negativity decays as a power of the separation between $A, B$, which differs from the exponential decay of ground states of (unitary) conformal field theories~\cite{cardy2012EN_a, cardy2012EN_b}. Furthermore, there is a significant difference between the scaling of mutual negativity and mutual information, which contrasts sharply with that of holographic stabilizer tensor network models and relatedly, large $d$ hybrid circuits.  We illustrate these aspects and intuition in a measurement-only model~\cite{nahum1911majorana, hsieh_sang_2004_protected, buechler2006projectiveTFIM}, in which we analytically derive the scaling dimension for mutual negativity.  For critical points involving both measurements and unitaries, we find that the mutual negativity appears to be remarkably ``super-universal'': in regimes where the mutual information scaling dimension changes continuously, the negativity scaling dimension is constant.  Mutual negativity thus serves as a valuable diagnostic unveiling more detailed entanglement structures and {potentially} distinguishing different measurement-driven critical universality classes, including large vs. small $d$. 

The rest of this paper is organized as follows.
In Sec.~\ref{sec:EN_stab} we introduce the definition of entanglement negativity, with a focus on ``mutual negativity'' of stabilizer states.
In Sec.~\ref{sec:moc_majorana} where a measurement-only stabilizer circuit that maps to critical percolation is considered, we show how mutual information and mutual negativity can be mapped to known geometrical observables on the boundary of percolation, and expose their different powerlaw decays at long distances.
In Sec.~\ref{sec:hybrid_circuit}, we consider several ``hybrid circuits'', which may be obtained from the measurement-only circuit by introducing additional unitary gates.
The unitaries are ``relevant'' and take the system to new critical points, {for which we numerically compute the scaling dimension of the mutual negativity.}
In Sec.~\ref{sec:discussion}, we discuss possible future directions that might be taken along these lines,
and mention a few questions that arise from this work.

\section{Entanglement negativity in general and in stabilizer states \label{sec:EN_stab}}


Given a state $\rho_{A \cup B}$ defined on $A\cup B$, the logarithmic entanglement negativity (or simply EN) of subsystem $A$ is defined as~\cite{vidal_werner_2001}
\env{align}{
	N_A(\rho_{A \cup B}) = \ln \lVert \rho_{A \cup B}^{\Gamma_A} \rVert_1 = \ln \sum_i |\lambda_i|.
}
where $\Gamma_A$ is the partial transpose operation of $\rho$ with respect to $A$, and $\lambda$ are the eigenvalues of $\rho_{A \cup B}^{\Gamma_A}$.



The EN is a quantity that is in general nontrivial to compute, but great simplifications occur when $\rho_{A \cup B}$ is a \emph{stabilizer state}~\cite{gottesman1997thesis, gottesman9807heisenberg}, whose density matrix takes the form~\cite{Klappenecker2002stabilizer, Fattal2004stabilizer}
\env{align}{
    \label{eq:stab_state}
	\rho_{A \cup B} = \frac{1}{2^{|{A \cup B}|}} \sum_{g \in \mc{S}} g.
}
Here $\mc{S}$ is an abelian subgroup of the Pauli group on ${A \cup B}$, known as the \emph{stabilizer group}.
With details in Appendix~\ref{app:EN_stab}, we quote the result that
\env{align}{
    \label{eq:EN_group_theoretic}
	N_A(\rho_{A \cup B}) = \frac{1}{2} \ln \lv \frac{\mathrm{proj}_A(\mc{S})}{Z(\mathrm{proj}_A(\mc{S}))} \rv.
}
Here $\mathrm{proj}_A(\mc{S})$ is the group obtained from $\mc{S}$ by ``restricting'' Pauli strings in $\mc{S}$ on $A$, and is in general non-abelian; and $Z(\mathrm{proj}_A(\mc{S}))$ is the \emph{central subgroup} of $\mathrm{proj}_A(\mc{S})$, i.e. the subgroup of elements that commutes with every element of $\mathrm{proj}_A(\mc{S})$.
{This result is the basis of our numerical computation of the EN in stabilizer circuits.}

We will focus on stabilizer states for the rest of this section, and provide a detailed characterization of the EN.
We will return to the general case in Sec.~\ref{sec:hybrid_circuit} with our numerics on hybrid circuits with Haar unitaries, and in Sec.~\ref{sec:discussion} with discussions.

\subsection{Mutual negativity bounds (half) the mutual information from below for stabilizer states}

Thoughout much of this work, we take $A$ and $B$ to be subsystems of a larger system (compare Fig.~\ref{fig:struct_thm}), and $\rho_{A \cup B}$
is a reduced density matrix on $A\cup B$. 
In this context, we find it convenient to define \emph{mutual negativity} (MN) between $A,B$ as 
\env{align}{
    \label{eq:MN_def}
	N_{A,B}
	\coloneqq N_A(\rho_{A \cup B})
 	= N_B(\rho_{A\cup B}).
}



We want to compare $N_{A,B}$ with $I_{A,B}$, the von Neumann mutual information (or simply MI) between $A$ and $B$, defined as 
\env{align}{
    \label{eq:def_IvN}
    I_{A,B}
    \coloneqq 
    S_A + S_B - S_{A \cup B},
}
where $\rho_A = \mathrm{Tr}_{B} \rho_{A \cup B}$,  $\rho_B = \mathrm{Tr}_{A} \rho_{A \cup B}$, and $S_A$ (and similarly for $S_B$ and $S_{A \cup B}$) is the von Neumann entropy of $\rho_A$,
\env{align}{
    \label{eq:def_SvN}
    S_A \coloneqq S_{\rm vN}(\rho_A) = - \mathrm{Tr}(\rho_A \ln \rho_A).
}
For a stabilizer state of the form in Eq.~\eqref{eq:stab_state}, the MI is related to the stabilizer group as follows (see Appendix~\ref{app:EN_stab})
\env{align}{
    \label{eq:MI_stab}
    \frac{1}{2} I_{A,B}
    = \frac{1}{2} \ln \lv \frac{\mathrm{proj}_A (\mc{S})}{\mathrm{Ker}\, \mathrm{proj}_B(\mc{S})} \rv.
}
Comparing Eqs.~(\ref{eq:EN_group_theoretic}, \ref{eq:MN_def}, \ref{eq:MI_stab}), and noticing that
\env{align}{
    \mathrm{Ker}\, \mathrm{proj}_B(\mc{S}) \subseteq Z(\mathrm{proj}_A(\mc{S})),
}
we have
\env{align}{
    \label{eq:ineq_NAB_IAB}
    N_{A,B} \le \frac{1}{2} I_{A,B}.
}
This inequality holds for all stabilizer density matrices $\rho_{A \cup B}$; and when $\rho_{A\cup B}$ is a pure density matrix, the two sides are equal.
However, a general density matrix does not obey this inequality; a counterexample can be found in Ref.~\cite{sherman2016negativityarealaw}.

A few more inequalities for $N_{A,B}$ can be obtained in a similar way, which we collect in Appendix~\ref{app:EN_stab}.

\subsection{Mutual negativity detects bipartite entanglement}

\begin{figure}[b]
    \centering
    \includegraphics[width=.5\textwidth]{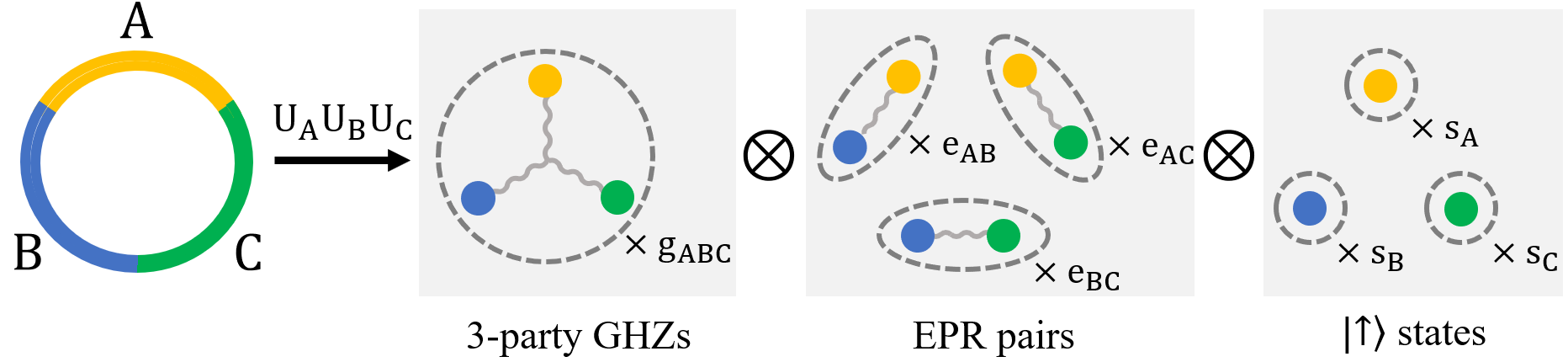}
    \caption{Illustration of the decomposition of a pure stabilizer state $\ket{\Psi_{Q = A\cup B\cup C}}$ in Eq.~\eqref{eq:structural_thm} for the tripartition $(A,B,C)$.
    For the ease of graphical presentation, we have chosen $A$, $B$, and $C$ to be contiguous segments; however, the state decomposition works generally for an arbitrary tripartition, and we will be mostly interested in the case when $A$ and $B$ are two contiguous, distant segments separated by $C = \ovl{A \cup B}$ (see inset of Fig.~\ref{fig:perc}(b)).
    }
    \label{fig:struct_thm}
\end{figure}

The physical significance of 
{MN} and 
{MI}, as well as the bound Eq.~\eqref{eq:ineq_NAB_IAB}, is made clear by the following ``structure theorem'' of pure stabilizer states.

We first introduce a purification of the stabilizer state $\rho_{A \cup B}$, that is, we embed $A \cup B$ into a larger system $Q$, which supports a pure stabilizer state $\rho_Q$ such that $\rho_{A \cup B} = \mathrm{Tr}_{\ovl{A \cup B}} \rho_Q$.
In general, $\rho_{A \cup B}$ is a mixed state.
We may write
\env{align}{
    \rho_Q = \ket{\Psi_Q} \bra{\Psi_Q},
}
where $\ket{\Psi_Q}$ is a stabilizer wavefunction.
Let $C \coloneqq \ovl{A \cup B} = Q - (A \cup B)$.
A stabilizer wavefunction $\ket{\Psi_Q}$ always admits the following representation, with suitable choices of $U_{A}$, $U_{B}$, and $U_{C}$~\cite{gottesman0504extraction} (see Fig.~\ref{fig:struct_thm}),
\env{align}{
    \label{eq:structural_thm}
  \ket{\Psi_Q}
    =&\ U_A U_B U_{C}
    \ket{\mathrm{GHZ}_{ABC}}^{\otimes g_{ABC}}\nn &\ 
    \otimes
    \ket{\mathrm{EPR}_{AB}}^{\otimes e_{AB}} \otimes
    \ket{\mathrm{EPR}_{BC}}^{\otimes e_{BC}} \otimes
    \ket{\mathrm{EPR}_{CA}}^{\otimes e_{CA}}\nn
    &\ 
    \otimes \ket{\ua_A}^{\otimes s_{A}}
    \otimes \ket{\ua_B}^{\otimes s_{B}}
    \otimes \ket{\ua_C}^{\otimes s_{C}}.
}
Here $U_A$, $U_B$, $U_C$ are \emph{local} Clifford unitaries supported on $A$, $B$, and $C$, respectively;
{$\ket{\ua_{A,B,C}}$ is a single-qubit ``up'' state in $A$, $B$, or $C$ that does not contribute entanglement;}
$\ket{\mathrm{EPR}_{AB}}$ denotes a two-qubit EPR pair with one qubit in $A$ and another in $B$ (and similarly for $\ket{\mathrm{EPR}_{BC}}$ and $\ket{\mathrm{EPR}_{CA}}$); and $\ket{\mathrm{GHZ}_{ABC}}$ is a {multi}-qubit GHZ state with one qubit in each of $A$, $B$, and $C$.


Since local unitary gates do not affect entanglement measures, we may read off the MN and the MI from the decomposition into EPR and GHZ states on the RHS of Eq.~\eqref{eq:structural_thm}.
We have
\env{align}{
    \label{eq:NAB_nEPR}
    N_{A,B} =&\, e_{AB} \ln 2, \\
    \label{eq:IAB_nEPR_nGHZ}
    \frac{1}{2}I_{A,B} =&\, \(e_{AB} + \frac{1}{2} g_{ABC} \) \ln 2,
}
{which manifests the bound in Eq.~\eqref{eq:ineq_NAB_IAB}.}
{Here, the MN receives contributions from the ``\emph{bi}partite'' entanglement, and filters out the GHZ ``\emph{tri}partite'' entanglement; in contrast, the MI receives contributions from both.}
The difference between Eq.~\eqref{eq:NAB_nEPR} and Eq.~\eqref{eq:IAB_nEPR_nGHZ} is precisely the amount of tripartite entanglement for this tripartition (up to a constant factor).
Notice that in general the GHZ-type entanglement cannot be captured by the ``tripartite mutual information'' or the ``topological entanglement entropy''~\cite{Hayden2016, barkeshli2004symmetric, huse1911tripartite}.

Eq.~\eqref{eq:NAB_nEPR} also implies the equality between $N_{A,B}$ 
and ``distillable entanglement''~\cite{bennett1996concentrating} for stabilizer states, {while in general the negativity is an upper bound of the distillable entanglement~\cite{vidal_werner_2001}}.

A related entanglement measure is the ``entanglement of purification''~\cite{terhal0202eofpurification}, which in general upper bounds $\frac{1}{2} I_{A,B}$, and for stabilizer states can be computed as follows~\cite{nguyen1709stabilizerTN},
\env{align}{
\label{eq:EPAB_nEPR_nGHZ}
    (E_P)_{A,B} =&\, \(e_{AB} + g_{ABC} \) \ln 2.
}
Comparing $(E_P)_{A,B}$ and $\frac{1}{2} I_{A,B}$ also leads to the separation of irreducible tripartite entanglement, and can be useful in other contexts~\cite{nguyen1709stabilizerTN, zaletel2011tripartite}.


\section{A measurement-only stabilizer circuit model that maps to percolation \label{sec:moc_majorana}}


\begin{figure*}[t]
    \centering
\subfigure[]{
    \includegraphics[width=.3\textwidth]{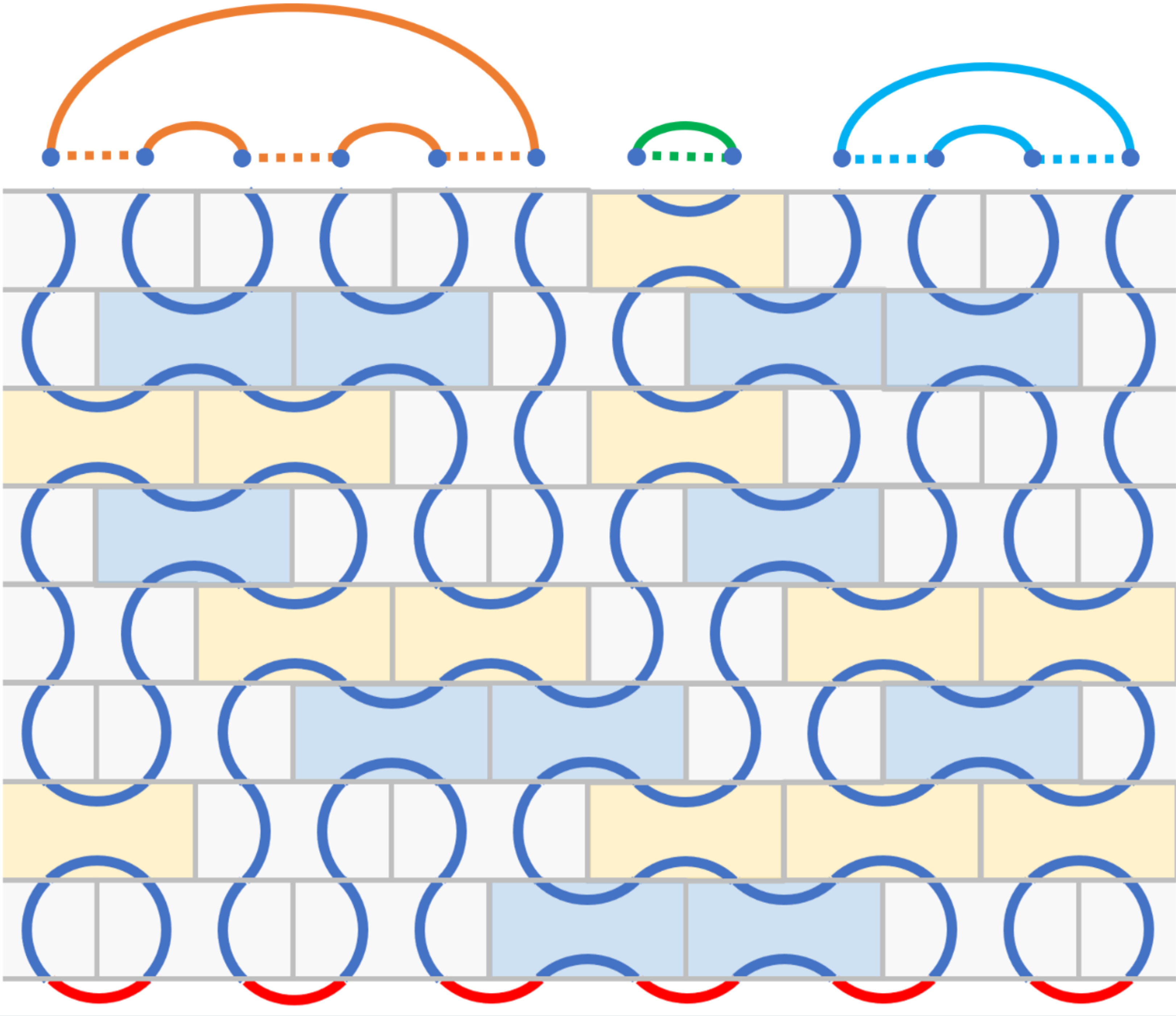}
}
\subfigure[]{
    \includegraphics[width=.3\textwidth]{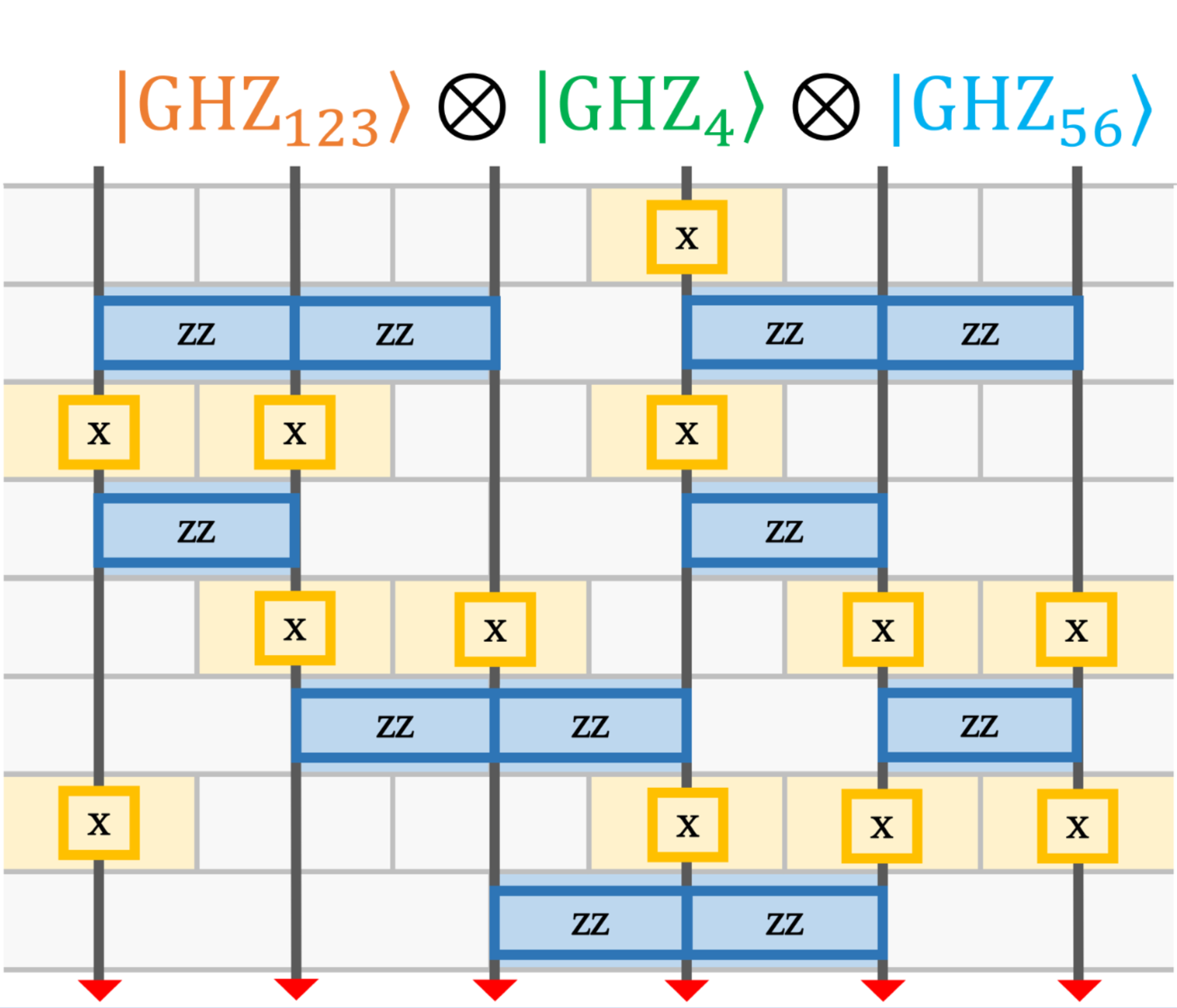}
}
\subfigure[]{
    \includegraphics[width=.3\textwidth]{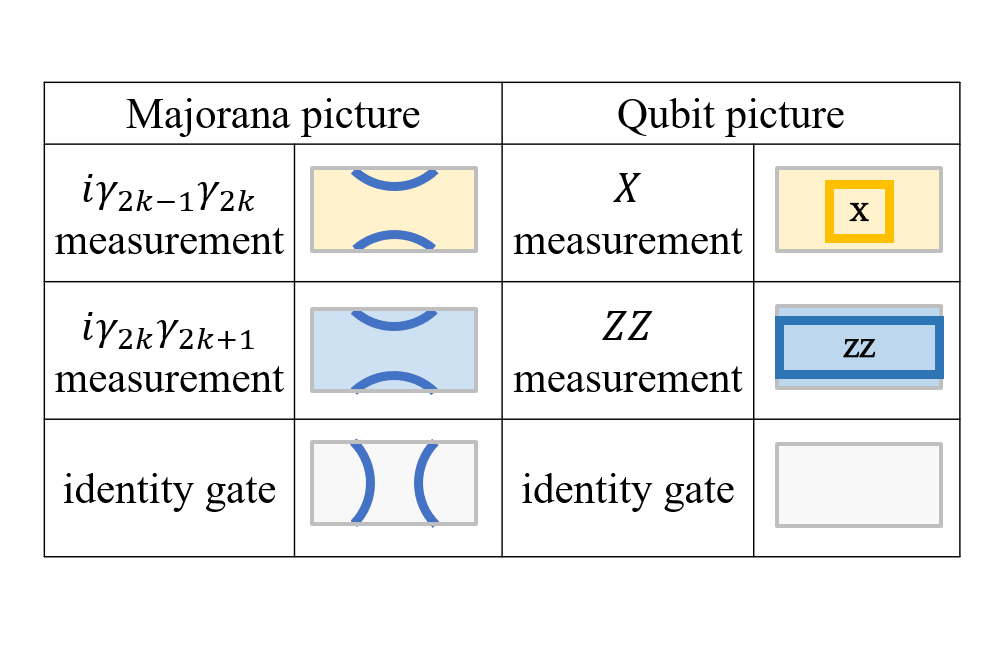}
}
\subfigure[]{
    \includegraphics[width=.3\textwidth]{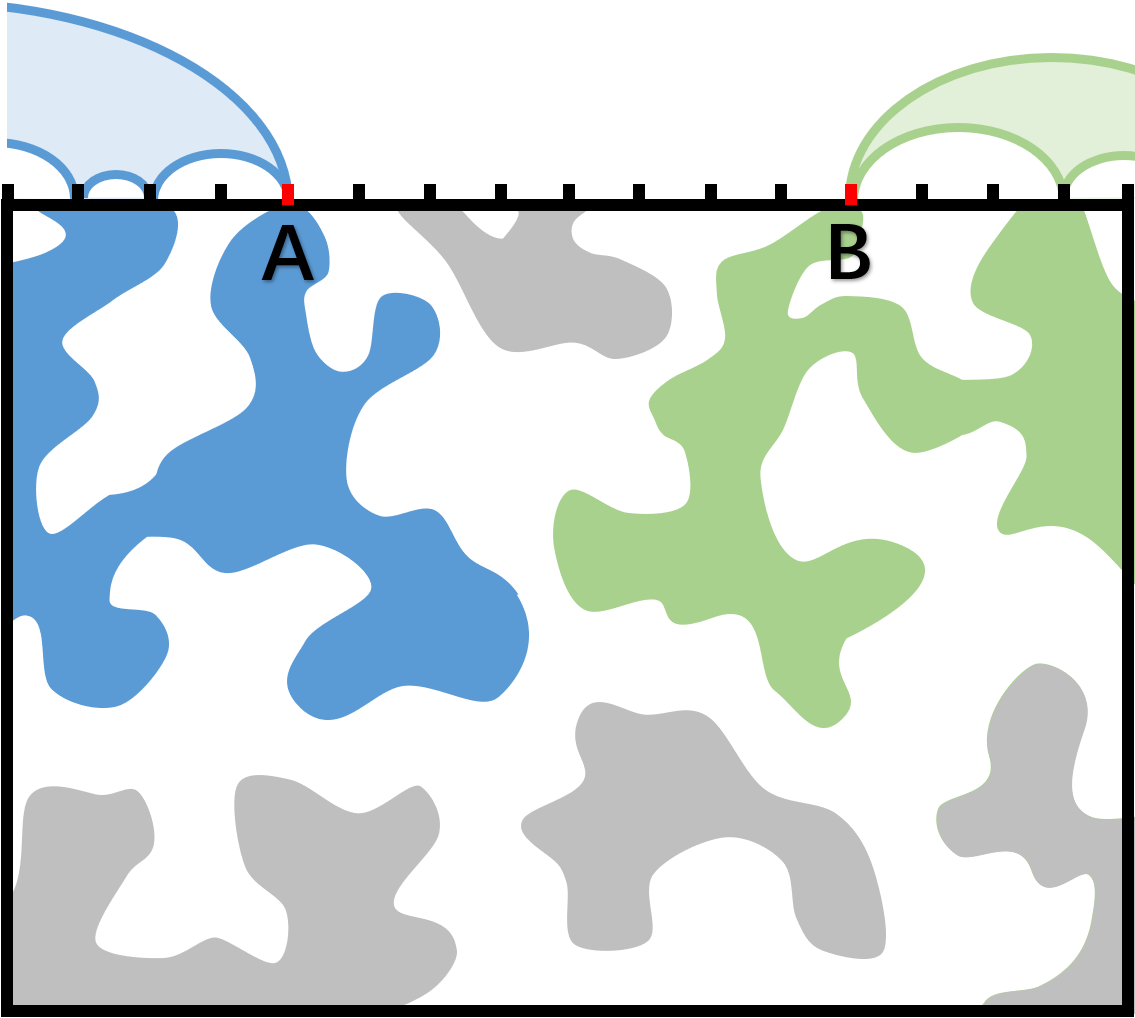}
}
\subfigure[]{
    \includegraphics[width=.3\textwidth]{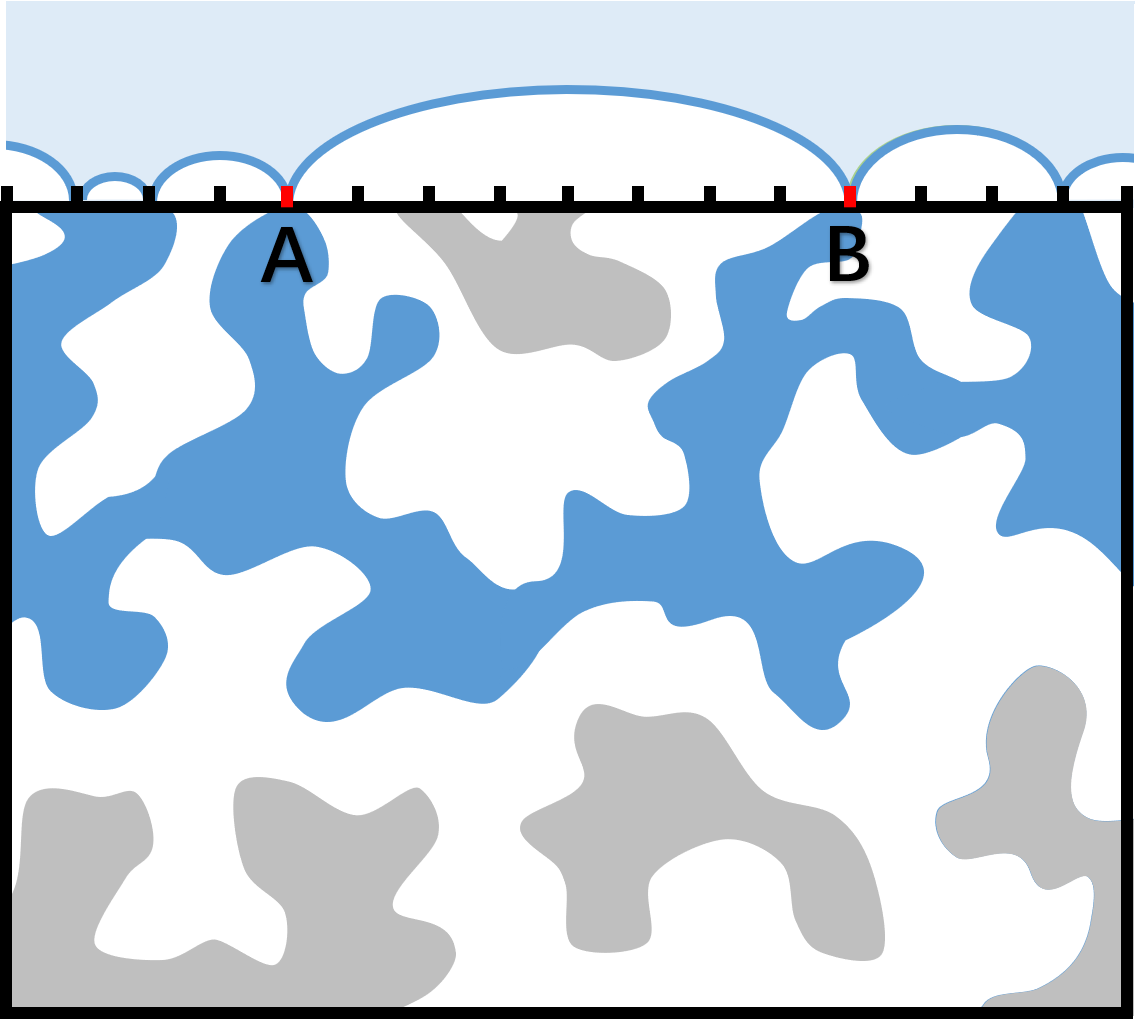}
}
\subfigure[]{
    \includegraphics[width=.3\textwidth]{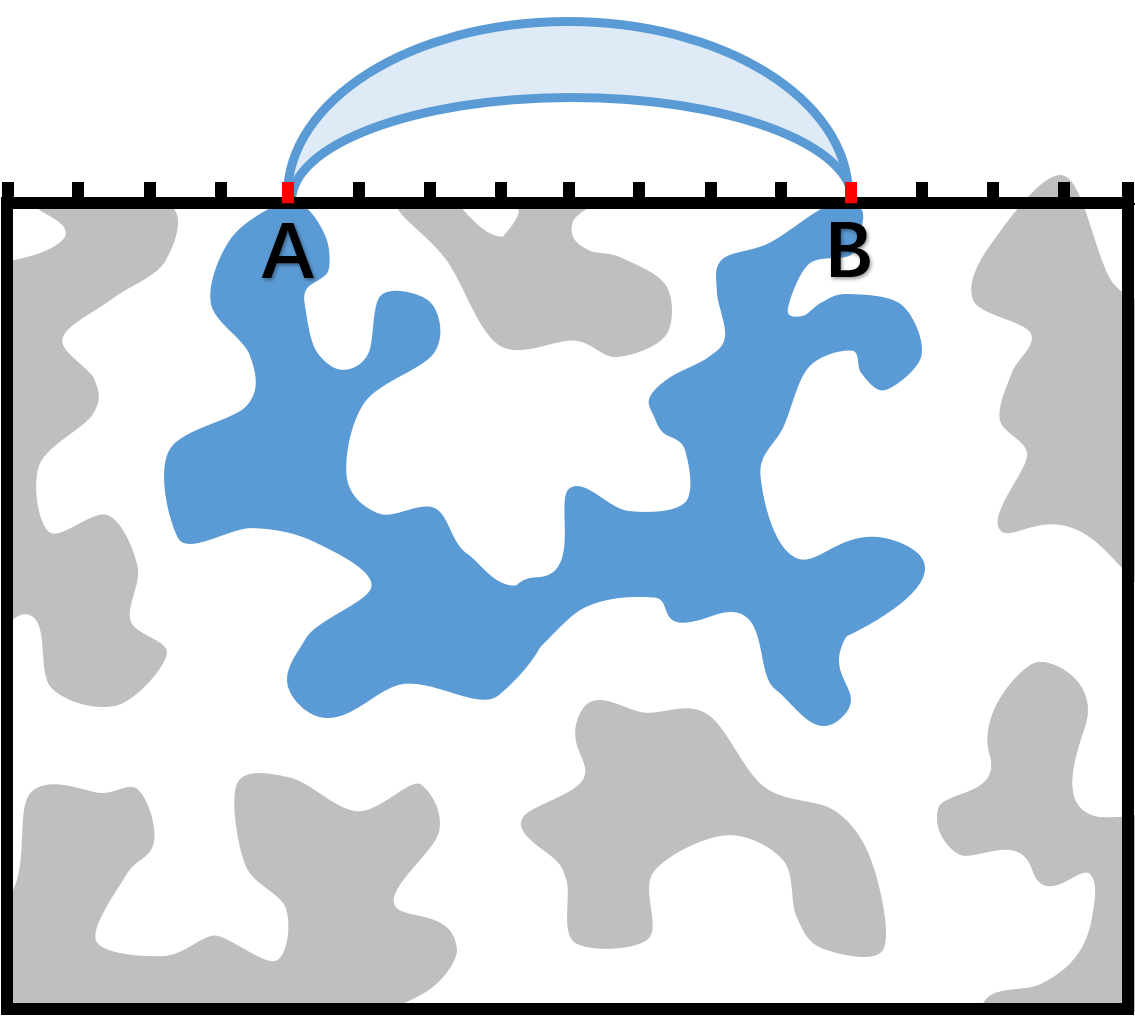}
}
\caption{(a) shows an instance of the measurement-only circuit (Sec.~\ref{sec:moc_majorana}) in the Majorana loop representation while (b) is the same circuit in the qubit representation.
In both figures the red part represents the initial product state as defined in Eq.~(\ref{eq:init_state}), and each rectangular corresponds to an operation happening at that spacetime location, which can either be a measurement ``$\TLE$'' or an identity operator ``$\TLI$''.
The possible operations in two representations are one-to-one related by the Jordan-Wigner transformation, as shown in (c). 
In (a), the output state can be determined by the pairing pattern of the end-points on the upper-boundary. By ``fusing'' Majorana pairs on odd bonds (dashed lines), pairing arcs form several disjoint loops with disjoint supports, each of which corresponds to a GHZ cluster in the output state in qubit representation.
{
The Majorana loops are cluster boundaries in percolation, and we provide several examples in (d-f).
Two qubits on the boundary (highlighted $A$ and $B$ here) belong to the same GHZ cluster if and only if they are connected to the same percolating cluster (colored regions) of the bulk.
In (d), $A$ and $B$ are disconnected, and $\frac{1}{2}I_{A,B} = N_{A,B} = 0$.
In (e), $A$ and $B$ belong to the same GHZ cluster which also touches the boundary at other qubits, and $\frac{1}{2}I_{A,B} = \frac{1}{2} \ln 2$, $ N_{A,B} = 0$.
In (f), $A$ and $B$ belong to the same GHZ cluster which does not contain other boundary qubits, and $\frac{1}{2}I_{A,B} = N_{A,B} = \ln 2$.
}
}
    \label{fig:loop}
\end{figure*}

In this section, we consider a stabilizer circuit model in $(1+1)$ space-time dimensions, composed of one-qubit $X_j$ and two-qubit $Z_j Z_{j+1}$ Pauli measurements~\cite{hsieh_sang_2004_protected, buechler2006projectiveTFIM}.
Under a Jordan-Wigner transformation, the dynamics is equivalently described by a model of Majorana fermions undergoing measurements of local fermion parity (with dimerized probabilities on even and odd links), and the circuit dynamics is mapped to a loop ensemble of non-crossing Majorana worldlines~\cite{nahum1911majorana}.
The critical point is in the universality class of critical percolation,
and we map MI and MN to known boundary correlation functions in order to illustrate the difference between bipartite and tripartite entanglement of the critical wavefunctions.

As in previous models~\cite{nahum2018hybrid, li1808hybrid} that exhibit measurement-induced criticality, we compute entanglement measures -- von Neumann entanglement entropy and mutual information, and logarithmic negativity -- for each quantum trajectory (as specified by the circuit geometry and measurement outcomes), and average that quantity against the ensemble of trajectories, weighted by their classical and quantum (Born) probabilities.
This applies to all models in this paper.
However, for the most part we will consider stabilizer circuits, for which the entanglement measures do not depend on the R\'{e}nyi indicies or the measurement outcomes.
Thus, we only need to average over circuit geometries and gates, and may well assume measurement outcomes all take a fixed value (say $+1$) whenever nondeterministic, for all Pauli strings measured.


\subsection{The Majorana loop representation and the Jordan-Wigner transformation}


For the sake of presentation, we start from the simpler Majorana representation (see Fig.~\ref{fig:loop}(a)), as introduced in Ref.~\cite{nahum1911majorana}.
Consider an even number of Majorana fermions in one dimension, $\gamma_{1}, \ldots, \gamma_{2L}$, placed at sites $1, \ldots, 2L$, and we take periodic boundary condition for the chain.
Initially, the system is prepared in a ``product'' state 
\env{align}{
    \label{eq:init_state}
    \ket{\psi_{\rm in}} = \ket{n_{12}} \otimes \ket{n_{34}} \otimes \ldots \otimes \ket{n_{2L-1, 2L}},
}
where $n_{ij} = (1 + i \gamma_i \gamma_j)/2 = 0, 1$ is the fermion number operator.
We say the initial state is defined by the ``pairing pattern'' $\{(1,2), (3,4), \ldots, (2L-1, 2L)\}$, and the occupation of each pair is irrelevant for our purposes.
As we will see, throughout the dynamics we are interested in, the state is always defined by a pairing pattern of the Majoranas.

The dynamics is defined so that at the $t$-th time step, we consider all nearest neighbor bonds depending on the parity of $t$, and measure the fermion number operator defined on this bond, at \emph{dimerized} probability.
Specifically, if $t$ is odd, each of $n_{12}, n_{34},  \ldots, n_{2L-1, 2L}$ is measured independently with probability $q$; and if $t$ is even, each of $n_{23}, n_{45},  \ldots, n_{2L-2, 2L-1}, n_{2L, 1}$ is measured indepedently with probability $(1-q)$.

This dynamics can be succinctly described in the Majorana representation, in particular using the pairing diagram,
for which we have the following update rule,
\env{align}{
\label{eq:pairing_pattern}
    \{\ldots, (i, j), (k, l), \ldots\} \xrightarrow[\text{measure $n_{jk}$}]{}
    \{\ldots, (j, k), (i, l), \ldots\}.
}

In Fig.~\ref{fig:loop}(a), we introduced a graphical representation of the evolution of the pairing pattern in the circuit spacetime.
At each spacetime site, the two possible events -- no measurement ``$\TLI$'' or measurement ``$\TLE$'' -- corresponds to ``rearrangement vertices'' of the pairing pattern for two Majoranas involved (see Fig.~\ref{fig:loop}(c)).
With this mapping, the circuit dynamics now generates an ensemble of completely-packed non-crossing loops.
We have both \emph{closed} loops (each with weight $1$, as a result of the stochastic process) in the bulk of the circuit, as well as exactly $L$ \emph{open} loops\footnote{In the rest of this paper, we will refer to open loops as ``arcs''.} with endpoints at the upper boundary of the circuit, where Majorana fermions at the endpoints of the same arc are paired.
This ensemble has the same distribution as cluster boundaries in two-dimensional percolation, and becomes {critical} at $q = q_c = 1/2$.
When $q$ is detuned from criticality, the loops will acquire a finite typical size, and corresponds to the super- or sub-critical phase of percolation.

\begin{figure}
    \centering
    \includegraphics[width=.49\textwidth]{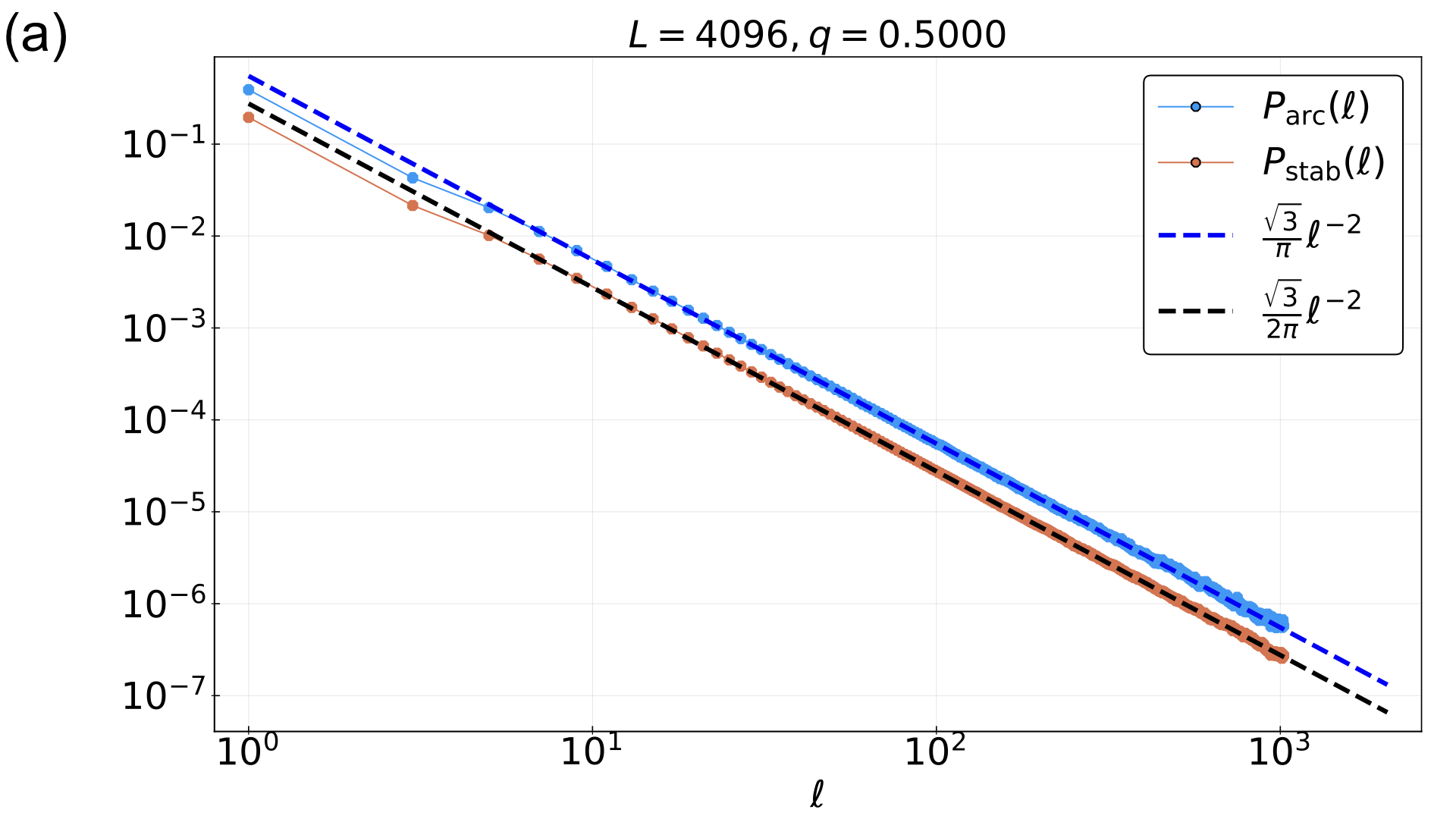}
    \includegraphics[width=.49\textwidth]{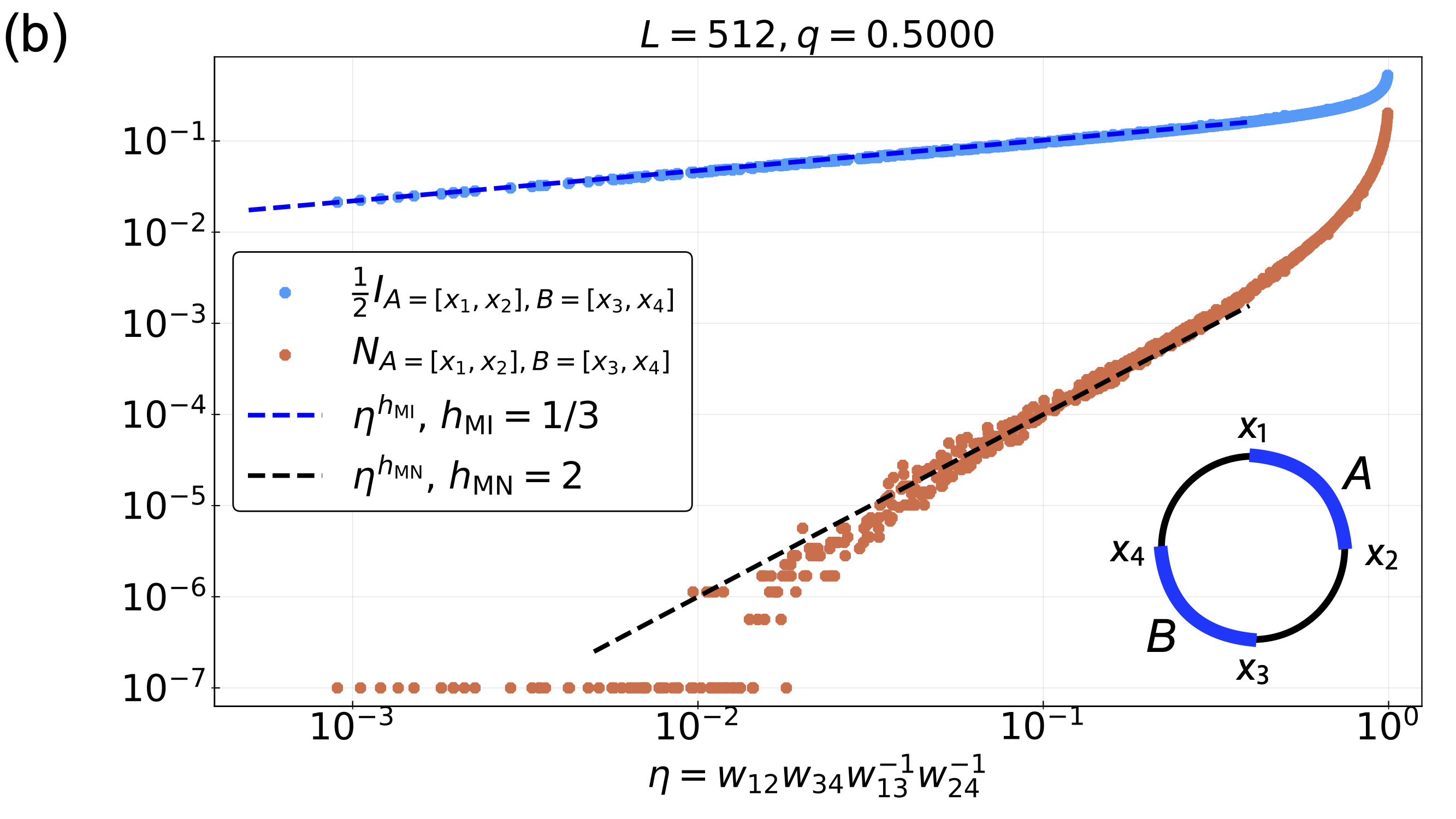}
    \caption{Numerical simulation results for the measurement-only circuit with periodic boundary condition. (a) The equilibrium length distributions of Majorana arcs and stabilizers. (b) Mutual information and mutual negativity between two disjoint, distant intervals $A$ and $B$ (separated by $C = \ovl{A \cup B}$) as a function of the cross ratio $\eta=w_{12}w_{34}w^{-1}_{13}w^{-1}_{24}$.}
    \label{fig:perc}
\end{figure}

In the rest of this section, we will focus on the critical point, $q = q_c$, where the critical loop ensemble is described by a conformal field theory (CFT).
Here, of particular interest is the universal length distribution of the Majorana arcs.
In particular, the \emph{density} of arcs of length $\ell$ follows an inverse square law with a universal coefficient~\cite{nahum1911majorana, cardy9911linking, cardy0103lecture, buechler2006projectiveTFIM, saleur2007VBEE},
\env{align}{
    \label{eq:P_arc_perc}
    P_{\rm arc}(\ell) \approx K \ell^{-2}, \text{ where } K = \frac{\sqrt{3}}{\pi} \approx 0.55.
}
We reproduce this result numerically in Fig.~\ref{fig:perc}(a).

Under a Jordan-Wigner transformation, Majorana fermions become Pauli string operators,
\env{align}{
    \gamma_{2j-1} =& \( \prod_{i < j} X_i \) Z_j, 1 \le j \le L,\\
    \gamma_{2j} =& \( \prod_{i < j} X_i \) Y_j, 1 \le j \le L.
}
We have qubits indexed by $Q = \{1, \ldots, L\}$, and the measured fermion density operators on ``odd'' and ``even'' bonds are respectively
\env{align}{
    i \gamma_{2j-1} \gamma_{2j} =&\, X_j, \\
    i \gamma_{2j} \gamma_{2j+1} =&\, Z_j Z_{j+1}.
}
The circuit model thus maps to a stabilizer circuit, with interleaving layers of single-site $X$ measurements, and nearest-neighbor $ZZ$ measurements (see Fig.~\ref{fig:loop}(b)).
The initial state has stabilizers $\{X_1, \ldots, X_L\}$;
and at each time step the stabilizers are just $i\gamma_{i} \gamma_{j}$ for each pair $(i,j)$ of Majoranas connected by an arc.
There is thus a one-to-one correspondence between the stabilizers and the Majorana arcs.
Moreover, the stabilizers are already in the ``clipped gauge''~\cite{nahum2017KPZ, li1901hybrid}.
We have the ``stabilizer length distribution'' (see Fig.~\ref{fig:perc}(b))
\env{align}{
    \label{eq:P_stab_arc}
    P_{\rm stab}(\ell) \approx P_{\rm arc}(2\ell - 1) + P_{\rm arc}(2\ell) \approx \frac{K}{2} \ell^{-2}.
}
This immediately leads to the entanglement entropy of a contiguous segment of qubits $A \subseteq Q$ when $1 \ll |A| \ll L$~\cite{nahum2017KPZ, li1901hybrid},
\env{align}{
\label{eq:h_ab_GHZ}
    S_A \coloneqq S_{\rm vN}(\rho_A) \approx \(\frac{K}{2} \ln 2\)\ln |A|.
}
Here we can idenfity the coefficient of the logarithmic entanglement entropy as \emph{twice}\footnote{In particular, the boundary conformal fields here are ``domain wall operators'' that signify the change of boundary conditions, and there is one such operator on each endpoint of $A$, accounting for the factor of two~\cite{li2003cft}.}
the scaling dimension of a boundary conformal field~\cite{vasseur2018rtn, andreas2019hybrid, li2003cft}, which we call $h_{\rm EE}$
\env{align}{
    h_{\rm EE} = \frac{\sqrt{3}}{4\pi} \ln 2 \approx 0.096.
}

In the following sections it will be useful to have a qubit representation of the dynamical states.
It can be shown -- most easily by following the stabilizer evolution -- that the state is alway a direct product of GHZ states (see Fig.~\ref{fig:loop}(b)), each supported on a subset of qubits, where the subsets form a partition (which changes over time) of $Q$~\cite{buechler2006projectiveTFIM}.
Two qubits belong to the same GHZ cluster if and only if they belong to the same connected component in percolation (see more configurations in Fig.~\ref{fig:loop}(d-f), where connected components are colored); a procedure of finding connected components is detailed in Fig.~\ref{fig:loop}.
For the example in Fig.~\ref{fig:loop}(b) we have 
\env{align}{
    \label{eq:perc_state_GHZ}
    \ket{\psi_{\rm fin}} =& \frac{\ket{\ua\ua\ua}_{123} + \ket{\da\da\da}_{123} }{\sqrt{2}} \nn
    & \otimes  \frac{ \ket{\ua}_{4} + \ket{\da}_{4} }{\sqrt{2}} \otimes  \frac{ \ket{\ua\ua}_{56} + \ket{\da\da}_{56} }{\sqrt{2}}.
}
This state has the following stabilizers
\env{align}{
    \{ X_1 X_2 X_3, Z_1 Z_2, Z_2 Z_3 \}
    \cup \{X_4\}
    \cup \{X_5 X_6, Z_5 Z_6\}.
}
Unlike the Majoranas which are supported on the interval endpoints, the stabilizers can have nontrivial contents between the endpoints, due to the ``strings attached'' in the Jordan-Wigner transformation.
Since it is a stabilizer state, it follows the general pattern in Eq.~\eqref{eq:structural_thm} for any tri-partition of $Q$, with appropriately chosen $U_{A}, U_B$ and $U_C$.

\subsection{Mutual information between segments \label{sec:moc_majorana_mi}}

We consider the mutual information between two disjoint subregions $A, B \subset Q$, where $A = [x_1, x_2]$, $B = [x_3, x_4]$, and $x_1 \le x_2 \le x_3 \le x_4$ (see inset of Fig.~\ref{fig:perc}(b)).

We first consider the case when $A$ and $B$ both contain only one qubit ($x_{12} = x_{34} = 1$), and when they are far apart ($x_{13} = x_{24} \gg 1$).
From the state decomposition in Eq.~\eqref{eq:perc_state_GHZ}, we see that $I_{A,B}$ is 
{nonzero}
if they belong to the same cluster (i.e. when they are in the same GHZ state; see Fig.~\ref{fig:loop}(e, f)), but zero otherwise (see Fig.~\ref{fig:loop}(d)).
{More precisely, $I_{A,B}$ is $\ln 2$ if the GHZ state has more than two qubits (Fig.~\ref{fig:loop}(e)), and $2\ln 2$ if the GHZ state has exactly two qubits (Fig.~\ref{fig:loop}(f)).
As we will see in Sec.~\ref{sec:moc_majorana_mn}, the latter contribution is subdominant, and to leading order the MI will be proportional to the probability that $A$ and $B$ belong to the same cluster (the crossing probability),
}
\env{align}{
    & I_{A = [x_1, x_2 = x_1 + 1],B = [x_3, x_4 = x_3 + 1]}\nn
    \approx &\, (\ln 2) \times  \mathbb{P}(A, B \text{ belong to the same GHZ cluster}) \nn
    \propto&\, \eta^{1/3}, \text{ where } \eta = \frac{w_{12} w_{34}}{w_{13} w_{24}} \to 0.
}
Here $w_{ij} \coloneqq \sin(\frac{\pi}{L} x_{ij})$ is the chord distance, as appropriate for a system with periodic boundary condition.
The last line follows from Cardy's formula for crossing probabilities in critical percolation~\cite{cardy0103lecture}, and we are quoting the leading order behavior at long distances.
To compare the MI for different models, we adopt the following more intuitive notation, which reads ``the mutual information exponent'',
\env{align}{
\label{eq:h_MI_GHZ}
    h_{\rm MI} = 1/3.
}
This result can also be obtained from knowledge of the loop ensemble~\cite{buechler2006projectiveTFIM}: $A$ and $B$ belong to the same GHZ cluster if and only if no Majorana arc has one endpoint inside the region bounded by $A$ and $B$ and the other endpoint outside.

In general, $A$ and $B$ can have more than one qubit, and the MI becomes the expected number of disjoint connected components spanning $A$ and $B$, up to a factor $\ln 2$.
It remains a function of the cross ratio, when the endpoints of $A$ and $B$ are now varied arbitrarily,
\env{align}{
    & I_{A = [x_1, x_2],B = [x_3, x_4]}\nn
    =&\, (\ln 2) \sum_{n=1}^{\infty} n \cdot \mathbb{P}(\text{exactly $n$ GHZ clusters span $A$ and $B$}) \nn
    =& F_I(\eta),
}
where the cross ratio $\eta$ can take its value in $[0, 1]$.
This fact reflects the conformal invariance of critical percolation.
When $A \cup B$ is almost the entire system $Q$ (or equivalently, when $\eta \to 1$), we have 
\env{align}{
    & I_{A = [x_1, x_2],B = [x_3, x_4]}\nn
    \approx&\, 2 S_A \nn
    \approx&\, 4 h_{\rm EE} \ln|A| \nn
    \approx& -2 h_{\rm EE} \ln(1-\eta),
}
where we used $1-\eta \propto |A|^{-2}$ and $S_A \approx S_B$, $S_{A\cup B} \approx S_Q = 0$ as $\eta \to 1$.

Summarizing,
\env{align}{
    \label{eq:def_F_I}
    & \frac{1}{2} I_{A = [x_1, x_2],B = [x_3, x_4]} \nn
    =&\, F_I(\eta)
    \sim \env{cases}{
        \eta^{h_{\rm MI}}, & \eta \to 0\\
        - h_{\rm EE} \ln(1-\eta), & \eta \to 1
    }.
}
Our numerical results for $I_{A,B}$ with a varying $\eta$ is shown in Fig.~\ref{fig:perc}(b).

\subsection{Mutual negativity between segments \label{sec:moc_majorana_mn}}


Again, we start by focusing on the simple case when $|A| = |B| = 1$.
From the state decomposition in Eq.~\eqref{eq:perc_state_GHZ}, we see that $A$ and $B$ has MN $N_{A,B} = \ln 2$ if these two qubits constitute a two-qubit GHZ state (i.e. an EPR pair; see Fig.~\ref{fig:loop}(f)) but unentangled from everything else; but zero otherwise -- either when $A$ and $B$ are not in the same cluster (Fig.~\ref{fig:loop}(d)), or they are in a cluster with at least three qubits (Fig.~\ref{fig:loop}(e)).
Thus
\env{align}{
    & N_{A = [x_1, x_2 = x_1 + 1],B = [x_3, x_4 = x_3 + 1]}\nn
    =&\, (\ln 2) \times  \mathbb{P}(A, B \text{ form an EPR pair}).
}
Since Majorana arcs represent cluster boundaries, qubits $i$ and $j$ are in an EPR pair if and only if arcs $(2i-1, 2j)$ and $(2i, 2j-1)$ are both present in the pairing pattern in the Majorana representation; in other words, there is one bit of nonzero MN if and only there is a ``double arc'' configuration, with two Majorana arcs nested together (see Fig.~\ref{fig:loop}(f)).
This probability may be related to a boundary correlation function between stress-energy tensors on $A$ and on $B$ in the underlying CFT (see Appendix~\ref{app:TLalgebra})
and therefore
\env{align}{
    & N_{A = [x_1, x_2 = x_1 + 1],B = [x_3, x_4 = x_3 + 1]}\nn
    \propto&\, \(x_{13}\)^{-4} \nn
    \propto&\, \eta^2, \text{ where } \eta \to 0.
}
Here the scaling dimension $2$ is that of the stress-energy tensor, which we identify as the ``mutual negativity exponent'',
\env{align}{
    \label{eq:h_MN_GHZ}
    h_{\rm MN} = 2.
}
For those familiar with Ref.~\cite{nahum2018hybrid}, this is the same exponent for the decay of ``first-passage mutual information'' at long distances.


It is clear from this consideration that the inequality Eq.~\eqref{eq:ineq_NAB_IAB} holds for this model (as it must); consequently $h_{\rm MN} \ge h_{\rm MI}$, and here they are not equal.
This reflects different operator contents in the leading conformal blocks of four-point functions $I_{A,B}$ and $N_{A,B}$.

In the more general situation where $A$ and $B$ can contain more than one qubit, $N_{A,B}$ remains a function of the cross ratio $\eta$ due to conformal invariance.
In particular, it is the expected number of EPR pairs spanning $A$ and $B$.
Furthermore, it has the same asymptotics as $\frac{1}{2} I_{A,B}$ as $\eta \to 1$, since then $\rho_{A \cup B}$ is approaching a global pure state for which  equality in Eq.~\eqref{eq:ineq_NAB_IAB} is saturated.
Summarizing,
\env{align}{
    \label{eq:def_F_N}
    & N_{A = [x_1, x_2],B = [x_3, x_4]} \nn
    =&\, F_N(\eta) \sim \env{cases}{
        \eta^{h_{\rm MN}}, & \eta \to 0\\
        - h_{\rm EE} \ln(1-\eta), & \eta \to 1
    }.
}
Our numerical results for $N_{A,B}$ with a varying $\eta$ is shown in Fig.~\ref{fig:perc}(b).

\subsection{Summary of this section and a few technical comments}

The purpose of this section was to analytically demonstrate and illustrate how MN and MI can be different.
Along the lines of the structure theorem in Sec.~\ref{sec:EN_stab}, but focusing on the percolation circuit, we see explicitly (using Eq.~\eqref{eq:perc_state_GHZ}) how MN ``filters out'' multi-partite entanglement and detects only direct EPR pairs while MI does not.
This difference is reflected in different geometrical conditions in percolation that respectively contributes to MN and MI, which leads to different $h_{\rm MI}$ and $h_{\rm MN}$.

Note that 
while in the Majorana fermion representation there seems to be only bipartite entanglement (where Majoranas are grouped up in pairs, and an arc can be drawn within each pair), the Jordan-Wigner transformation itself is non-local and introduces multi-partite entanglement into the qubit representation.
As a consequence, the MI and MN become slightly more complicated ``loop observables''.
In the next section, we will see how the same observables naturally defined for qubits become rather unnatural and difficult to treat in the loop model, upon introduction of additional ``crossing'' vertices.


\section{Relevant perturbations to percolation: hybrid circuit models \label{sec:hybrid_circuit}}

In this section we consider ``hybrid'' quantum circuit models, composed of \emph{both} unitary gates and measurements.
Again, we stay in $1+1$ spacetime dimensions, where we have $L$ qubits arranged on a regular array with periodic boundary condition.

\subsection{Completely-packed Majorana loop model with crossings \label{sec:CPLC}}

\begin{figure}[b]
    \centering
    \subfigure[]{
        \includegraphics[width=.35\textwidth]{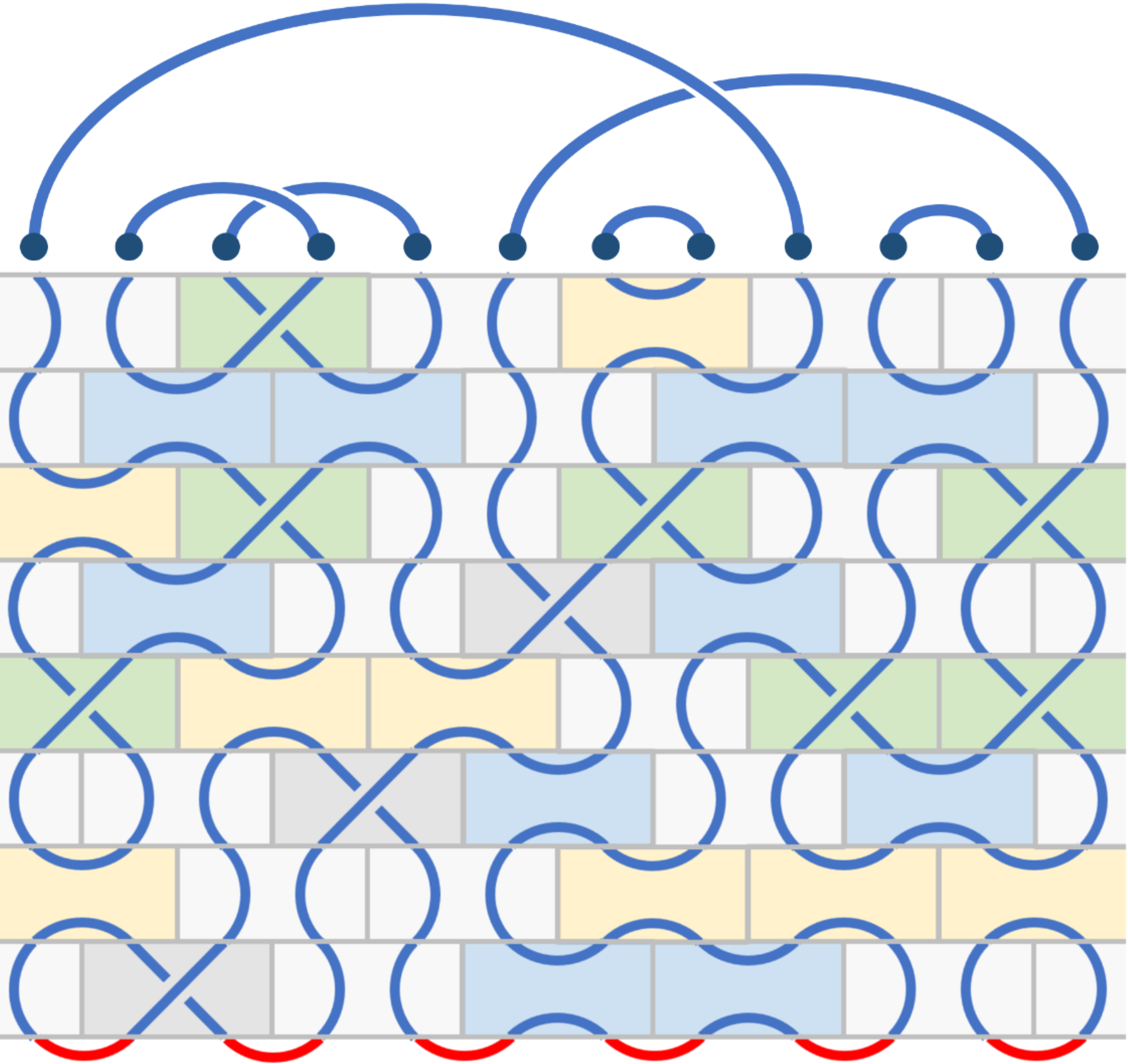}
        }
    \subfigure[]{
        \includegraphics[width=.35\textwidth]{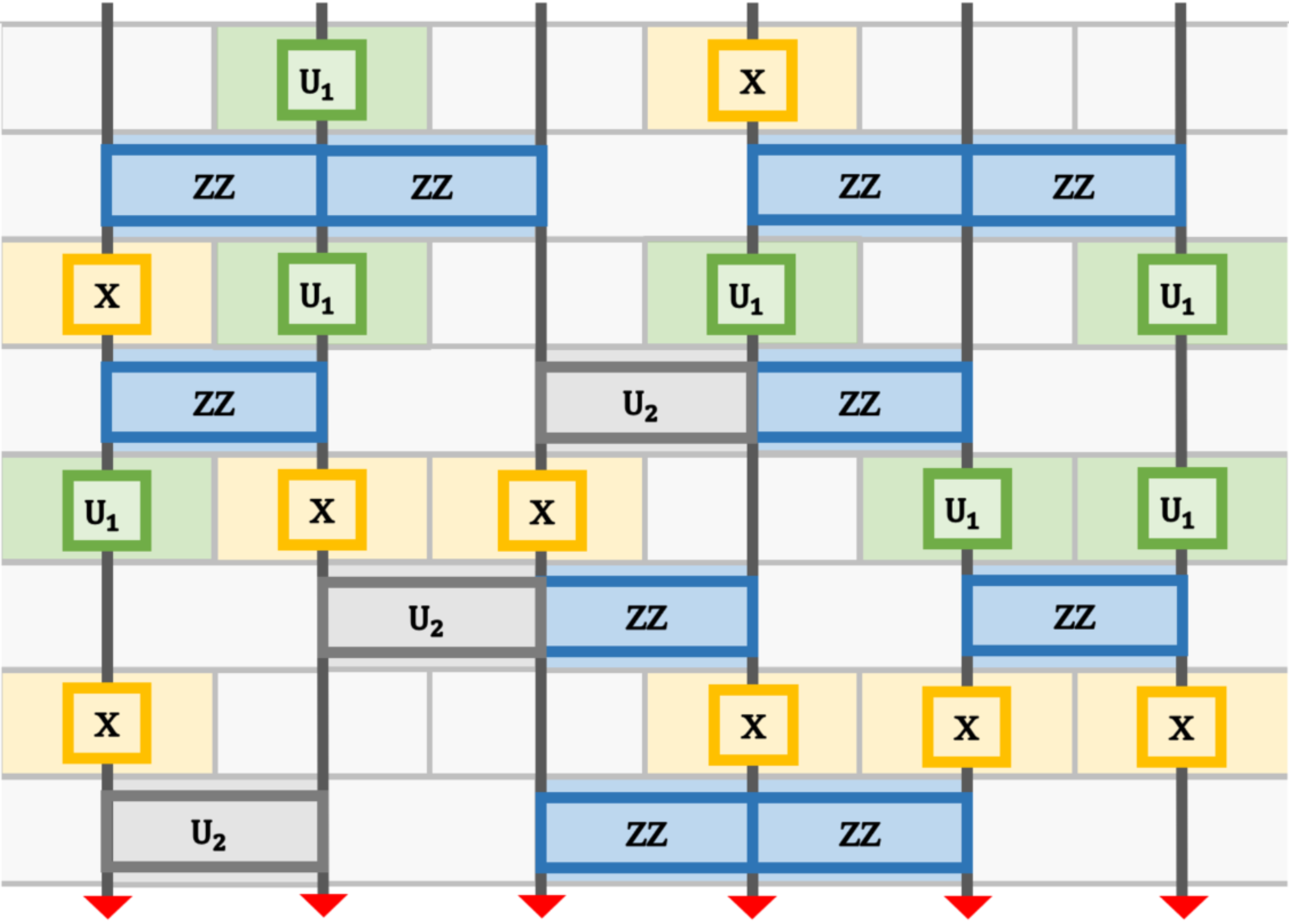}
    }
    \subfigure[]{
        \includegraphics[width=.25\textwidth]{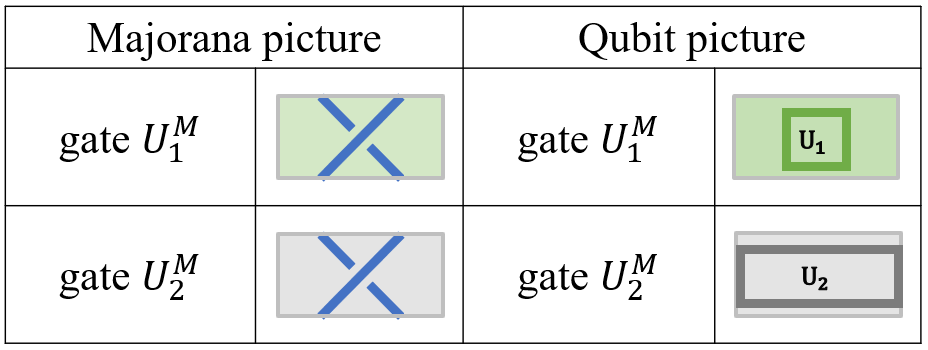}
        }
    \caption{(a) An circuit instance of the completely-packed Majorana loop model with crossings, and (b) its counterpart in the qubit representation after the Jordan-Wigner transformation. }
    \label{fig:crossing_loop}
\end{figure}

The first hybrid circuit is again a stabilizer circuit, and is a generalization of the measurement-only circuit in Sec.~\ref{sec:moc_majorana}.
Again, it is simpler to start from the Majorana representation.
In addition to measurements of one-qubit or two-qubit Paulis that represent ``swapping'' events,
we now have ``crossing'' events~\cite{jacobsen_read_saleur_2003_denseloop, nahum1303crossing} between neighboring Majorana fermions (see Fig.~\ref{fig:loop}(c) and Fig.~\ref{fig:crossing_loop}), which correspond to certain unitaries from the Clifford group.
Following how Majoranas transform under crossing, we have
\env{align}{
    U_1^{M}: \gamma_{2j-1} \leftrightarrow \gamma_{2j}\quad \Leftrightarrow_{JW}&  \quad U_1^{S}: Z_j \leftrightarrow Y_j; \\
    U_2^{M}: \gamma_{2j} \leftrightarrow \gamma_{2j+1}\quad \Leftrightarrow_{JW}&  \quad U_2^{S}: Y_j I_{j+1} \leftrightarrow  X_j Z_{j+1}.
}
Moreover, in the second case, we have Majoranas $\gamma_{2j-1} ( =Z_j I_{j+1})$ and $\gamma_{2j+2} ( =X_j Y_{j+1})$ left unchanged by the unitary gate.
These  are sufficient for uniquely specifying the Clifford unitaries $U_{1,2}$.
The loop ensemble will then be generated by these ``swapping'' and ``crossing'' vertices, as well as the ``identity'' vertex, with each closed loop assigned weight $1$.

Furthermore, we arrange the dynamics such that the Majorana fermion loop ensemble is exactly the completely-packed loop model with crossings (CPLC) from Ref.~\cite{nahum1303crossing}.
The stochastic process that generates the CPLC is precisely defined as follows (compare Fig.~\ref{fig:crossing_loop}), 
    At the $t$-th time step, we consider all nearest neighbor bonds depending on the parity of $t$. 
    Specifically, 
    \env{itemize}{
        \item
        If $t$ is odd, each of the pairs $(\gamma_1, \gamma_2)$, 
        $(\gamma_3, \gamma_4)$,  $\ldots$, $(\gamma_{2L-1}, \gamma_{2L})$ goes through the unitary ``crossing'' vertex with probability $p$, or the ``swapping'' vertex (a measurement of $\gamma_{2j-1} \gamma_{2j}$) with probability $(1-p)(1-q)$, or the ``identity'' vertex with probability $(1-p)q$;
        \item
        If $t$ is even, each of the pairs $(\gamma_2, \gamma_3)$, $(\gamma_4, \gamma_5)$,  $\ldots$, $(\gamma_{2L}, \gamma_{1})$ goes through the unitary ``crossing'' vertex with probability $p$, or the ``swapping'' vertex (a measurement of $\gamma_{2j} \gamma_{2j+1}$) with probability $(1-p)q$, or the ``identity'' vertex with probability $(1-p)(1-q)$.
    }
Notice the dimerization of the measurement probability on even and odd links.
This model reduces to the one in Fig.~\ref{fig:loop} when $p = 0$.

\begin{figure}
    \centering
    \includegraphics[width=0.4\textwidth]{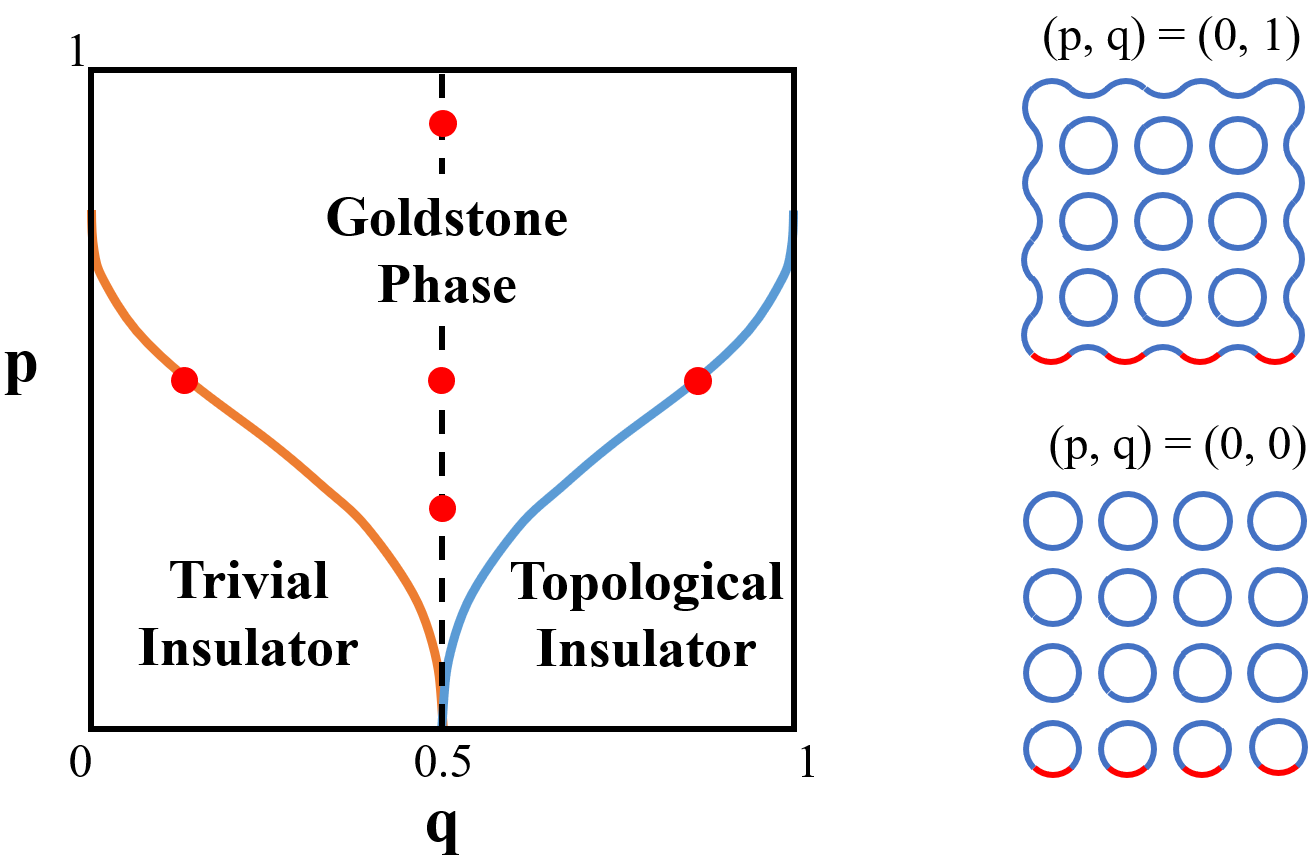}
    \caption{The phase diagram for the completely packed Majorana loop model with crossings (CPLC), which is composed of a Goldstone phase, an topological insulating phase, and an trivial insulating phase. Red dots mark the points that we studied numerically in \ref{sec:CPLC}. The way we identity Majorana pairs into qubits breaks the full lattice translation symmetry as well as the equivalence of the two short loop phases. Particularly, the loop configurations correspond to points $(p, q) = (0, 0)$ and $(p, q) = (0, 1)$ have the same bulk pattern, while the $(p, q) = (0, 1)$ one has an additional large macroscopic loop (``edge mode'') on the its boundary.
   }
    \label{fig:phase_diagram}
\end{figure}

The phase diagram of the CPLC is known from Ref.~\cite{nahum1303crossing}, and is reproduced in Fig.~\ref{fig:phase_diagram}.
Away from the boundaries of the phase diagram, the CPLC is described by the $\mathbb{RP}^{n-1}$ sigma model in the replica limit $n \to 1$.
Here, the loop crossings are relevant perturbations to the percolation critical point at $p=0, q=1/2$, which broadens into a critical phase when $p > 0$ near $q = 1/2$ -- the low-temperature ``Goldstone phase'' of the sigma model, wherein the spin stiffness flows to infinity in the infrared.
As we increase the dimerization of $q$ on the two sublattices, the Goldstone phase goes into two ``insulating'' phases where the loops have a finite characteristic size.
The phase transitions to the insulating phases are driven by the proliferation of $\mb{Z}_2$ vortices of the $\mathbb{RP}^{n-1}$ spins.

Ref.~\cite{nahum1303crossing} provides  much analytic understanding and numerical results of the CPLC.
Here, we slightly extend the CPLC by interpreting the endpoints of open loops as Majorana fermions, and compute MI and MN in the qubit representation.
As a result, we obtain new critical exponents of the CPLC, both inside the Goldstone phase and at the phase transitions.

\subsubsection{The Goldstone phase \label{sec:CPLC_goldstone}}

\begin{figure}
    \centering
    \includegraphics[width=.49\textwidth]{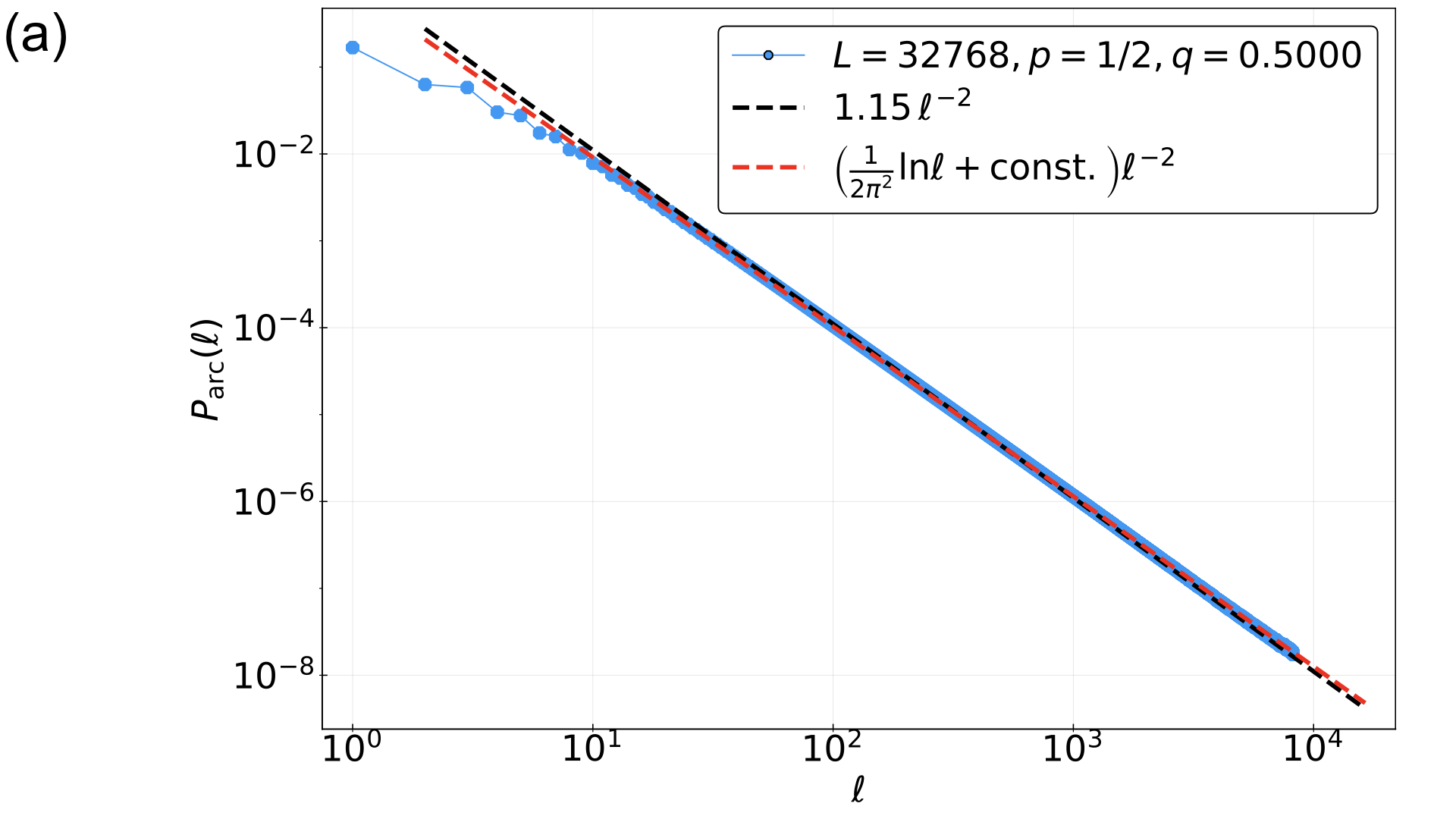}
    \includegraphics[width=.49\textwidth]{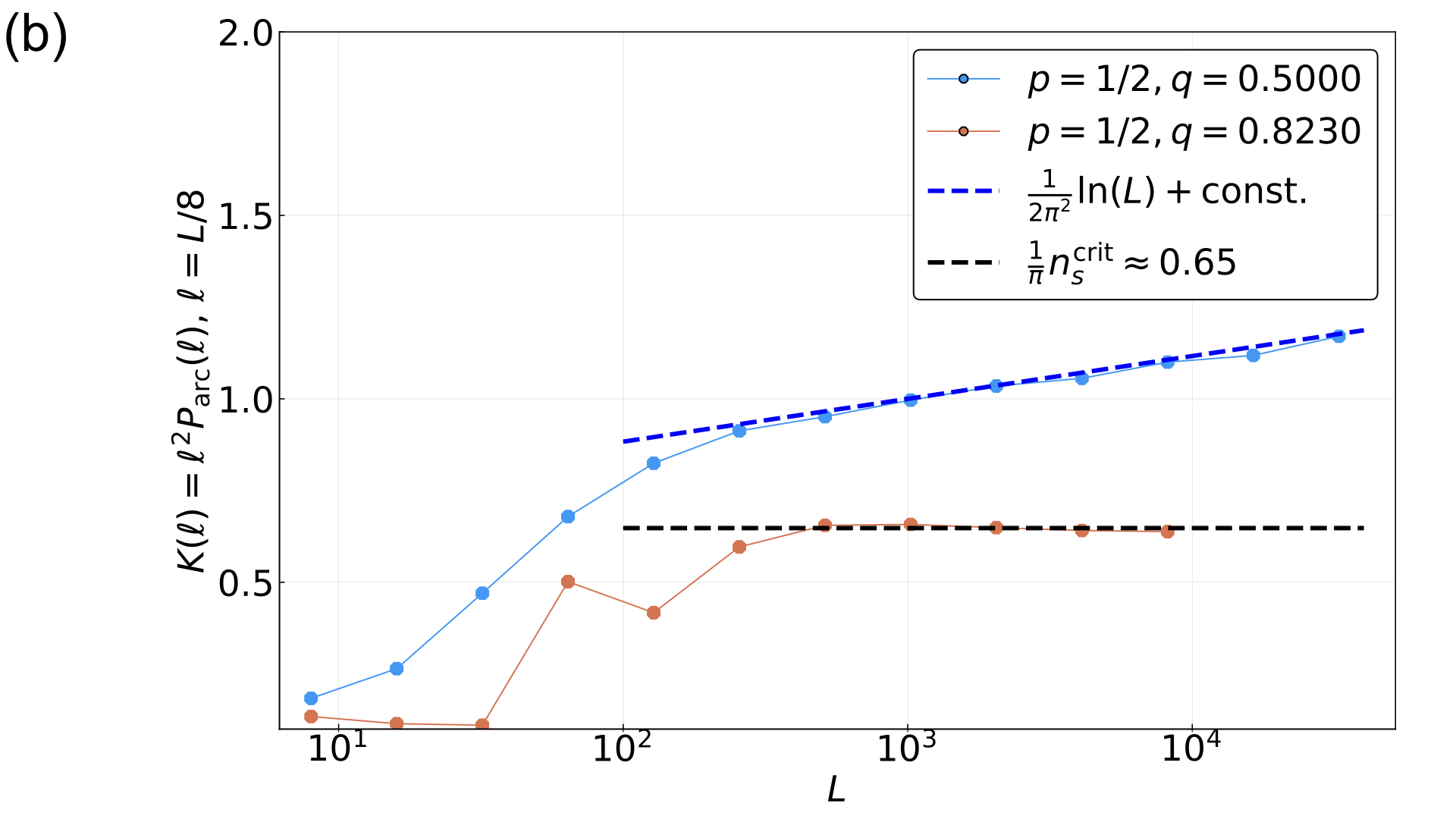}
    \caption{(a) The arc length distribution $P_{\rm arc}$ for CPLC model in the Goldstone phase $(p, q) = (0.5, 0.5)$. 
    {Here we fit $P_{\rm arc}$ in two ways, and notice that $K(\ell) \ell^{-2}$ (see Eqs.~(\ref{eq:P_arc_CPLC}, \ref{eq:K_ell_flow})) works slightly better than $\ell^{-2}$ for the system size $L = 2^{15}$ we have here.
    Their difference is made clearer in (b), where we plot}
    the $L$ dependence of $K(\ell = L/8)$ 
    {at different points in the phase diagram.}
    Within the Goldstone phase at $(p, q) = (0.5, 0.5)$,  $K(\ell = L/8)$ depends logarithmically on $L$. While on the critical line $(p, q) = (0.5, 0.823)$, $K(\ell = L/8)$ is {a universal constant proportional to the critical spanning number~\cite{nahum1303crossing} (see also Eq.~\eqref{eq:h_EE_CPLC}).}
    }
    \label{fig:cplc_arc}
\end{figure}

\begin{figure}
    \centering
    \includegraphics[width=.49\textwidth]{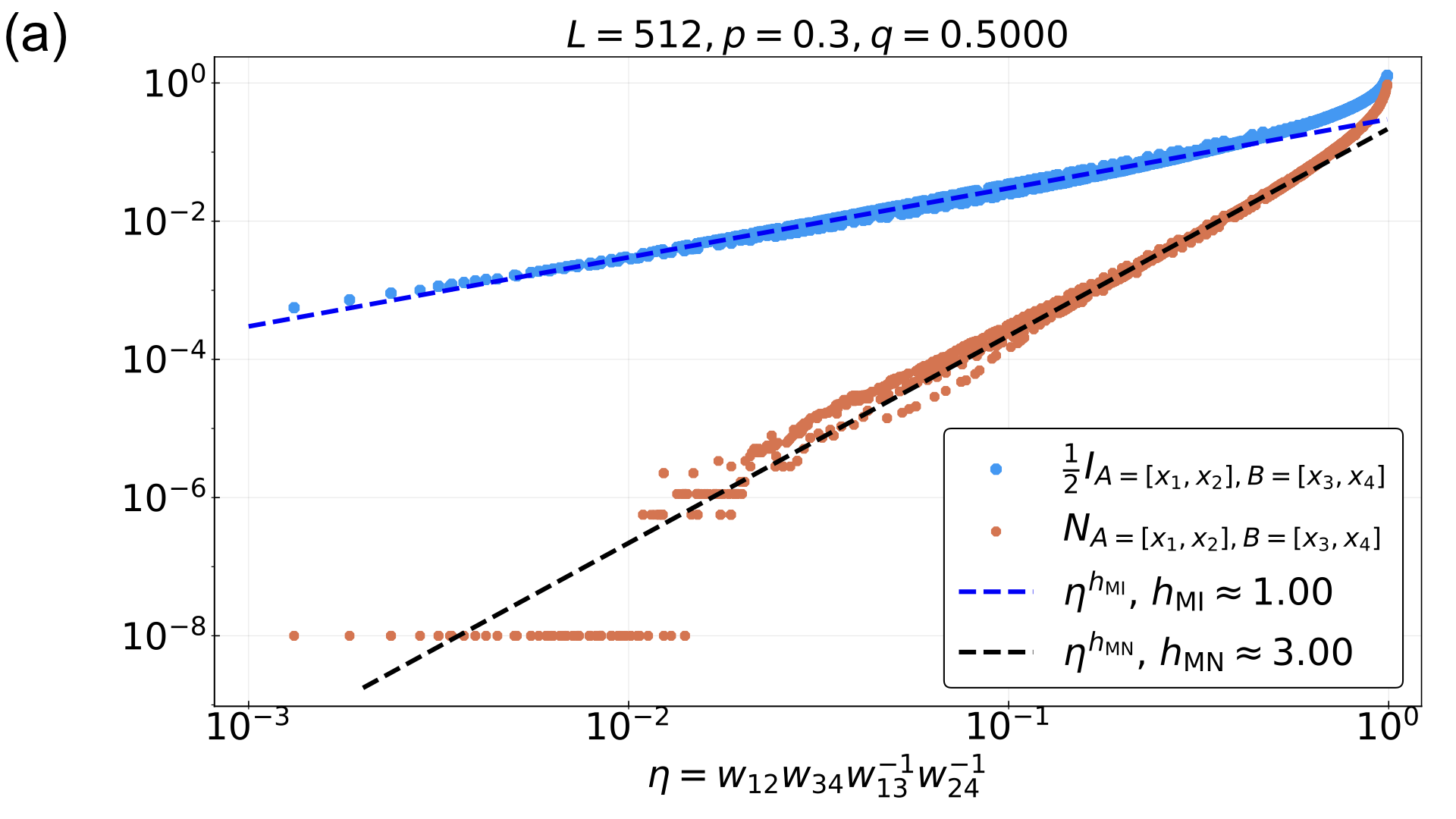}
    \includegraphics[width=.49\textwidth]{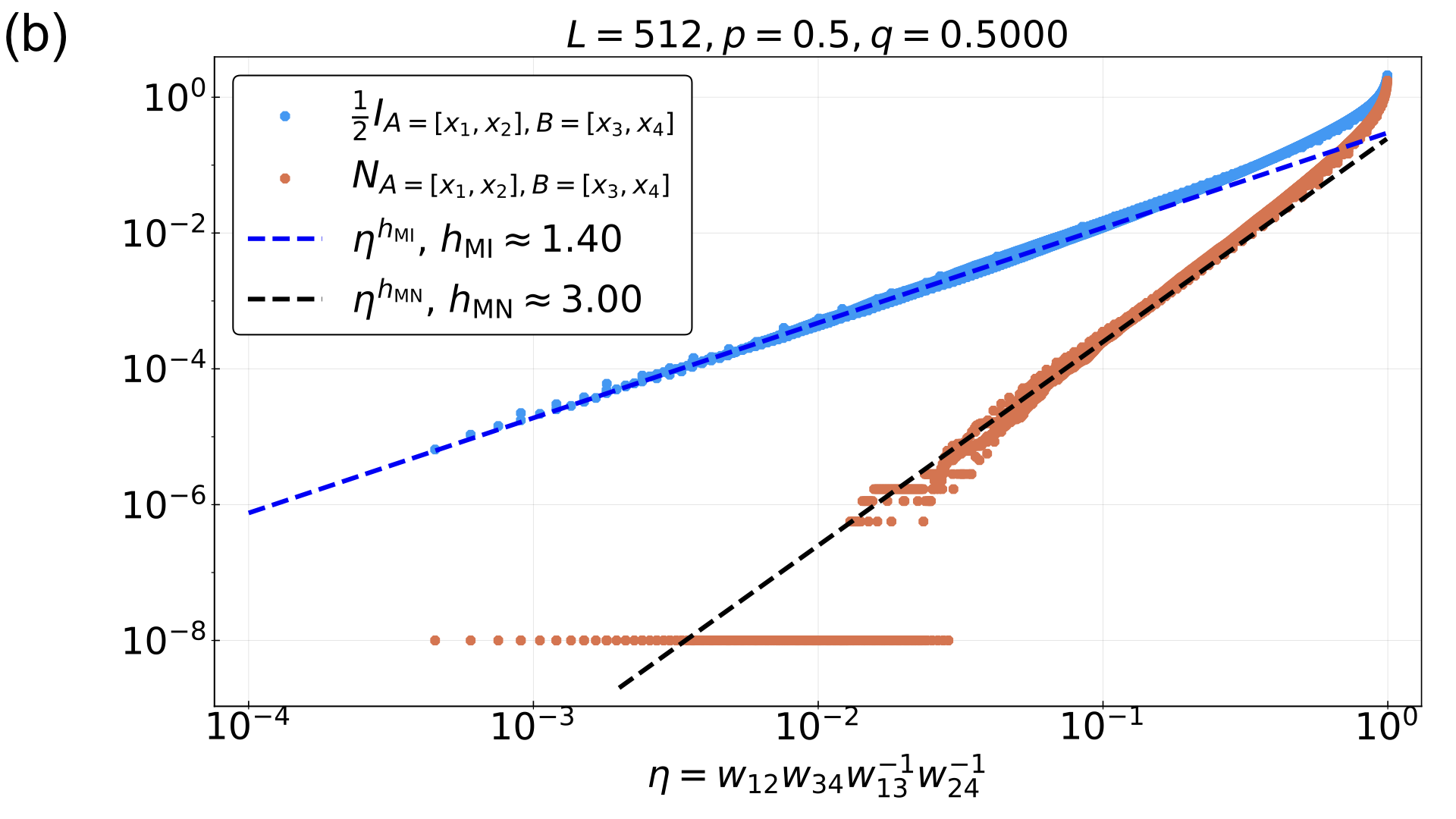}
    \includegraphics[width=.49\textwidth]{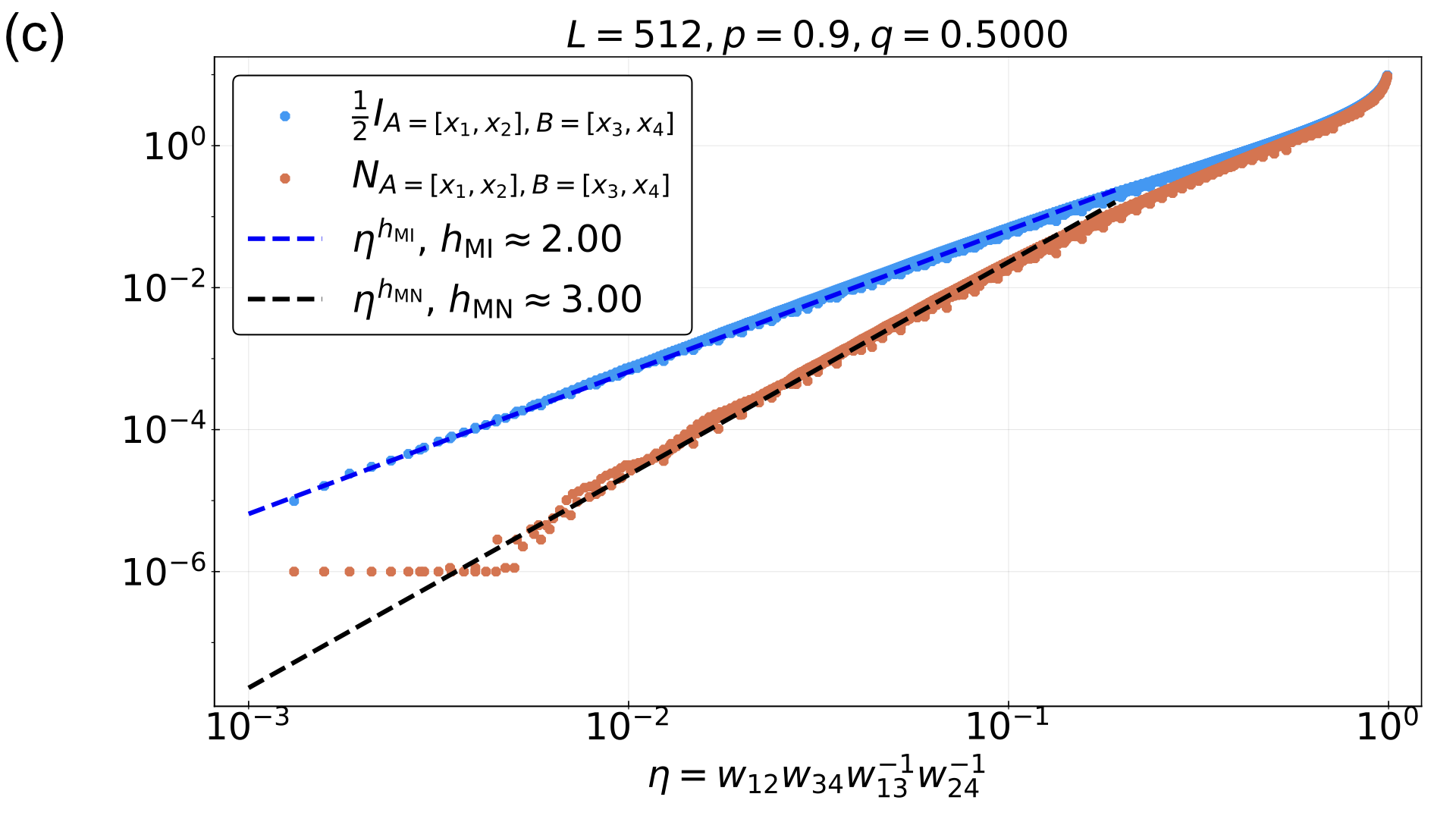}
    \caption{Mutual information and mutual negativity between two disjoint segments $A$ and $B$ for several choices of $(p, q)$ in the Goldstone phase of the CPLC model.}
    \label{fig:cplc_metal}
\end{figure}

In the Goldstone phase, we first argue for an \emph{approximate} scale invariance of the system, that will be useful in extracting critical exponents.
To do so, we consider the entanglement entropy of a contiguous segment of qubits.
The general comment on clipped gauge for stabilizer states still holds, and so is the general relation between the arc length distribution and the stabilizer length distribution in Eq.~\eqref{eq:P_stab_arc},
\env{align}{
    P_{\rm stab}(\ell) \approx P_{\rm arc}(2 \ell - 1) + P_{\rm arc}(2 \ell). 
}
From a direct simulation of the CPLC with system size $L$ and depth $T \gg L$, we observe (see Fig.~\ref{fig:cplc_arc}(a))
\env{align}{
    \label{eq:P_arc_CPLC}
    P_{\rm arc}(\ell) \approx K(\ell) \ell^{-2}, \text{ when } \ell \ll L.
}
Different from Eq.~\eqref{eq:P_arc_perc} where the coefficient is a constant, here $K(\ell)$ fits to the following function of $\ell$, which diverges as $\ell \to \infty$ (Fig.~\ref{fig:cplc_arc}(b)):
\env{align}{
    \label{eq:K_ell_flow}
    K(\ell) \approx \frac{1}{2\pi^2} \ln \ell + \rm{const.}
}
In particular, $\pi K(\ell)$ is the ``spanning number'' of the loops in an $\ell \times \ell$ cylinder~\cite{nahum1303crossing}, that is proportional to the flowing spin stiffness of the sigma model.
Similarly to Eq.~\eqref{eq:h_ab_GHZ},
for a subregion $A$ we have~\cite{nahum1911majorana}
\env{align}{
    S_A
    \approx& \frac{\ln 2}{2} \(
    \# \ln |A| + \frac{1}{4 \pi^2} (\ln |A|)^2 \).
}

Because of the weak (logarithmic) scale dependence of $K(\ell)$ and the rather small coefficient $(2\pi^2)^{-1} \approx 0.05$, for system sizes of interest (see below) we may treat $K(\ell)$ as a constant, whence the loop ensemble is approximately scale invariant.
We take a step further and assume the system has approximate conformal invariance, and use data collapse in extracting $h_{\rm MI}, h_{\rm MN}$, as we did before for the measurement-only case.
This approach will be justified by our numerics, below.


Before presenting results on MI or MN in the CPLC, we briefly explain our method of computing them.
While for many observables (including $K(\ell)$) one can access system sizes of $O(10^6)$ with a simulation of the CPLC~\cite{nahum1303crossing}, the MI and MN of the qubits, as it turns out, are not simple observables in the CPLC.\footnote{\label{footnote:3}For example, when $|A| = |B| = 1$, the MN is $\ln 2$ if the double arc is present (as before in Sec.~\ref{sec:moc_majorana}), \emph{and} they do not intersect any other arcs; but zero otherwise.
The MN in general cases becomes too complicated to describe, while the MI is always complicated.
Compared to the circuit in Sec.~\ref{sec:moc_majorana}, these complications are direct consequences of the ``crossing'' unitaries in the hybrid circuit.
}
Given a instance of the loop ensemble, the MI and MN of two regions on the upper boundary cannot be naively computed in $O(1)$ time, and this fact limits system sizes that can be accessed.
We find it most convenient to work with the Jordan-Wigner transformed stabilizer circuit in Fig.~\ref{fig:crossing_loop}(b), which allows us to access system sizes of $O(10^3)$.
This turns out to work well for our purposes.

Focusing on the symmetry axis of the Goldstone phase, $p > 0, q = 1/2$, and for $L = 512, T/L \gg 1$, we choose two subregions $A = [x_1, x_2]$, $B = [x_3, x_4]$, and compute $\frac{1}{2}I_{A,B}$ and $N_{A,B}$ against the cross ratio $\eta$ for varying values of $x_{j}$.
The results are plotted in Fig.~\ref{fig:cplc_metal}, where we see the data points form a narrow band, suggesting that the approximate conformal symmetry is at work.
We can read off the powerlaws when $\eta \to 0$, which we identify as $h_{\rm MI}$ and $h_{\rm MN}$ (compare Eqs.~(\ref{eq:def_F_I}, \ref{eq:def_F_N})).\footnote{If one does not use the approximate conformal symmetry, such a powerlaw can be still extracted by a direct computation of MI and MN of two finite regions at long distances.
The flowing spin stiffness might introduce a multiplicative logarithmic correction to the powerlaw, which will be difficult to numerically resolve in any case.
}

We find that $h_{\rm MN}$ appears nearly constant for the points we take, all taking the value $\approx 3.0$.
This observation holds throughout the Goldstone phase (see Appendix~\ref{app:CPLC_numerics} for more results) when we detune from the $q = 1/2$ line.
This turns out to be also the case for two other hybrid circuits considered in  Sec.~\ref{sec:rand_Clifford_Haar}.
On the other hand, $h_{\rm MI}$ seems to vary continuously as we move within the Goldstone phase.
From the CPLC perspective, this constrast is perhaps related to observation in footnote~\ref{footnote:3}, that MN admits a relatively simple description in terms of the loops in the $\eta \to 0$ limit whereas MI does not.

%

\subsubsection{Critical lines \label{sec:CPLC_critical}}

\begin{figure}
    \centering
    \includegraphics[width=.49\textwidth]{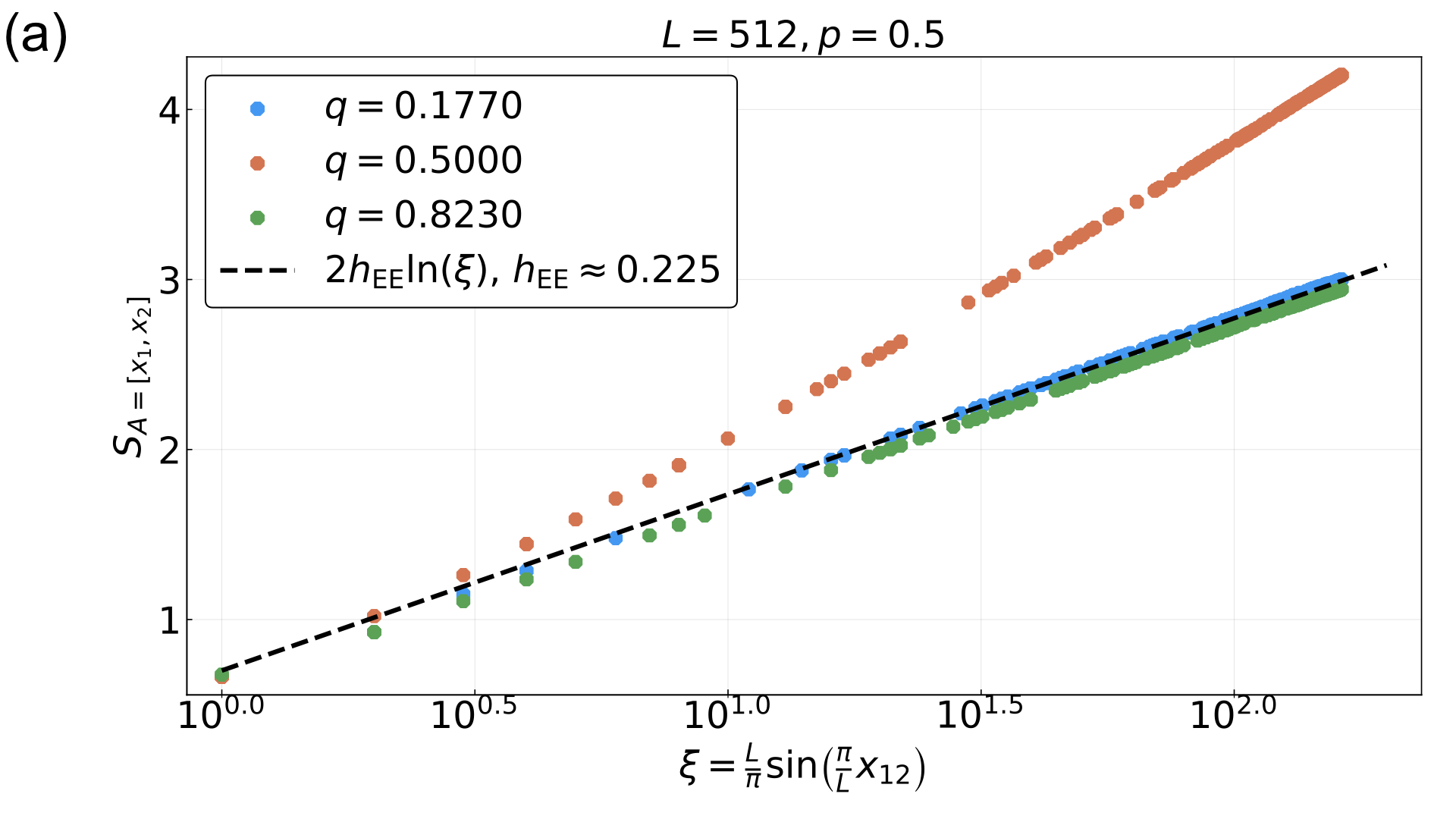}
    \includegraphics[width=.49\textwidth]{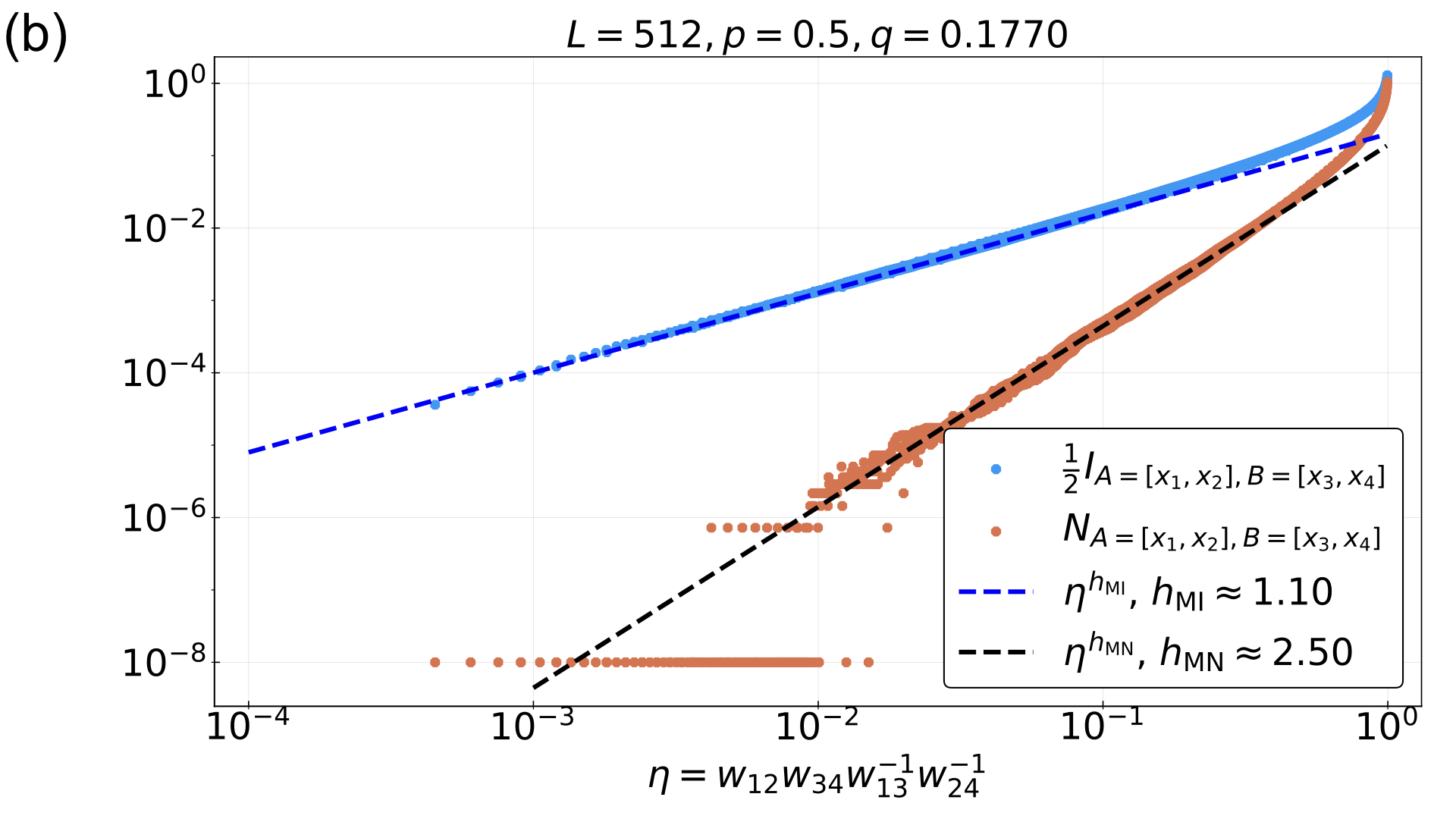}
    \includegraphics[width=.49\textwidth]{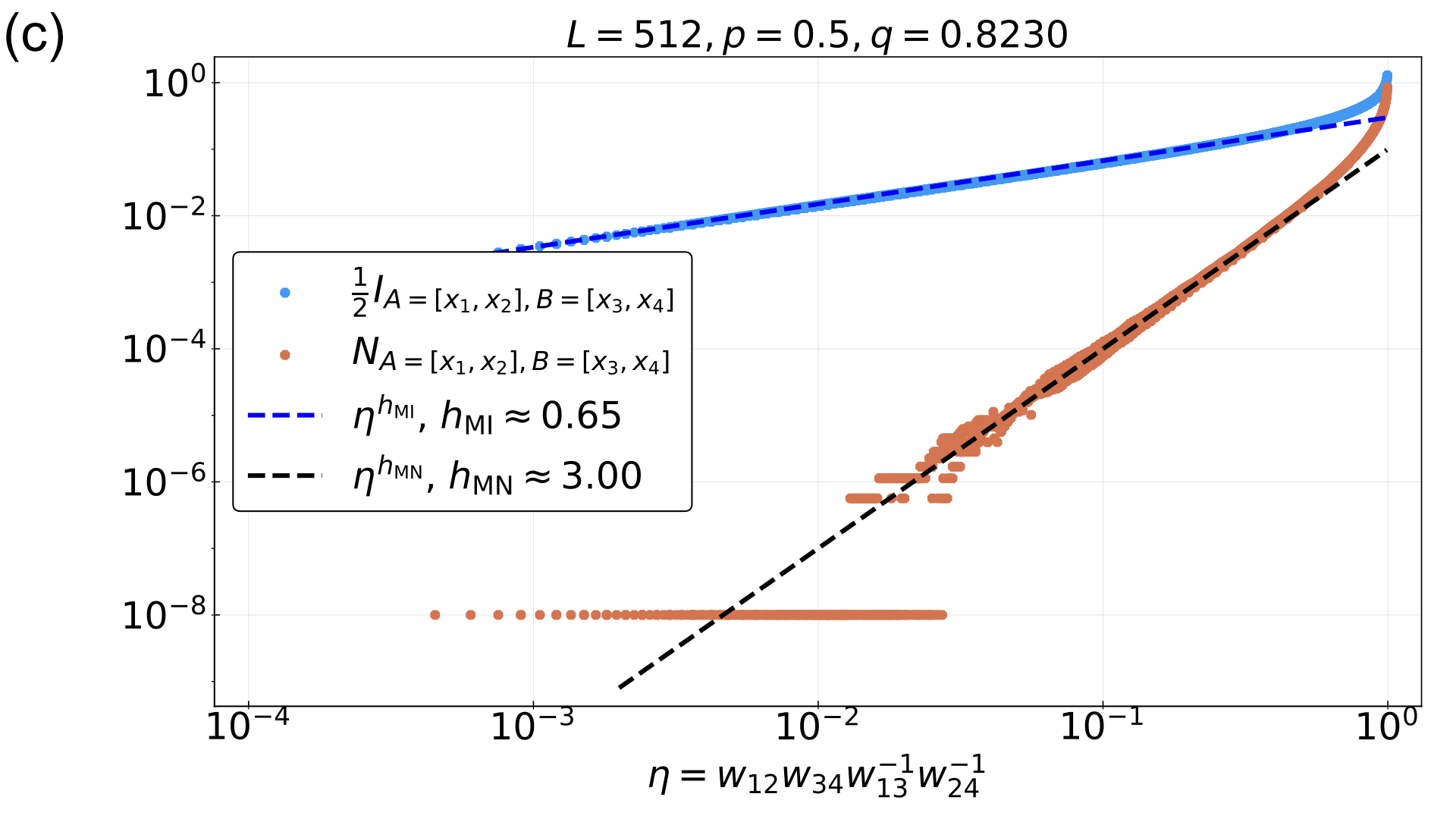}
    \caption{Numerical results for CPLC model on critical lines. (a) Entanglement entropy of a contiguous segment $A = [x_1, x_2]$ within the Goldstone phase and on two critical lines. (b) Mutual information and mutual negativity  between two disjoint segments $A$ and $B$ on the Goldstone-trivial critical line. (c) Mutual information and mutual negativity  between two disjoint segments $A$ and $B$ on the Goldstone-topological critical line.}
    \label{fig:cplc_critical}
\end{figure}

Here we briefly mention numerical results for the two critical lines separating the Goldstone phase from the two insulating phases.
The bulk transitions on these two critical lines should be the same, as they are related by a lattice translation of one Majorana fermion.
Characterization of the bulk transition is detailed in Ref.~\cite{nahum1303crossing}.
As a consistency check, we find the values of $h_{\rm EE}$ on the two critical lines are both consistent with the critical spanning number, up to a constant factor (see Fig.~\ref{fig:cplc_arc}(b) and Fig.~\ref{fig:cplc_critical}(a)),
\env{align}{
    \label{eq:h_EE_CPLC}
    h_{\rm EE} = \frac{\ln 2}{2} K = \frac{\ln 2}{2\pi} n_s^{\rm crit} \approx 0.225.
}

However, the boundary critical behavior -- specifically $h_{\rm MI}$ and $h_{\rm MN}$ -- can be different, as we see from numerical results in Fig.~\ref{fig:cplc_critical}(b,c).
The difference might be due to the presence or absense of an ``edge mode'' in the two insulating phases (see Fig.~\ref{fig:phase_diagram}):
in the insulating phase where $q > 1/2$, there is a Majorana loop of infinite size running against the upper boundary, hence making the insulator ``topological'';
while the other insulating phase where $q < 1/2$ is ``ordinary'', with only short loops near the boundary.
Moreover, the edge mode appears to introduce long range, multipartite entanglement, as evidenced by the smaller value of $h_{\rm MI}$ but larger value of $h_{\rm MN}$ at the transition on the ``topological'' side.  In the qubit language, this phase corresponds to more $ZZ$ than $X$ measurements, thus generating GHZ clusters and more tripartite entanglement. 


The last phenomenon is reminiscent of the metal to spin Hall insulator transition~\cite{obuse0805boundarycriticality}, where the topology of the insulating phase can affect critical properties on the boundary but not in the bulk.

\subsection{The random Clifford circuit and the random Haar circuit at local Hilbert space dimension $d=2$ \label{sec:rand_Clifford_Haar}}

\begin{figure}[b]
    \centering
    \includegraphics[width=.4\textwidth]{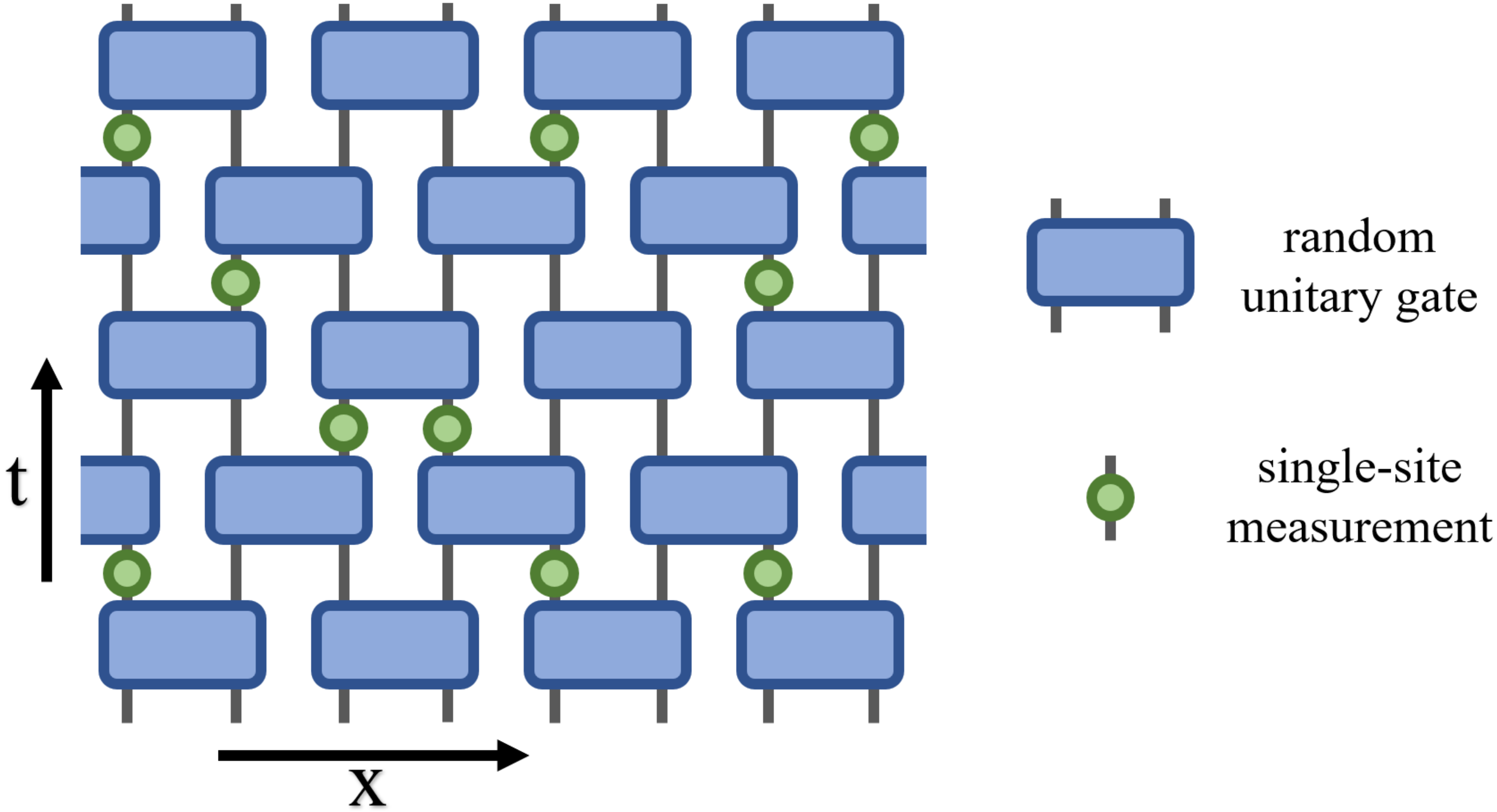}
    \caption{An instance of the hybrid random circuits. Two-site unitary gates (blue box) are arranged in a brick-wall pattern.
    On each vertical link a single-site measurement of $X$ (green circle) may be applied independently with probability $p_{\rm meas}$.
    The unitaries are either sampled uniformly from the two-qubit Clifford at $d=2$ (results in Fig.~\ref{fig:hybrid_numerics}(a)), or from the Haar measure on the two-qubit unitary group $\mathsf{SU}(4)$ (results in Fig.~\ref{fig:hybrid_numerics}(b)).
    }
    \label{fig:hybrid}
\end{figure}
\begin{figure}
    \centering
    \includegraphics[width=.49\textwidth]{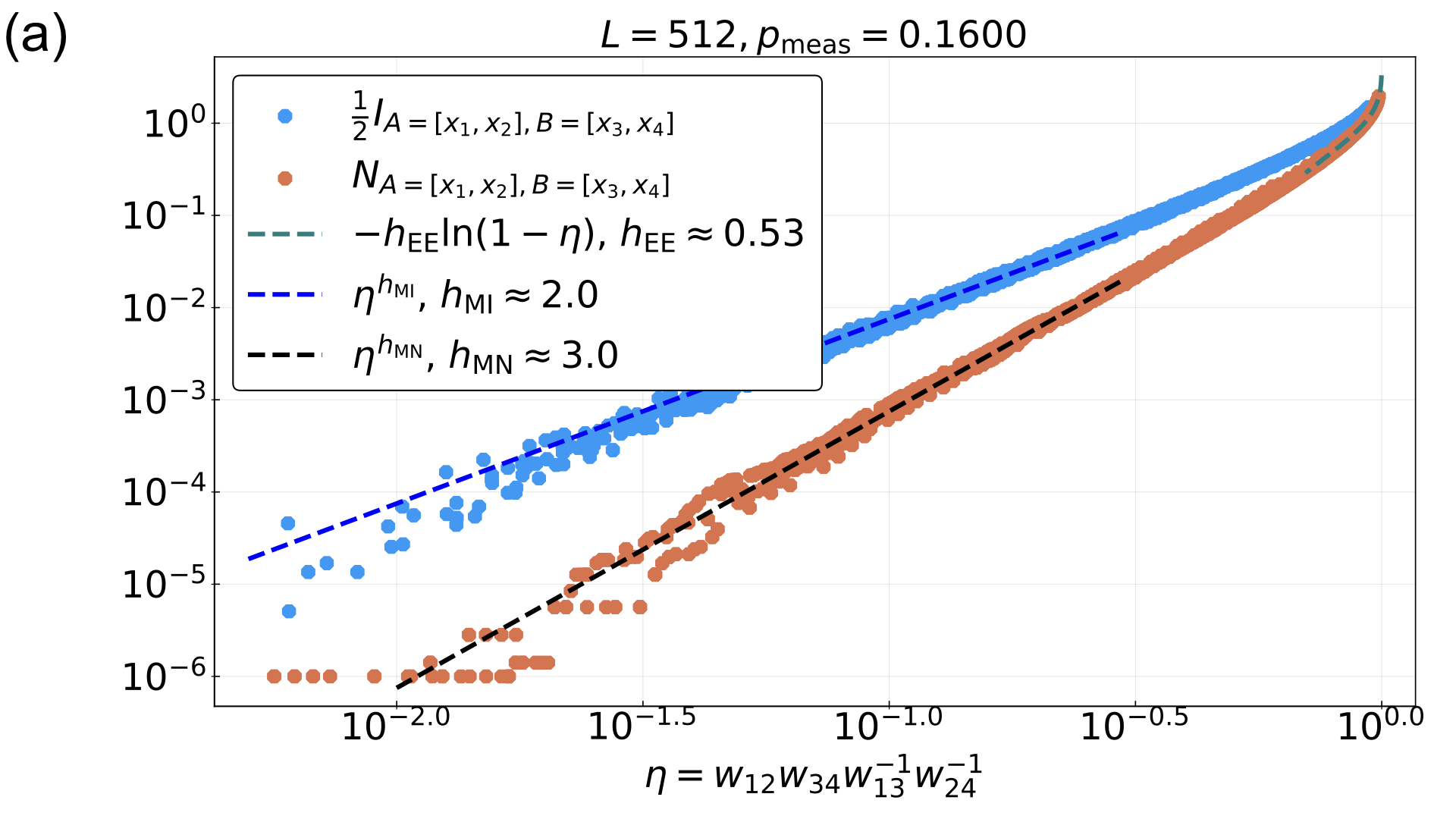}
    \includegraphics[width=.49\textwidth]{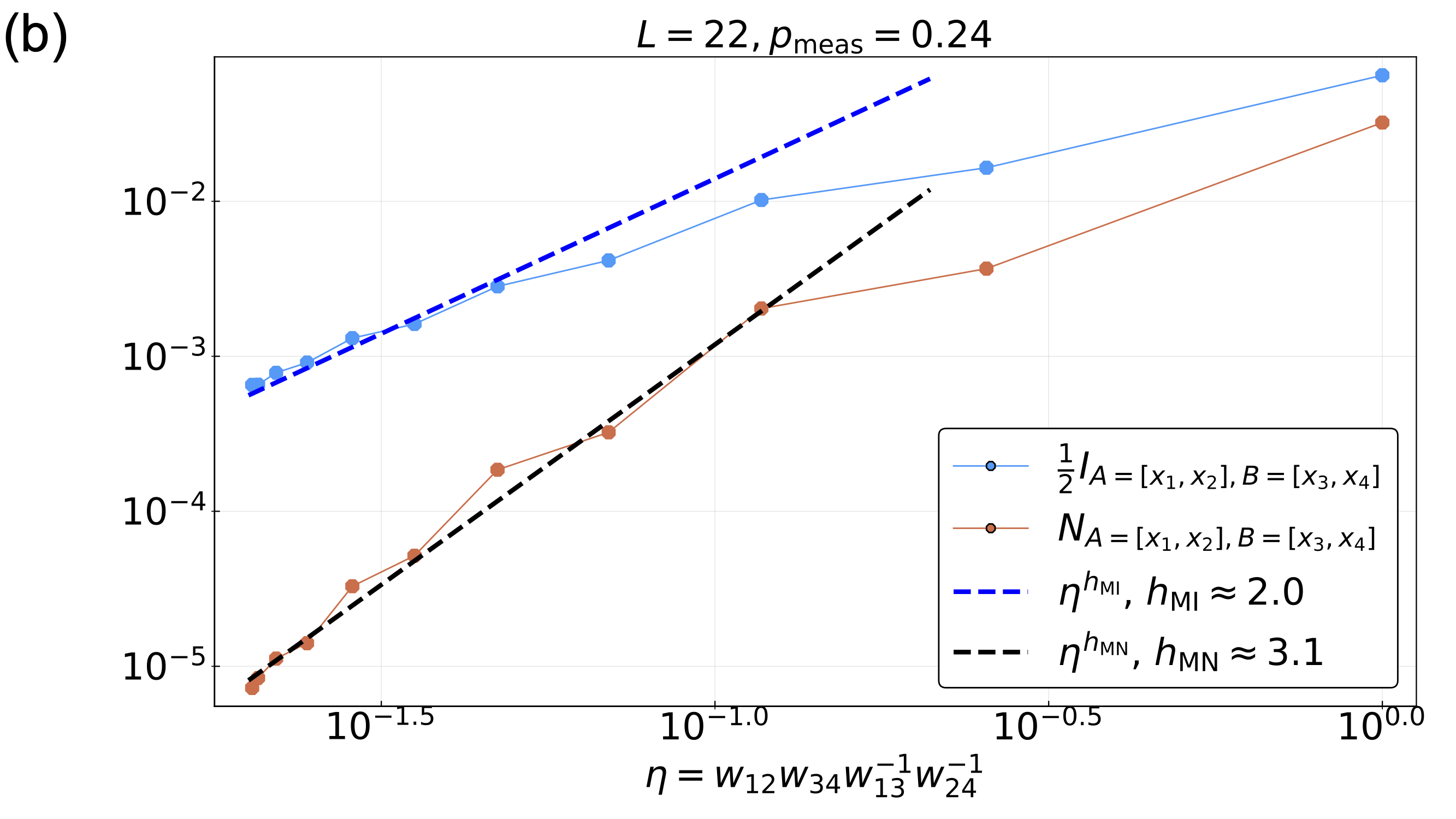}
    \caption{Mutual information and mutual negativity between two disjoint segments $A$ and $B$ for hybrid Clifford and Haar random unitary circuits. 
    }
    \label{fig:hybrid_numerics}
\end{figure}

We next consider more familiar models of hybrid circuits, consisting of single-qubit measurements and two-qubit unitaries, which can be either random Clifford unitaries~\cite{li1901hybrid} or random Haar unitaries~\cite{nahum2018hybrid} (see Fig.~\ref{fig:hybrid}).
{In either case, the unitary gates form a brickwork structure, and the measurements are performed at probability $p_{\rm meas}$.}
These circuits are apparently structurally similar to the previous Majoarana CPLC in that they are all ``hybrid'',
and as numerical results suggest (see below and discussions in Sec.~\ref{sec:discussion}), the similarity might be beyond merely structural.

In Fig.~\ref{fig:hybrid_numerics}, we plot data collapses of MI and MN at the critical points for the two models, repsectively.
In the random Clifford circuit (Fig.~\ref{fig:hybrid_numerics}(a)),
a clear separation between these two observables is seen, and we can fit for
\env{align}{
    h_{\rm MI}^{\rm Clifford} \approx 2.0,\quad h_{\rm MN}^{\rm Clifford} \approx 3.0.
}
The latter ($h_{\rm MN}^{\rm Clifford}$) agrees with that of the Majorana CPLC inside the Goldstone phase and on one of the critical lines.
The former ($h_{\rm MI}^{\rm Clifford}$) is consistent with results discussed extensively elsewhere~\cite{li1901hybrid, li2003cft}.

For the random Haar circuit, the data for smaller system sizes ($L \le 22$) gives us the following fitting results (Fig.~\ref{fig:hybrid_numerics}(b))
\env{align}{
    h_{\rm MI}^{\rm Haar} \approx 2.0,\quad h_{\rm MN}^{\rm Haar} \approx 3.1.
}
The value of $h_{\rm MI}^{\rm Haar}$ 
is consistent with Refs.~\cite{nahum2018hybrid, li1901hybrid}, whereas the value of $h_{\rm MN}^{\rm Haar}$ is close in value to other hybrid circuits.

{
Another hybrid model related to the random Clifford circuits and the Majorana CPLC circuit is the one considered in Ref.~\cite{hsieh_sang_2004_protected}, where Clifford unitaries with a global ``fermion parity'' $\mb{Z}_2$ symmetry are introduced on top of the measurement-only Majorana circuit (Sec.~\ref{sec:moc_majorana}).
These include the ``Majorana crossing'' unitaries of the CPLC, but can also represent interactions between Majorana fermions.
The system is no longer a free fermion model, and can have a volume-law phase (like the random Clifford circuit) as well as a critical phase (like the CPLC).
On the other hand, due to the symmetry, 
these unitaries are 
more constrained
as compared to the random Clifford unitaries.
This will be an interesting middleground for testing generalities and perculiarities of MIC in hybrid circuits.}
We include related results for this model in Appendix~\ref{app:z2_numerics}.


\subsection{Mixed phase of the random Clifford circuit}

\begin{figure}[t]
    \centering
    \includegraphics[width=.49\textwidth]{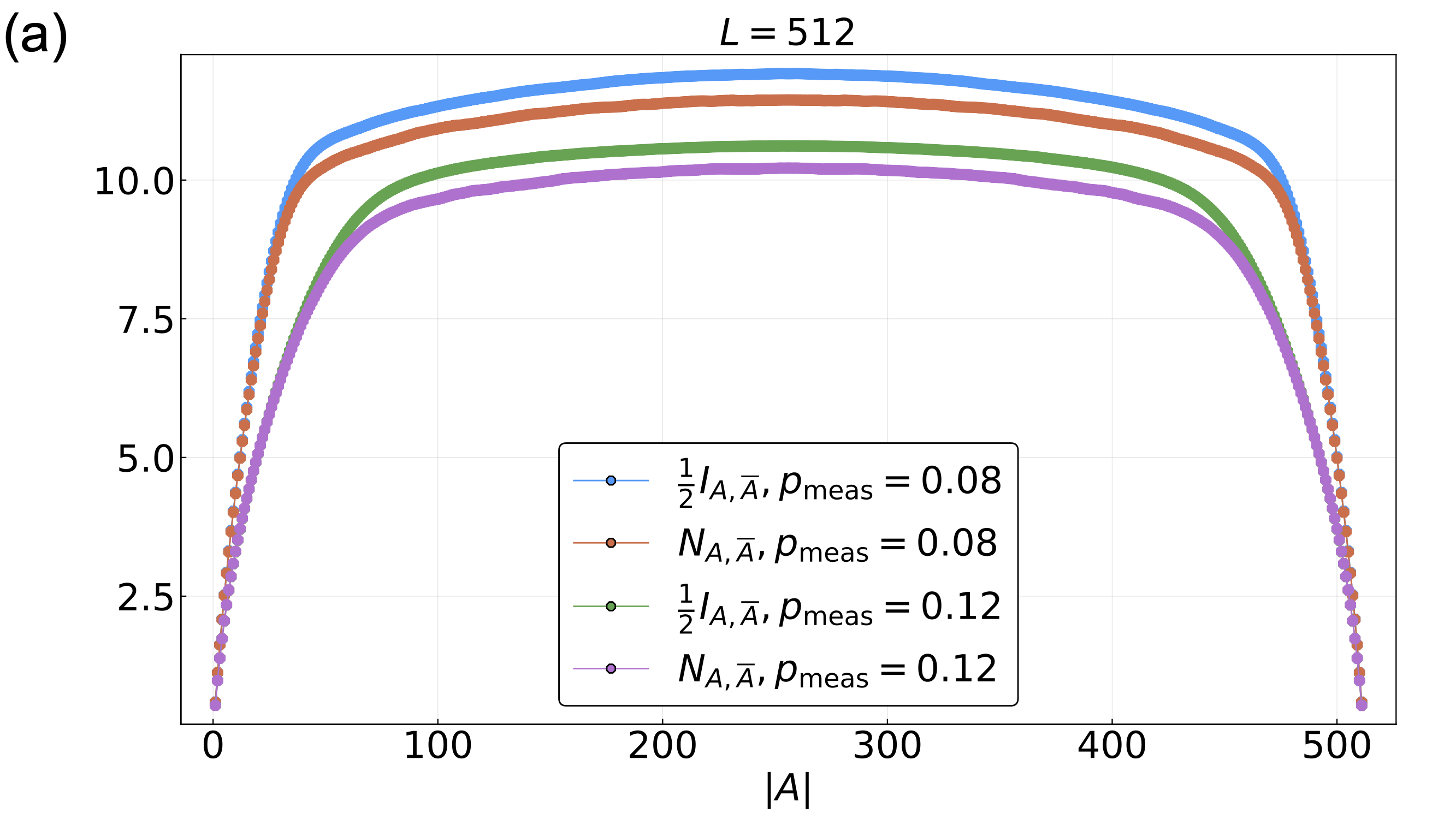}
    \includegraphics[width=.49\textwidth]{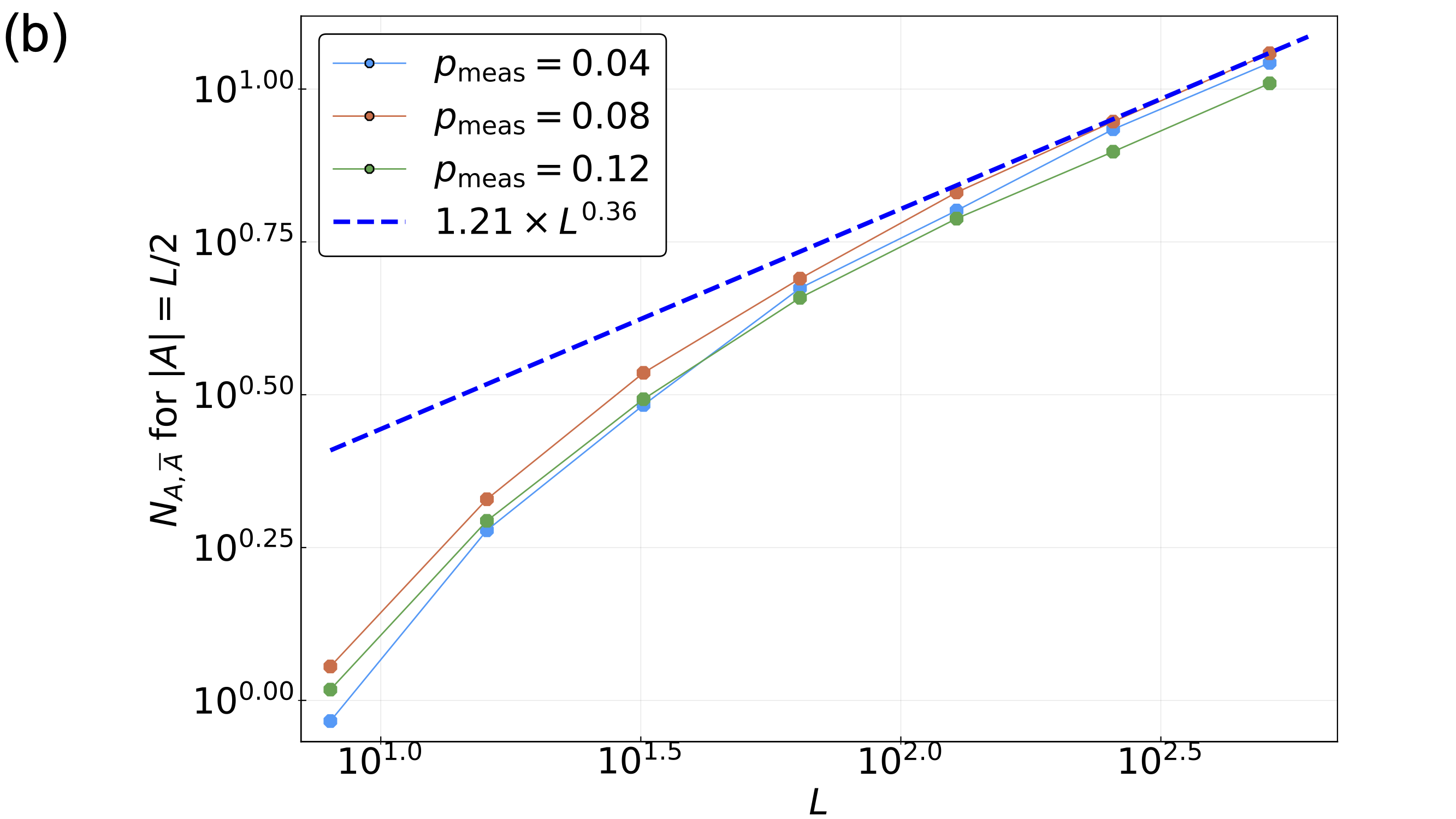}
    \includegraphics[width=.49\textwidth]{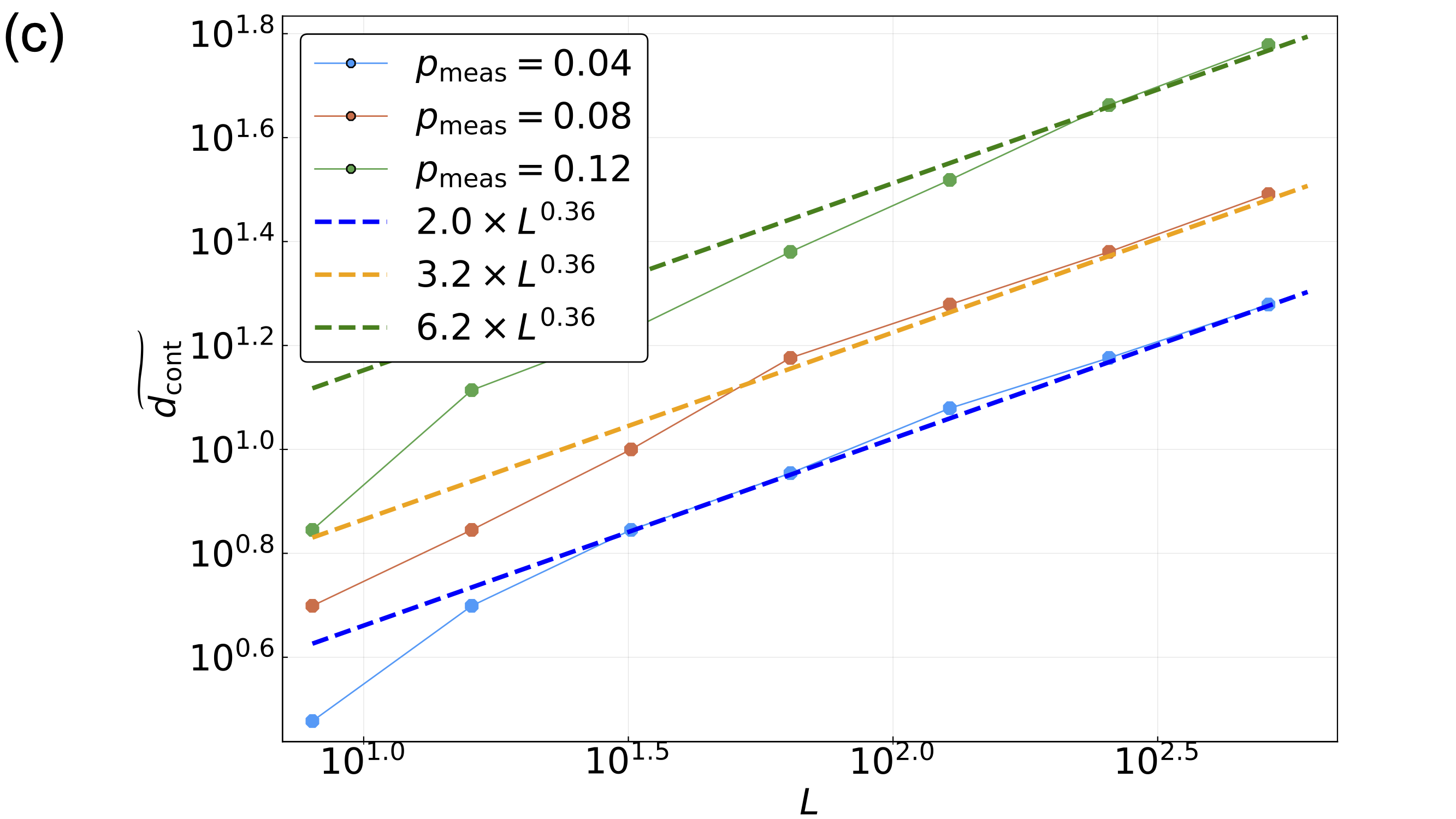}
\caption{
Numerical results for the random Clifford circuit in its mixed phase.
(a) Comparison between $\frac{1}{2}I_{A,\ovl{A}}$ and $N_{A,\ovl{A}}$ for a varying $A$, where we find they differ by at most an $O(1)$ constant.
(b) The halfcut mutual negativity of $N_{A,\ovl{A}}$ for $|A| = |\ovl{A}| = L/2$ as a function of $L$, where we find $N_{A,\ovl{A}} \propto L^{0.36}$.
(c) The quantity $\widetilde{d_{\rm cont}}$, defined as $|A|$ where $N_{A,R} = \epsilon (\ln 2)$ (compare Eq.~\eqref{eq:d_cont_tilde}).
We see that $\widetilde{d_{\rm cont}}$ is proportional to $d_{\rm cont}$~\cite{li2007capillary_qecc}.
}
    \label{fig:hybrid_mixed}
\end{figure}

In this subsection we briefly deviate from the main focus of this paper, and discuss the scaling of entanglement negativity in the ``mixed'' phase~\cite{gullans1905purification} of the random Clifford circuit in Fig.~\ref{fig:hybrid}.
We take a maximally mixed initial state for the circuit, and choose $p_{\rm meas} < 0.16$ such that the state $\rho_Q$ retains a finite entropy density for a time exponential in $L = |Q|$.
Given that the negativity is best known as a mixed state entanglement measure, it is natural to explore this.

We first compare $N_{A,\ovl{A} = Q-A}$ and $\frac{1}{2} I_{A,\ovl{A}}$ while varying $|A|$,\footnote{Here the geometry is different from our focus elsewhere in this paper, $N_{A,B}$, where $A$ and $B$ are small distant regions.
Since $N_{A,B} \le \frac{1}{2} I_{A,B}$, and the latter decays exponentially with their distance~\cite{li1901hybrid}, a similar exponential decay is expected for $N_{A,B}$.
}
and plot the results in Fig.~\ref{fig:hybrid_mixed}(a).
We find that they differ by at most an $O(1)$ amount for all values of $|A|$.
We further consider the ``half-cut mutual negativity'', defined as $N_{A,\ovl{A}}$ when $|A| = L/2$ (Fig.~\ref{fig:hybrid_mixed}(b)), and find that this quantity scales with the system size as $L^\gamma$ where $\gamma \approx 0.36$.
This is the same exponent for the bipartite mutual information $I_{A, \ovl{A}}$ when $|A| = L/2$~\cite{li2007capillary_qecc}.


Second, we introduce a reference system (``the environment'') $R$ that purifies $Q$, and consider the mutual negativity $N_{A,R}$, which for stabilizer states satisfies the following equation
\env{align}{
    N_{A,R} = \frac{1}{2} I_{A, R} - \( \frac{1}{2} I_{A, \ovl{A}} - N_{A, \ovl{A}} \) \ge 0,
}
where in particular (see Appendix~\ref{app:EN_stab})
\env{align}{
    \frac{1}{2} I_{A, R} \ge \frac{1}{2} I_{A, \ovl{A}} - N_{A, \ovl{A}}  \ge 0.
}
The quantity $I_{A,R}$ vanishes -- i.e. the subsystem $A$ and the reference $R$ ``decouple'' -- when $|A| \le d_{\rm cont}$, where $d_{\rm cont}$ is the ``contiguous code distance'' of the dynamical state~\cite{li2007capillary_qecc} that also diverges with $L$ as $L^\gamma$.
The decoupling condition ($I_{A,R} = 0$) clearly implies the vanishing of bipartite entanglement between $A$ and $R$ ($N_{A,R} = 0$) (see Fig.~\ref{fig:hybrid_mixed}(c)); but the latter condition itself may be used to define a distance $\widetilde{d_{\rm cont}}$, where
\env{align}{
    \label{eq:d_cont_tilde}
    |A| \le \widetilde{d_{\rm cont}} \quad \Leftrightarrow \quad  N_{A,R} = 0.
}
In Fig.~\ref{fig:hybrid_mixed}(d), we see that $\widetilde{d_{\rm cont}} \propto L^{\gamma}$, and therefore proportional to $d_{\rm cont}$.

Overall, we have found that in the mixed phase of the random Clifford circuit, the MN and MI -- and quantities that derive from these -- behave qualitatively the same.
Colloquially, one may say for the $(A, \ovl{A}, R)$ tripartition, most of the entanglement is EPR-like.
In this sense, the mixed phase is similar to the ``unitary limit'' when $p_{\rm meas} = 0$~\cite{shapourian2011diagrammatic}.

Of course, it is more interesting when MN and MI are qualitatively different.
One such instance is the MIC; other examples may be found by considering different tripartitions and/or different types of dynamics~\cite{tarun2008transition}.
It would be also interesting if negativity
can be used to identify novel phase transitions inside the mixed phase~\cite{hsieh_sang_2004_protected}.


\section{Discussions \label{sec:discussion}}

\subsection{Summary and discussion}

\begin{table*}[t]
\centering
\begin{tabular}{c || c | c | c}
\hline
 ~ & $h_{\rm EE}$ & 
 $h_{\rm MI}$ 
 &
 $h_{\rm MN}$ 
 \\
\hline \hline
 1) Measurement-only Majorana circuit (Sec.~\ref{sec:moc_majorana})
 & $\frac{\sqrt{3}}{4\pi} \ln 2 \approx 0.096$~\cite{nahum1911majorana}
 & $\frac{1}{3} \approx 0.333$~\cite{buechler2006projectiveTFIM}
 & $2$
 \\ \hline
 2) First-passage percolation ($S_0$ in random Haar circuit)
 & $\frac{\sqrt{3}}{2\pi} \ln 2 \approx 0.191$~\cite{nahum2018hybrid}
 & $2$~\cite{nahum2018hybrid}
 & $2$\footnote{The mutual negativity in FPP for two single sites here is defined as the probability of $I_{A,B} = 2$~\cite{nahum2018hybrid}.
 This exponent is the same as row~(1).
 }
 \\ \hline
 3) Random Haar circuit as $d \to \infty$
 & $\frac{1}{6} \approx 0.166$ (for $S_{n \ge 1}$)~\cite{andreas2019hybrid}
 & -
 & -
 \\ \hline \hline
 4a) Majorana CPLC: Goldstone phase (Sec.~\ref{sec:CPLC_goldstone})
 & -
 & varying
 & $3.0$ 
 \\ \hline
 4b) Majorana CPLC: transition to trivial insulator (Sec.~\ref{sec:CPLC_critical})
 & $\frac{\ln 2}{2\pi} n_s^{\rm crit} \approx 0.225$~\cite{nahum1303crossing}
 & $1.1$ 
 & $2.5$ 
 \\ \hline
 4c) Majorana CPLC: transition to topological insulator (Sec.~\ref{sec:CPLC_critical})
 & $\frac{\ln 2}{2\pi} n_s^{\rm crit} \approx 0.225$~\cite{nahum1303crossing}
 & 0.65 
 & 3.0 
 \\ \hline
 5) Random Clifford circuit at $d = 2$ (Sec.~\ref{sec:rand_Clifford_Haar})
 & $0.53$~\cite{li2003cft}
 & $2.0$~\cite{li2003cft, li1901hybrid}
 & $3.0$ 
 \\ \hline
 6) Random Haar circuit at $d = 2$ (Sec.~\ref{sec:rand_Clifford_Haar})
 & -
 & $2.0$ (for $S_{1}$)~\cite{nahum2018hybrid, li1901hybrid}
 & $3.1$ 
\end{tabular}
\caption{Comparison of critical exponents in several circuit models.
See the main text for discussions.
}
\label{table:compare_exponents}
\end{table*}

In Table~\ref{table:compare_exponents}, we summarize the main critical exponents 
in several circuit models -- some in this work, some in previous works -- whenever available.
We see that entanglement negativity is a useful diagnosis that provides additional information of measurement-induced criticality.
{
This Table is, however, in no sense exhaustive.
For example, one can examine MN in various boundary conditions~\cite{li2003cft} for models we considered in this paper, and will likely find new critical exponents.
}

It is worth noting that rows (1-3) of Table~\ref{table:compare_exponents} are all critical percolation, but in different guises.
In each of these guises entanglement measures can correspond to different observables and can have different exponents, such as $h_{\rm EE}$ and $h_{\rm MI}$.
This fact makes it a tricky business to compare exponents between different models, and to classify measurement-induced transitions based on this information alone.
On the other hand, unlike $h_{\rm EE}$ and $h_{\rm MI}$, the mutual negativity exponent $h_{\rm MN}$ seems to exhibit a somewhat simpler pattern: 
it is plausible that $h_{\rm MN}$ equals $2$ for all three occurences of percolation (e.g. rows (1-3)), while always taking a different value for models that are not percolation (e.g. rows (4-6)), thereby making it a ``fingerprint'' of percolation. 

To make the last point, we need to obtain $h_{\rm MN}$ in row (3), i.e. the 
random Haar circuit at infinite Hilbert space dimension.
It is described by a replica spin model~\cite{choi2019spin, andreas2019hybrid}, where entanglement entropies are free energy costs subject to a boundary condition change.
As it turns out, entanglement negativity also corresponds to a free energy cost subject to a different -- but known -- set of boundary condition change, where the relevant domain wall operators might also be identified similarly as in Ref.~\cite{andreas2019hybrid}.
Whether this is a viable path of reasoning and gives the value of $h_{\rm MN}$ as expected will be tested in future works.

Going away from the infinite $d$ limit of random Haar circuits (row (3)), we obtain row (6) where $d$ is finite.
The $1/d$ corrections to row (3) are known to be ``two-hull'' perturbations to the percolation fixed point~\cite{andreas2019hybrid}, and are relevant under renormalization group (RG) transformations.
Meanwhile, rows (4a-4c) (Majorana CPLC) are obtained from row (1) upon introduction of loop crossings, the latter also known to be two-hull perturbations to the percolation fixed point~\cite{nahum1303crossing, vasseur2018rtn}.
It is thus at least suggestive that generic critical point of random Haar circuits at finite values of $d$ {may} have something in common with the CPLC, particularly on one of the critical lines where the values of $h_{\rm MN}$ are close.

In any case, in the Goldstone phase of the CPLC, the result $h_{\rm MN} \approx 3.0$ is interesting and might be obtained by  calculating the corresponding loop observable of the MN, perhaps making use of the CPLC sigma model, which becomes Gaussian in the infrared.
The ``metal-insulator'' transition in the CPLC with different boundary critical behaviors is also interesting in its own right.
The bulk transition belongs to a class of non-unitary CFTs that remain to be understood, and here topology on the edge appears important.

Having mentioned that {hybrid circuits in} rows (4) and (6) are both related to percolation by a relevant perturbation, we comment that row (5) -- the random Clifford circuit at finite $d$ -- also becomes percolation-like in a certain limit.
Upon taking $d \to \infty$ for a fixed $L$, we observe 
that entanglement entropies can be well approximated by ``minimal cuts'' in percolation (as in row (2)),\footnote{This result might be related to random stabilizer tensor networks in the limit of infinite bond dimension~\cite{Hayden2016,nezami1608stabilizerTN, nguyen1709stabilizerTN,shinsei1808EN_holoraphy}.
}
and more importantly, the value of $h_{\rm MN}$, being approximately $3.0$ at small $d$, approaches $h_{\rm MI}$ at large $d$, which remains approximately $2.0$ for all values of $d$. 
Notice that this is also the value of $h_{\rm MI}$ and $h_{\rm MN}$ for row (2), the first-passage percolation description of $S_0$ in random Haar circuits.
It is therefore conceivable that by varying $d$ in the Clifford circuit, one observes a crossover~\cite{nahum2018hybrid} from the percolation fixed point at infinite $d$ (when $L \ll \xi_\ast(d)$) to the generic fixed point at finite $d$~\cite{li2003cft} (when $L \gg \xi_\ast(d)$), where $\xi_\ast(d)$ is a length scale that diverges with $d$.
Details of these results will be reported elsewhere.



\subsection{Outlook}

We mention a few other possible directions of exploration where the negativity might be useful. 
\env{itemize}{


\item
Recent work has shown negativity is particularly good at detecting topological order at finite temperatures, in particular whether ``quasiparticle poisoning'' spoils the braiding statistics~\cite{tarun_tim_1912_topo_order}.
In the CPLC phase diagram, we see 
another example where negativity is sensitive to topology, albeit in a very different context. 
It should be interesting to explore these cases in other topological phases of matter and/or topological phase transitions.

\item
{
We remark on a subtle distinction between the ensemble averaged mutual negativity considered in this paper and the mutual negativity of a CFT ground state.
The former, as exemplified by analytic results in Sec.~\ref{sec:moc_majorana}, corresponds to \emph{boundary} correlation functions of a CFT and decays as a powerlaw of the distance, where the power is usually given by the leading primary field in the relevant OPE.
The latter, however, is a correlation function between twist fields in the \emph{bulk}, and depends on the full spectrum of the CFT~\cite{cardy2012EN_a, cardy2012EN_b}.
In particular, the latter has a striking essential singularity at long distances, in sharp contrast with the simple powerlaws we obtained here.
Although for some critical non-unitary dynamics~\cite{chenxiao2004nonunitary} the analogy between the dynamical state and a CFT ground state can be fruitful, this is not the case here.
Understanding their distinction will be an interesting topic for future work.
}

\item
{
We briefly describe the ``purification'' of a maximally mixed initial state under the free-fermion dynamics in Sec.~\ref{sec:moc_majorana} and Sec.~\ref{sec:CPLC}.
Here, it is believed~\cite{cao2018monitoring} -- even proven in some cases~\cite{fidkowski2008forget} -- that a ``mixed phase''~\cite{gullans1905purification} cannot be sustained as long as the measurement rate is finite.
This has a simple heuristic explanation in the loop picture.
Here, the initial state has $2L$ vertical Majorana strands, and the entropy of the system at some later circuit time is proportional to the number of strands that thread through the upper and lower boundary.\footnote{This is precisely the ``spanning number'' defined in Ref.~\cite{nahum1303crossing}.}
A measurement decreases the entropy by one unit if and only if it is performed on two ``spanning'' strands.
Assuming that the spanning strands are distributed evenly and independently across the system, we have the following equation for the entropy density $s \coloneqq L^{-1} S_{\rm vN}(\rho_Q)$,
\env{align}{
    \Delta s \propto -  L^{-1} s^2.
}
This agrees well with Ref.~\cite{fidkowski2008forget}.
In particular, at a finite aspect ratio of the circuit, Ref.~\cite{fidkowski2008forget} anticipates that the entropy can be either finite or logarithmic in $L$, as exemplified by the ``percolation'' and ``CPLC'' circuit, respectively.
The breakdown of this picture due to fermion interactions will be interesting to understand.
}


\item
{
It may be interesting to explore various entanglement measures in holographic random tensor networks~\cite{Pastawski2015,Hayden2016}. As pointed out in Refs.~\cite{vasseur2018rtn,vasseur2003mft}, an entanglement phase transition can be obtained precisely by tuning the bond dimension of the tensor network. Similar to the MIC in some hybrid circuits, the phase transition is described by some novel critical point caused by a relevant perturbation to percolation~\cite{vasseur2018rtn}.
It is interesting 
to construct a random stabilizer tensor network and investigate the bipartite/multipartite entanglement at the critical point~\cite{nezami1608stabilizerTN}.}

\item
{Besides the CPLC model considered in this paper, there are many other interesting loop models which can also exhibit unusual critical phenomena and exotic phases~\cite{Cardy_Lykke_1997,Martins_1998,nahum1303crossing,Gruzberg_1999,Chalker_1988,jacobsen_read_saleur_2003_denseloop,Dai_Nahum_2020,Vernier_2016}.
These models have a close connection with random spin models and disordered free fermion systems which are not fully understood due to strong randomness~\cite{Cardy_Lykke_1997,Chalker_1988}.
It is possible to map these systems to some quantum dynamics models, as in this paper, and then analyze the boundary wave function from the perspective of quantum information.}

\item
Recently in Ref.~\cite{zaletel2011tripartite}, the non-negative quantity $(E_P)_{A,B} - \frac{1}{2}I_{A,B}$ was proposed as a measure of irreducible tripartite entanglement for ground states of one-dimensional spin chains, where $(E_P)_{A,B}$ is the entanglement of purification between $A$ and $B$.
Although for stabilizer states this quantity and $\frac{1}{2}I_{A,B} - N_{A,B}$ are both proportional to the ``GHZ content'' $g_{ABC}$ of the state (see Eqs.~(\ref{eq:structural_thm}, \ref{eq:NAB_nEPR}, \ref{eq:IAB_nEPR_nGHZ}, \ref{eq:EPAB_nEPR_nGHZ})), the latter difference (while not necessarily positive) might serve as an independent measure of tripartite entanglement for general states, while also being easier to compute.
}

\emph{Note added}: We would like to bring the reader's attention to a related work by Xin Dai, Bowen Shi, and Yuan-Ming Lu (to appear in the same arXiv posting).

\section*{Acknowledgements}

We thank Adam Nahum and Andreas Ludwig for most helpful discussions.

{This work was supported in part by the Heising-Simons Foundation (YL and MPAF), and 
by the Simons Collaboration on Ultra-Quantum Matter, which is a grant from the Simons Foundation (651440, MPAF).}
SS and TH acknowledge support from the Natural Sciences and Engineering Research Council of Canada (NSERC) through a
Discovery Grant. Research at Perimeter Institute is supported in part by the Government of Canada through the Department of Innovation, Science and Economic Development Canada and by the Province of Ontario through the Ministry of Colleges and Universities.
TZ was supported by a postdoctoral fellowship from the Gordon and Betty Moore Foundation, under the EPiQS initiative, Grant GBMF4304, at the Kavli Institute for Theoretical Physics. 
This research is supported in part by the National Science Foundation under Grant No. NSF PHY-1748958.
This work
was made possible by the facilities of the Shared Hierarchical Academic Research Computing Network (SHARCNET) and Compute/Calcul Canada.
Use was made of computational facilities purchased with funds from the National Science Foundation (CNS-1725797) and administered by the Center for Scientific Computing (CSC). The CSC is supported by the California NanoSystems Institute and the Materials Research Science and Engineering Center (MRSEC; NSF DMR-1720256) at UC Santa Barbara.

\appendix

\begin{widetext}

\section{Entanglement negativity (EN) of stabilizer states : more details\label{app:EN_stab}}

\subsection{An algorithm for computing EN}

Let $A$ and $B$ each be a set of 
qubits, $\mc{S}$ be an abelian subgroup of the Pauli group $\mc{P}(A \cup B)$, and $\mc{G}(\mc{S})$ be a linearly independent generating set (or simply a ``basis'') of $\mc{S}$.
We have
\env{align}{
    \mc{G}(\mc{S}) =&\, \{g_1, \ldots g_{|\mc{G}(\mc{S})|} \}, \\
    \mc{S} =&\, \avg{\mc{G}(\mc{S})}, \quad \dim \mc{S} = \lv \mc{G}(\mc{S}) \rv = \log_2 \lv \mc{S} \rv.
}
Here $\dim \mc{S}$ is the dimension of $\mc{S}$ when viewed as a vector space.
We will henceforth denote
\env{align}{
    m \coloneqq \dim \mc{S} = \lv \mc{G}(\mc{S}) \rv = \log_2 \lv \mc{S} \rv.
}

The subgroup $\mc{S}$ defines the following physical density matrix on ${A \cup B}$,
\env{align}{
    \rho_{A \cup B}= \frac{1}{2^{|{A \cup B}|}} \sum_{g \in \mc{S}} g.
}

Given the bipartition $(A,B)$ of $A \cup B$, we define the following mappings
\env{itemize}{
\item
The ``restriction'' on $A$:
\env{align}{
    \mathrm{proj}_A : \hspace{.3in} \mc{P}(A \cup B) \quad \to&\quad \mc{P}(A) \nn
    g_A \otimes g_B \quad \mapsto&\quad g_A
}
\item
The ``partial transpose'' on $A$:
\env{align}{
    \Gamma_A : \hspace{.3in} \mc{P}(A \cup B) \quad \to&\quad \mc{P}(A \cup B) \nn
    g_A \otimes g_B \quad \mapsto&\quad (g_A)^\mathsf{T} \otimes g_B
}
Here $\mathsf{T}$ denotes the matrix transposition, say in the computational basis where
\env{align}{
    X = \env{pmatrix}{
        0 & +1\\
        +1 & 0
    }, \quad 
    Y = \env{pmatrix}{
        0 & -i\\
        +i & 0
    }, \quad 
    Z = \env{pmatrix}{
        +1 & 0\\
        0 & -1
    }.
}
}
The partial transpose of $\rho_{A \cup B}$ is as follows
\env{align}{
    \rho_{A \cup B}^{\Gamma_A} = 
    \frac{1}{2^{|{A \cup B}|}} \sum_{g \in \mc{S}} g^{\Gamma_A}.
}

Since $X^\mathsf{T} = X$, $Z^\mathsf{T} = Z$, while $Y^\mathsf{T} = -Y$,
the partial transpose introduces a factor $x_A(g) \coloneqq (-1)^{\#Y_A(g)}$ on $g \in \mc{P}({A \cup B})$, where $\#Y_A(g)$ counts the number of $Y$ factors in $\mathrm{proj}_A(g)$,
\env{align}{
    g^{\Gamma_A} = (-1)^{\#Y_A(g)} g = x_A(g) \cdot g.
}
Define a sign function $\mathrm{Sgn}_A (g, h) \in \{-1, +1\}$, such that 
\env{align}{
    \mathrm{proj}_A(g) \cdot \mathrm{proj}_A(h) = \mathrm{Sgn}_A(g, h)\,  \mathrm{proj}_A(h) \cdot \mathrm{proj}_A(g).
}
A key observation is the following ``cocycle condition'', that can be explicitly verified:
\env{align}{
    x_A(g) x_A(h) x_A(gh) = \mathrm{Sgn}_A (g, h).
}

Let us also define the following $m \times m$ ``commutator matrix'',
\env{align}{
    \(\mathsf{K}_A\)_{ij} = \env{cases}{
        0, \text{ if } \mathrm{proj}_A(g_i) \cdot \mathrm{proj}_A(g_j) = + \mathrm{proj}_A(g_j) \cdot \mathrm{proj}_A(g_i) \\
        1, \text{ if } \mathrm{proj}_A(g_i) \cdot \mathrm{proj}_A(g_j) = - \mathrm{proj}_A(g_j) \cdot \mathrm{proj}_A(g_i)
    }
}
for $1 \le i, j \le m$,
where the $g_i$'s on the RHS are elements of $\mc{G}(\mc{S})$.
It is obvious that $\mathsf{K}_A$ is symmetric, and has $0$'s on its diagonal.

For an arbitrary element $g \in \mc{S}$, we have the following representation
\env{align}{
    g \coloneqq g^u = \prod_{j = 1}^m g_j^{u_j},  \text{ where } u \in \{0, 1\}^{m} = \(\mathbb{F}_2\)^m, g_j \in \mc{G}(\mc{S}).
}
We then have
\env{align}{
    \mathrm{Sgn}_A(g^u, g^v) = (-1)^{\avg{u, \mathsf{K}_A v}}.
}

With these, we want to show that $\( \rho_{A \cup B}^{\Gamma_A} \)^2 \propto \( \rho_{A \cup B}^{\Gamma_A} \)^4$.
\env{align}{
    & \( \rho_{A \cup B}^{\Gamma_A} \)^2 \nn
    =& 
    \frac{1}{2^{2|{A \cup B}|}} \sum_{g, h \in \mc{S}} g^{\Gamma_A} h^{\Gamma_A} \nn
    =& 
    \frac{1}{2^{2|{A \cup B}|}} \sum_{g, h \in \mc{S}} x_A(g) x_A(h) \cdot gh \nn
    =& 
    \frac{1}{2^{2|{A \cup B}|}} \sum_{g, h \in \mc{S}} x_A(g) x_A(gh) \cdot h \nn
    =&
    \frac{1}{2^{2|{A \cup B}|}} \sum_{g, h \in \mc{S}} \mathrm{Sgn}_A (g, h) x_A(h) \cdot h \nn
    =&
    \frac{1}{2^{2|{A \cup B}|}} \sum_{u, v \in \(\mathbb{F}_2\)^m} \mathrm{Sgn}_A (g^u, g^v) x_A(g^v) \cdot g^v \nn
    =&
    \frac{1}{2^{2|{A \cup B}|}} \sum_{v \in \(\mathbb{F}_2\)^m} \( \sum_{u \in \(\mathbb{F}_2\)^m} (-1)^{\avg{u, \mathsf{K}_A v}} \) x_A(g^v) \cdot g^v \nn
    =&
    \frac{1}{2^{2|{A \cup B}|}} \sum_{v \in \(\mathbb{F}_2\)^m} 2^m \cdot \delta(\mathsf{K}_A v, \mathbf{0}) \cdot x_A(g^v) \cdot g^v \nn
    =&
    \frac{1}{2^{2|{A \cup B}|-m}} \sum_{v \in \mathrm{Ker}(\mathsf{K}_A)} x_A(g^v) \cdot g^v,
}
\env{align}{
    & \( \rho_{A \cup B}^{\Gamma_A} \)^4 \nn
    =&
    \lz
    \frac{1}{2^{2|{A \cup B}|-m}} \sum_{v \in \mathrm{Ker}(\mathsf{K}_A)} x_A(g^v) \cdot g^v
    \rz^2 \nn
    =&
    \frac{1}{2^{4|{A \cup B}|-2m}} \sum_{u, v \in \mathrm{Ker}(\mathsf{K}_A)} x_A(g^u) x_A(g^v) \cdot g^u g^v
    \nn
    =&
    \frac{1}{2^{4|{A \cup B}|-2m}} \sum_{u, v \in \mathrm{Ker}(\mathsf{K}_A)} (-1)^{\avg{u, \mathsf{K}_A(v)}} x_A(g^{u+v})  \cdot g^{u+v}
    \nn
    =&
    \frac{1}{2^{4|{A \cup B}|-2m}} \lv \mathrm{Ker}(\mathsf{K}_A) \rv \sum_{v \in \mathrm{Ker}(\mathsf{K}_A)}  x_A(g^{v})  \cdot g^{v}
    \nn
    =&
    \frac{1}{2^{2|{A \cup B}|-m}} \lv \mathrm{Ker}(\mathsf{K}_A) \rv \( \rho_{A \cup B}^{\Gamma_A} \)^2.
}
Thus,
\env{align}{
    & (\ln 2)^{-1} N_A(\rho_{A \cup B}) \nn
    =& \lim_{n\to \frac{1}{2}} \log_2 \mathrm{Tr} \( \rho_{A \cup B}^{\Gamma_A} \)^{2n} \nn
    =& \lim_{n\to \frac{1}{2}} \log_2 \ld
    \( \frac{1}{2^{2|{A \cup B}|-m}} \lv \mathrm{Ker}(\mathsf{K}_A) \rv\)^{n-1}
    \mathrm{Tr} \( \rho_{A \cup B}^{\Gamma_A} \)^2
    \rd \nn
    =& \lim_{n\to \frac{1}{2}} \log_2 \ld
    \( \frac{1}{2^{2|{A \cup B}|-m}} \lv \mathrm{Ker}(\mathsf{K}_A) \rv\)^{n-1}
    \frac{1}{2^{|{A \cup B}|-m}}
    \rd \nn
    =& \log_2 \( \frac{2^m}{\lv \mathrm{Ker}(\mathsf{K}_A) \rv}\)^{1/2} \nn
    =&\,
    \frac{1}{2} \lz m - \dim \mathrm{Ker}(\mathsf{K}_A) \rz \nn
    =&\,
    \frac{1}{2} \dim \mathrm{Im}(\mathsf{K}_A). 
}
It is perhaps obvious that 
\env{align}{
N_A(\rho_{A \cup B}) = N_B(\rho_{A \cup B}), 
}
since $\mathsf{K}_A = \mathsf{K}_B$.
This is indeed consistent with
\env{align}{
    \(\rho_{A \cup B}^{\Gamma_A}\)^{\mathsf{T}} = \rho_{A \cup B}^{\Gamma_{A \cup B}}.
}
We will henceforth adopt the notation in Eq.~\eqref{eq:MN_def}, and denote the negativity $N_A(\rho_{A \cup B}) = N_B(\rho_{A \cup B})$ by $N_{A,B}$.



\subsection{EN from the stabilizer group}

Define
\env{align}{
    m_A \coloneqq \dim \mathrm{proj}_A (\mc{S}) 
    \le m.
}
Without loss of generality, we assume
\env{align}{
    \mathrm{proj}_A(\mc{S}) = \avg{\mc{G}(\mathrm{proj}_A(\mc{S}))} = \avg{\{\mathrm{proj}_A(g_1), \mathrm{proj}_A(g_2), \ldots, \mathrm{proj}_A(g_{m_A})\}}, \text{ where } g_{j} \in \mc{G}(\mc{S}).
}
We have $\forall m_A < i \le m, \ \exists u\in (\mathbb{F}_2)^{m_A}$  such that
\env{align}{
    \mathrm{proj}_A(g_i) = \prod_{j=1}^{m_A} \(\mathrm{proj}_A(g_j)\)^{u_j} 
}
for which
\env{align}{
    \(\mathsf{K}_A\)_{ik} = \sum_{j=1}^{m_A} u_j \(\mathsf{K}_A\)_{jk},
    \quad
    \(\mathsf{K}_A\)_{ki} = \sum_{j=1}^{m_A} u_j \(\mathsf{K}_A\)_{kj}. 
}
Using these, we can always perform a change of basis, and assume without loss of generality that
\env{align}{
\label{eq:triviality_on_A}
    \forall m_A < i \le m, \  
    \mathrm{proj}_A(g_i) = \mathbbm{1}_A.
}

We define the following $m_A \times m_A$ matrix
\env{align}{
    \(\widetilde{\mathsf{K}}_A\)_{ij}
    = \env{cases}{
        0, \text{ if } \mathrm{proj}_A(g_i) \cdot \mathrm{proj}_A(g_j) = + \mathrm{proj}_A(g_j) \cdot \mathrm{proj}_A(g_i) \\
        1, \text{ if } \mathrm{proj}_A(g_i) \cdot \mathrm{proj}_A(g_j) = - \mathrm{proj}_A(g_j) \cdot \mathrm{proj}_A(g_i)
    }
}
for $1 \le i, j \le m_A$.
It is the upper-left $m_A \times m_A$ submatrix of $\mathsf{K}_A$.
As shown above, rows $m_A + 1 \ldots m$ and columns $m_A + 1 \ldots m$ of $\mathsf{K}_A$ are linearly dependent on rows $1 \ldots m_A$ and columns $1 \ldots m_A$, and can be eliminated, 
without changing the rank.
We thus have
\env{align}{
    \dim \mathrm{Im}(\mathsf{K}_A)
    =
    \dim \mathrm{Im}(\widetilde{\mathsf{K}}_A).
}

Next, consider the center subgroup of $\mathrm{proj}_A(\mc{S})$,
\env{align}{
    Z(\mathrm{proj}_A(\mc{S})) \coloneqq \{g \in \mathrm{proj}_A(\mc{S}) \ |\ gh = hg \text{ for all } h \in \mathrm{proj}_A(\mc{S})\}.
}
Without loss of generality, let
\env{align}{
    Z(\mathrm{proj}_A(\mc{S}))
    = \avg{ \{\mathrm{proj}_A(g_1), \ldots, \mathrm{proj}_A(g_{\mu_A})\} } \text{ where } \mu_A \le m_A, \quad g_j \in \mc{G}(\mc{S}).
}
By definition, the first $\mu_A$ columns and rows of $\widetilde{\mathsf{K}}_A$ are zero, thus do not contribute to $\dim \mathrm{Im}(\widetilde{\mathsf{K}}_A)$.
By induction one can show that $\widetilde{\mathsf{K}}_A$ can be brought into the following ``canonical form'' with congruence transformations~\cite{kim_two_2008},
\env{align}{
    \label{eq:KA_canonical}
    \widetilde{\mathsf{K}}_A = \env{pmatrix}{
        0_{\mu_A \times \mu_A} & 0_{\mu_A \times k_A} & 0_{\mu_A \times k_A} \\
        0_{k_A \times \mu_A} & 0_{k_A \times k_A} & \mathbbm{1}_{k_A \times k_A} \\
        0_{k_A \times \mu_A} & \mathbbm{1}_{k_A \times k_A} & 0_{k_A \times k_A}
    },
}
where $k_A$ is an integer, and $\dim \mathrm{Im}(\widetilde{\mathsf{K}}_A) = 2k_A$.
We then have
\env{align}{
    m_A =&\ \mu_A + 2k_A \nn
    \Leftrightarrow\quad 
    \dim \mathrm{proj}_A (\mc{S}) =&\ \dim Z(\mathrm{proj}_A (\mc{S})) + \dim \mathrm{Im}(\mathsf{K}_A),
}
and 
\env{align}{
    N_{A,B}
    = k_A \ln 2
    = \frac{\ln 2}{2} \dim \mathrm{Im}(\mathsf{K}_A)
    = \frac{1}{2} \ln \lv \frac{\mathrm{proj}_A(\mc{S})}{Z(\mathrm{proj}_A (\mc{S}))} \rv.
}

\subsection{Bounding the EN with entanglement entropies}

It is now clear that in the canonical basis of Eq.~\eqref{eq:KA_canonical},
\env{align}{
    \{\mathrm{proj}_A (g_1), \ldots, \mathrm{proj}_A (g_{\mu_A}) \} \cup \{\mathrm{proj}_A (g_{\mu_A + 1}), \ldots, \mathrm{proj}_A (g_{\mu_A + k_A}) \}
}
is a linearly independent, mutually commuting set of Pauli strings on $A$.
Thus we have
\env{align}{
    \mu_A + k_A \le |A|,
}
thus
\env{align}{
    (\ln 2)^{-1} 
    N_{A,B}
    = k_A = m_A - (\mu_A + k_A)
    \ge  m_A - |A|
    = (\ln 2)^{-1} \(S_B - S_{A \cup B} \).
}
{Here, we recall that~\cite{Fattal2004stabilizer}
$S_B = |B| - (m - m_A)$, and $S_{A \cup B} = |{A \cup B}| - m$.}
On the other hand, consider the following set of Pauli strings on $B$,
\env{align}{
    \{\mathrm{proj}_{B} (g_{m_A + 1}),
    \mathrm{proj}_{B} (g_{m_A + 2}),
    \ldots,
    \mathrm{proj}_{B} (g_m) \}
    \cup \{\mathrm{proj}_{B} (g_{\mu_A + 1}),
    \ldots,
    \mathrm{proj}_{B} (g_{\mu_A + k_A})
    \}.
}
Using Eq.~\eqref{eq:triviality_on_A}, one can see that it is a linearly independent, mutually commuting set of Pauli strings on $B$.\footnote{
They are obviously commuting.
The linear independence is perhaps less obvious.
Let us prove by contradiction.
Suppose the product of some of these is $\mathbbm{1}_{B}$.
If this product involves some $\mathrm{proj}_{B} (g_{\mu_A + j}) (1 \le j \le k_A)$, then $\mathrm{proj}_{B} (g_{\mu_A + j})$ cannot possibly anticommute with $\mathrm{proj}_{B} (g_{\mu_A + j + k_A})$, which it must.
Thus the product must involve only $\{
    \mathrm{proj}_{B} (g_{m_A + 1}),
    \mathrm{proj}_{B} (g_{m_A + 2}),
    \ldots,
    \mathrm{proj}_{B} (g_m)\}$.
But based on Eq.~\eqref{eq:triviality_on_A}, linear dependence among these would imply linear dependence of $\{
    g_{m_A + 1},
    g_{m_A + 2},
    \ldots,
    g_m\}$, which cannot be the case either.
}
Thus
\env{align}{
    m-m_A + k_A \le |B| = |{A \cup B}| - |A|,
}
or equivalently
\env{align}{
    (\ln 2)^{-1} N_{A,B} = k_A  \le m_A + |{A \cup B}| - |A| - m = (\ln 2)^{-1} S_B.
}

Therefore
\env{align}{
    \label{eq:A38}
    S_B - S_{A \cup B} \le N_{A,B} \le S_B.
}
Since $ N_{A,B} =  N_{B,A}$, we have
\env{align}{
    \label{eq:A39}
    & \max\{ S_B, S_A \} - S_{A \cup B}
    \le N_{A,B}
    \le
    \min\{ S_B, S_A \}.
}
In the case of a global pure state, $S_{A \cup B} = 0$, and
\env{align}{
    N_{A,B} = S_B = S_A.
}
This result can be also obtained from the general equality between $N_{A,B}$ and $S^{1/2}(\rho_A)$ (i.e. the $\frac{1}{2}$-th R\'{e}nyi entropy of $\rho_A$) when $\rho_{A \cup B}$ is pure, and the fact that stabilizer states have flat entanglement spectra.
The identification between the negativity and entanglement entropy does not hold for mixed states in general, for in general
\env{align}{
    S_B \neq S_A.
}


\subsection{Bounding the EN with mutual information}

We compute the mutual information between $A$ and $B$,
\env{align}{
    &  (\ln 2)^{-1} I_{A, B} \nn
    =&\ S_A + S_B - S_{A \cup B} \nn
    =&\ (|A| - \dim \mathrm{Ker}\, \mathrm{proj}_{B}(\mc{S}) )
    + (|B| - \dim \mathrm{Ker}\, \mathrm{proj}_{A} (\mc{S}) ) - (|{A \cup B}| - m) \nn
    =&\, \dim \mc{S} - \dim \mathrm{Ker}\, \mathrm{proj}_{B} (\mc{S}) - \dim \mathrm{Ker}\, \mathrm{proj}_{A}(\mc{S}) \nn
    =&\, \dim \mathrm{proj}_A(\mc{S})  - \dim \mathrm{Ker}\, \mathrm{proj}_{B} (\mc{S}),
}
thus, using  $\mathrm{Ker}\ \mathrm{proj}_{B}(\mc{S}) \subseteq Z(\mathrm{proj}_{A}(\mc{S}))$,
\env{align}{
    & (\ln 2)^{-1} \(2 N_{A,B} \) = \dim \mathrm{proj}_A(S) - \dim Z(\mathrm{proj}_A(S)) \nn
    \le& \dim \mathrm{proj}_A(\mc{S})  - \dim \mathrm{Ker}\, \mathrm{proj}_{B} (\mc{S}) = (\ln 2)^{-1} I_{A, B},
}
or
\env{align}{
     N_{A,B} \le \frac{1}{2} I_{A,B}.
}

We may upper bound their difference using the following observation~\cite{li2007capillary_qecc}
\env{align}{
\frac{
    Z(\mathrm{proj}_{A}(\mc{S}))
}{
    \mathrm{Ker}\ \mathrm{proj}_{B}(\mc{S})
}
\subseteq
\mc{L}_A
\coloneqq
\frac{\ld g \in \mc{C}(\mc{S}) \ |\  \mathrm{proj}_{B}(g) \in \mathrm{proj}_{B}(\mc{S}) \rd} {\mc{S}},
}
so that
\env{align}{
    0 \le
    \dim
    Z(\mathrm{proj}_{A}(\mc{S}))
    -
    \dim
    \mathrm{Ker}\ \mathrm{proj}_{B}(\mc{S})
    \le
    \dim \mc{L}_A = (\ln 2)^{-1} I_{A, C},
}
where $C$ is a purification of ${A \cup B}$, and 
\env{align}{
    I_{A,C} = S_A + S_C - S_{A\cup C} = S_A + S_{A \cup B} - S_B.
}
Thus\footnote{This result maybe alternatively derived as follows:
\env{align}{
    &  I_{A,C}\nn
    =& S_A + S_{A \cup B} - S_B \nn
    =& I_{A, B} - 2 ( S_B - S_{A \cup B} ) \nn
    \ge& I_{A, B} - 2 N_{A,B}, \nonumber
}
making use of Eqs.~(\ref{eq:A38}, \ref{eq:A39}).
}
\env{align}{
    0 \le \frac{1}{2} I_{A, B} - N_{A,B} \le \frac{1}{2} I_{A, C}.
}
In particular, it implies that $\frac{1}{2} I_{A, B} = N_{A,B}$ when the subsystem $A$ decouples from the environment ($I_{A,C} = 0$).

\subsection{A motivating example of the structure theorem}

\begin{figure*}
    \centering
    \includegraphics[width=0.7\textwidth]{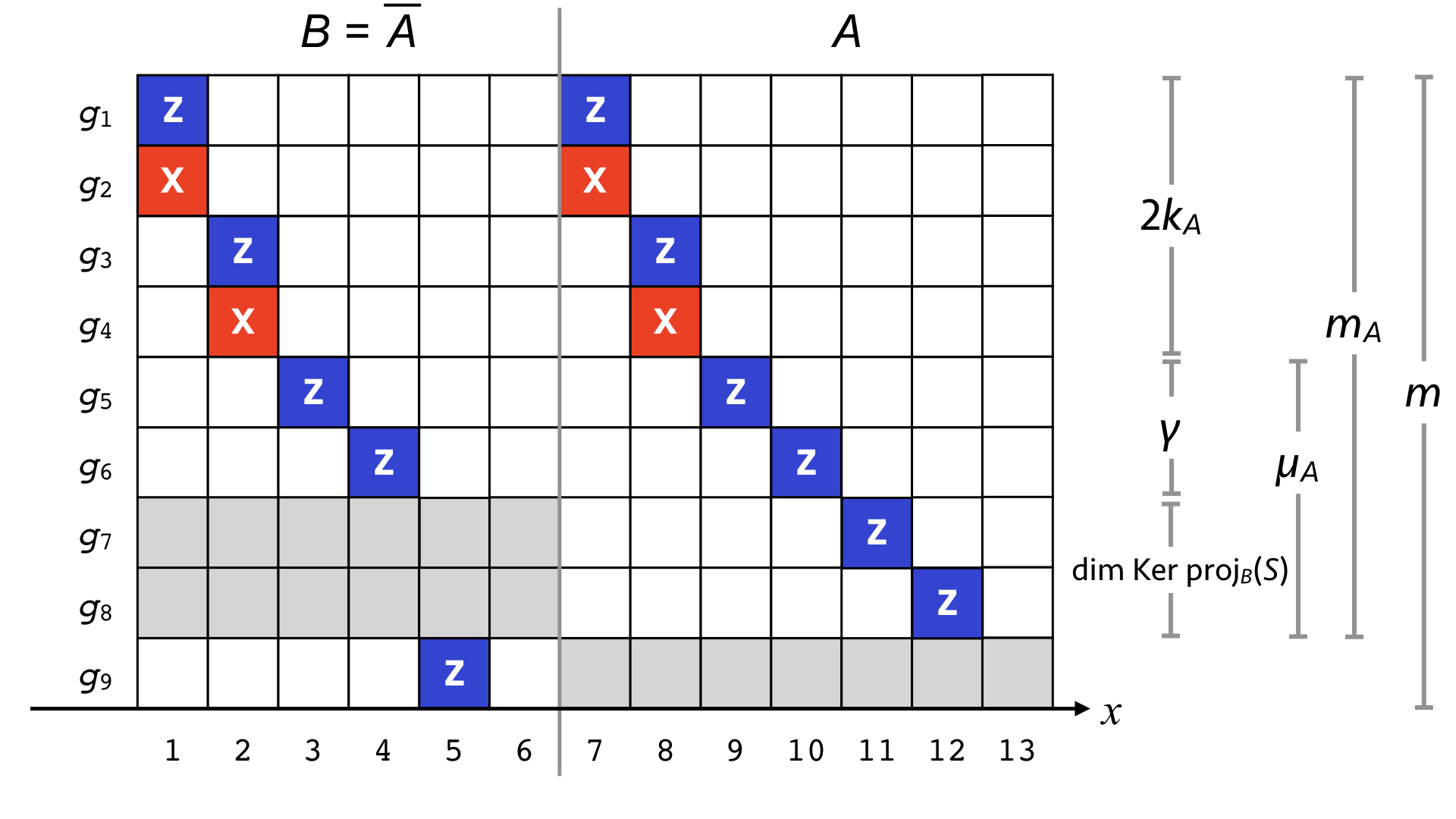}
\caption{An example of a bipartite mixed stabilizer state.
Here we have $|{A \cup B}| = 13$ lattice sites, where $|A| = 7$ and $|B| = 6$.
There are $m = 9$ stabilizers in the basis of the stabilizer group, represented by each row of $|{A \cup B}|$ squares.
Blue squares represent local Pauli $Z$ operators, red squares represent local Pauli $X$ operators, and white or gray squares represent identity operators.
Stabilizers in this basis can be classified into four sets, as detailed in Eq.~\eqref{eq:A50}.
When a stabilizer is supported nontrivially only on $A$ or $B$, they are called ``local'', and we can disregard their (trivial) content on the other subsystem (represented with gray color).
}
    \label{fig:stabilizer}
\end{figure*}

We illustrate our reasoning so far with an example, shown in Fig.~\ref{fig:stabilizer}.
This simple example represents a general pattern: with local Clifford unitaries on $A$ and $B$, we can always obtain the following ``canonical'' basis of the stabilizer group:
\env{align}{
\label{eq:A50}
    \mc{G}(\mc{S}) =& \{2 k_A \text{ stabilizers contributing to } k_A \text{ EPR pairs across $A$ and $B$}\} \nn
    &\quad \cup
    \{
    \gamma = \mu_A - \dim \mathrm{Ker}\, \mathrm{proj}_{B}(\mc{S}) 
    =
    \mu_{B} - \dim \mathrm{Ker}\, \mathrm{proj}_{A}(\mc{S}) \text{ ``classical'' stabilizers across $A$ and $B$}
    \} \nn
    &\quad \cup
    \{
    \dim \mathrm{Ker}\, \mathrm{proj}_{B}(\mc{S}) \text{ local stabilizers on $A$}
    \} \nn
    &\quad \cup
    \{
    \dim \mathrm{Ker}\, \mathrm{proj}_{{A}}(\mc{S}) \text{ local stabilizers on $B$}
    \}.
}
Here,
\env{itemize}{
\item
The first set of $2k_A$ stabilizers are those contributing to the nonzero entries of $\mathsf{K}_A$ (they pairwise anticommute when restricted to $A$ or $B$);
\item
The second set of $\gamma$ ``classical'' stabilizers and the third set of $\dim \mathrm{Ker}\, \mathrm{proj}_{B}(\mc{S})$ local stabilizers together generate $Z(\mathrm{proj}_{A}(\mc{S}))$; 
\item
The first three sets altogether generate $\mathrm{proj}_{A}(\mc{S})$.
}

The ``classical'' stabilizers take the form of $Z_i Z_j$, with $i \in A$ and $j \in B$, for which a stabilizer of the form $X_i X_j$ -- that anticommutes with it when restricted to either $A$ or $B$ -- is \emph{absent}.
They introduce classical correlations, and contribute to the mutual information but not the negativity,
\env{align}{
    \frac{1}{2}I_{A,B} =&\, \(k_A + \frac{\gamma}{2}\) \ln 2, \\
    N_{A,B} =&\, k_A \ln 2.
}
Upon the introduction of a ``reference system'' $C$ that purifies ${A \cup B}$, each of the $\gamma$ classical stabilizers extends to a tripartite GHZ state on $A$, $B$, and $C$.
With further considerations along these lines, one can obtain the ``structure theorem'' in Eq.~\eqref{eq:structural_thm}, with the following identifications
\env{align}{
    k_A = k_B \to e_{AB}, \quad \gamma \to g_{ABC}.
}
The structure theorem can be used to check the inequalities derived above.
\section{Calculating the mutual negativity in the loop representation of critical percolation \label{app:TLalgebra}}

Here we compute $h_{\rm MN}$ in the measurement-only circuit in Sec.~\ref{sec:moc_majorana}, making use of the loop representation.

We represent the loop ensemble at time $T$ as a wavefunction, which is a linear superposition of all possible pairing patterns (compare Eq.~\eqref{eq:pairing_pattern}) weighted by their probabilities
\env{align}{
    \ket{\Omega(T)} = \sum_{w} \lambda_w \ket{w}, \quad \lambda_w > 0.
}
This wavefunction has the following ``transfer matrix'' representation,
\env{align}{
    \ket{\Omega(T)} = \(e^{\hat{H}} \)^T \ket{\Omega(0)},  
}
where
\env{align}{
    e^{\hat{H}} =
    \lz 
    \prod_{j\, {\rm even}} \(\frac{\TLI\, +\, \TLE}{2}\)_j
    \rz
    \cdot
    \lz
    \prod_{j\, {\rm odd}} \(\frac{\TLI\, +\, \TLE}{2}\)_j
    \rz.  
}
Thus, $\TLE_j$ can be viewed as a ``Hamiltonian density'' at site $j$.
Formally, $\TLE_j \, (1 \le j \le L)$ are generators of a Temperley-Lieb algebra~\cite{temperley_lieb_1971, blote_nienhuis_1989, jacobsen2009book}, and $\TLI_j$ is the identity element of the algebra.
We have in particular
\env{align}{
    \label{eq:B4}
    \(\TLE_j\)^2 =&\ \sqrt{Q}\ \TLE_j,\\
    \TLE_j \cdot \TLE_{j+1} \cdot \TLE_j =&\ \TLE_j, \\
    \TLE_j \cdot \TLE_{j-1} \cdot \TLE_j =&\ \TLE_j, \\
    \TLE_j \cdot \TLE_i  =&\ \TLE_i \cdot \TLE_j, \text{ if $|j-i| \neq 1$}.
}
This loop ensemble is that of the critical $Q$-states Potts model, where each closed loop is assigned weight $\sqrt{Q}$ due to Eq.~\eqref{eq:B4}.
The $Q = 1$ case is critical percolation.

Next, we relate the desired ``double arc'' probability to boundary correlation functions.
Denote by $\avg{\cdot}$ the expectation value against the statistical ensemble encoded in $\ket{\Omega(T)}$ when the inserted operator is applied at the final time step.
For example, we have
\env{align}{
\label{eq:B8}
    \bigg\langle
        \TLE_i
    \bigg\rangle = \frac{\sum_w \lambda_w \(\sqrt{Q}\)^{\# \text{ new loops generated upon applying } \TLE_i}}{\sum_w \lambda_w}.
}
Using Eq.~\eqref{eq:B8} and a bit of combinatorics, we can show that
\env{align}{
    &  \mathbb{P}(\text{a double arc connects qubits } i, j) \nn
    \propto& \sum_{w: \text{ a double arc connects qubits } i, j \text{ in } \ket{w}} \lambda_w \nn
    \propto &\, \frac{d}{d\sqrt{Q}}\bigg|_{\sqrt{Q} = 1} \lz
        \bigg\langle \TLE_i \  \TLE_j \bigg\rangle
        -
        \bigg\langle
        \TLE_i
        \bigg\rangle
        -
        \bigg\langle \TLE_j
        \bigg\rangle
    \rz.
}
The last two terms are constants and do not depend on $x_{ij}$, whereas the first is a correlation function between Hamlitonian densities or stress-energy tensors thus decays as $\(x_{ij}\)^{-4}$ in a two-dimensional conformal field theory.
We have the same powerlaw for $1\le Q \le 4$, thus
\env{align}{
    \mathbb{P}(\text{a double arc connects qubits } i, j) \propto  \(x_{ij}\)^{-4},
}
or equivalently,
\env{align}{
    N_{[x_i, x_i+1], [x_j, x_j+1]} \propto \eta^{h_{\rm MN}}, \text{ where } \eta \propto \(x_{ij}\)^{-2}, h_{\rm MN} = 2.
}

\section{Addtional numerical results in the CPLC Goldstone phase\label{app:CPLC_numerics}}

\begin{figure}
    \centering
    \includegraphics[width=.45\textwidth]{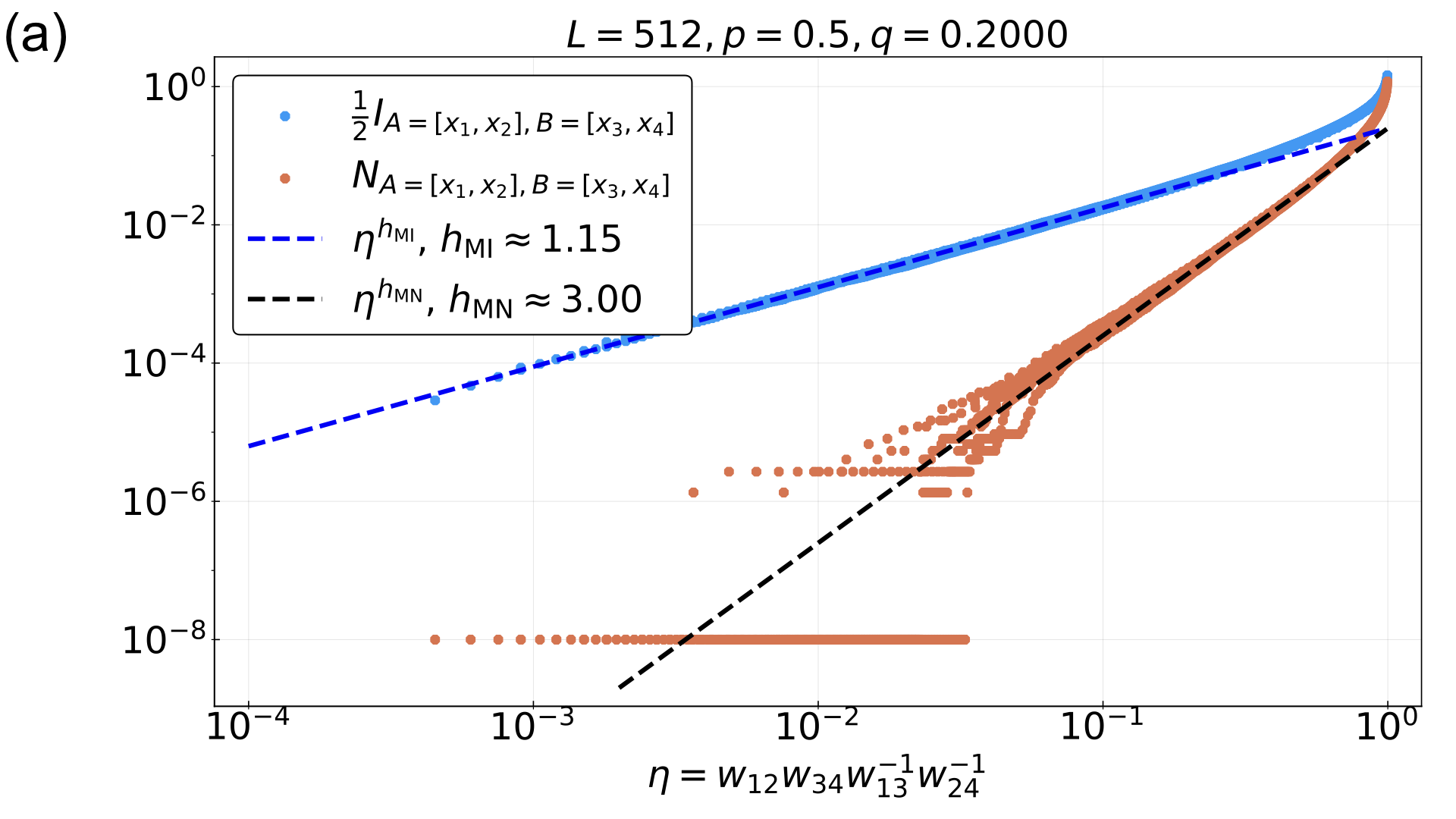}
    \includegraphics[width=.45\textwidth]{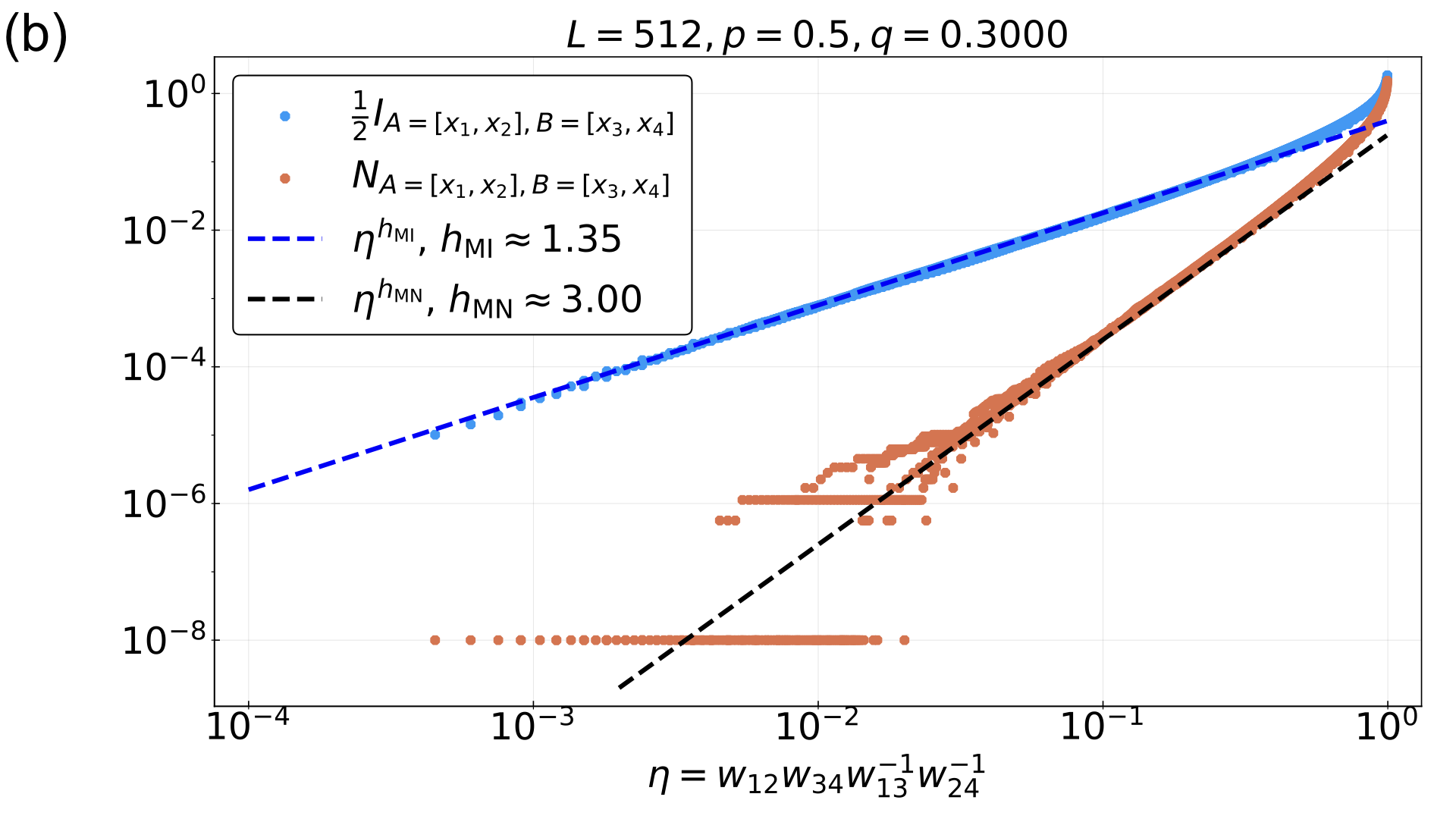}
    \includegraphics[width=.45\textwidth]{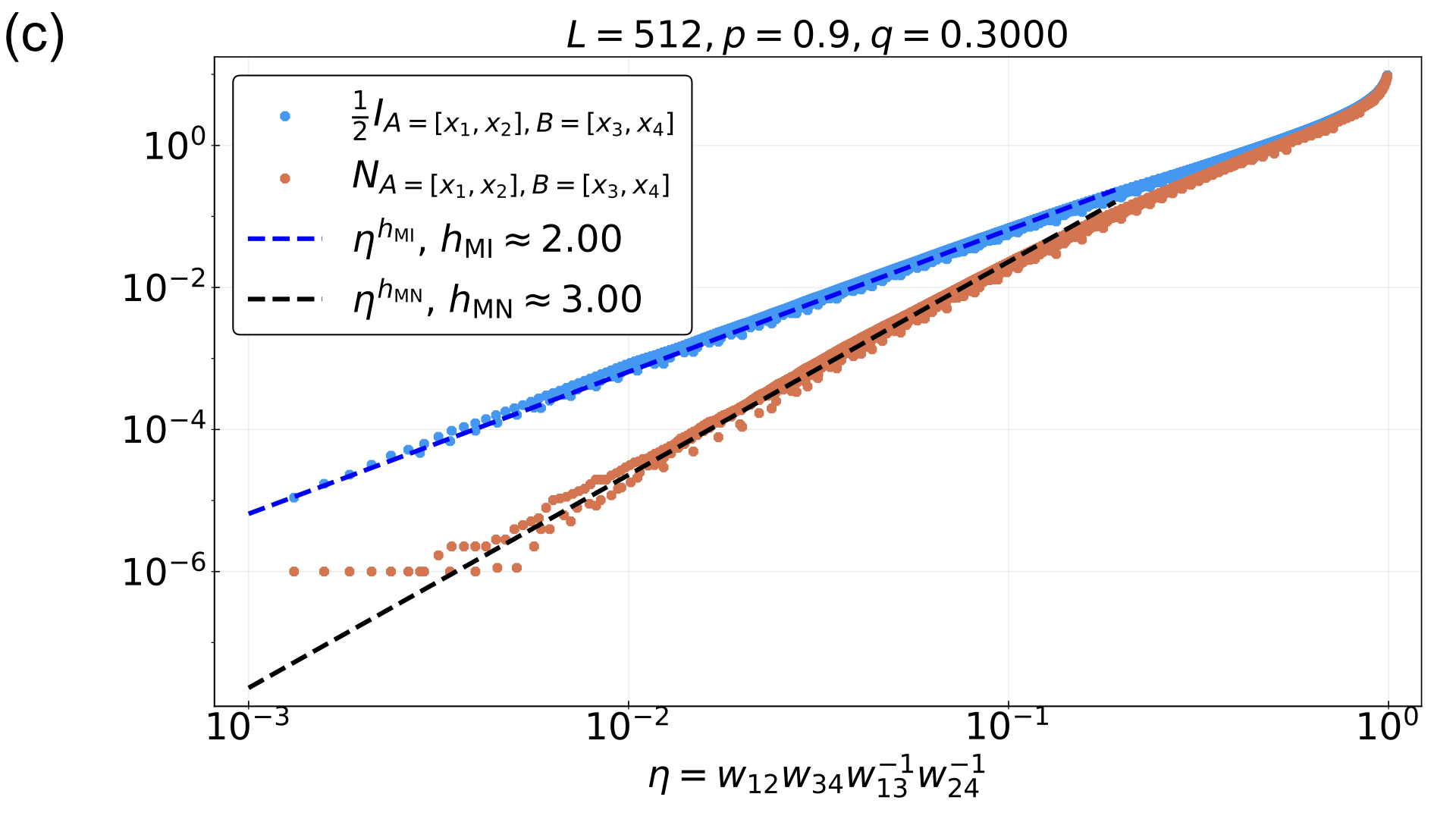}
    \includegraphics[width=.45\textwidth]{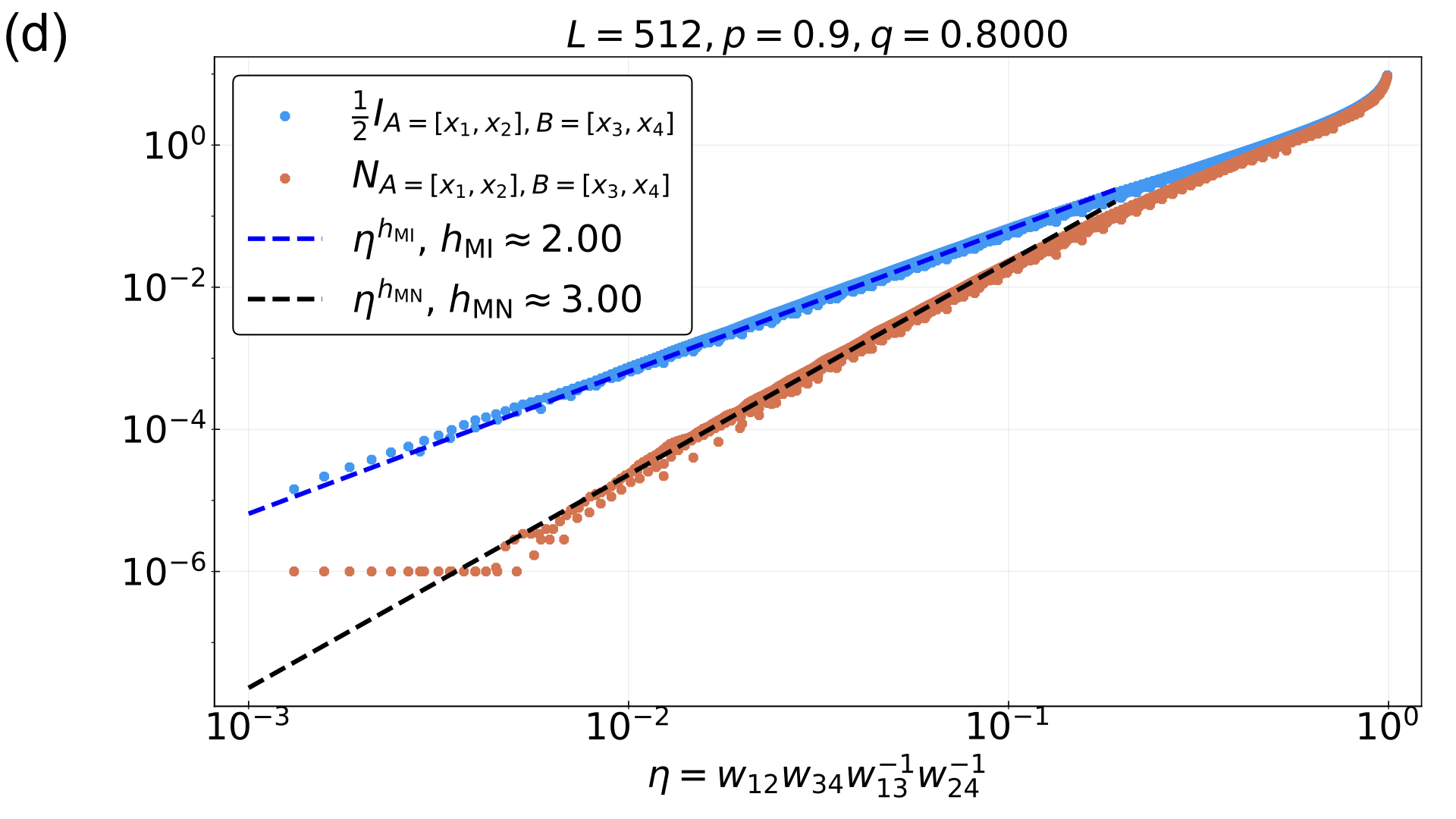}
    \caption{Numerical results for MI and MN at a few more points in the CPLC Goldstone phase; compare Figs.~\ref{fig:phase_diagram}, \ref{fig:cplc_metal}.}
    \label{fig:CPLC_metal_app}
\end{figure}

In Fig.~\ref{fig:CPLC_metal_app} we show numerical results for MI and MN at a few more points in the CPLC Goldstone phase.
These plots supplement those in Fig.~\ref{fig:cplc_metal}.
We see a varying $h_{\rm MI}$, but a rather robust $h_{\rm MN}$.

Moreover, we notice that as the crossing probability $p$ approaches $1$, the exponents become close to those in random Clifford and Haar circuits (compare Fig.~\ref{fig:hybrid_numerics}).
\section{Numerical results for hybrid circuit with $\mathbb{Z}_2$ symmetry \label{app:z2_numerics}}

In this appendix we show numerical results for MI and MN 
for two hybrid circuit models that preserves the $\mathbb{Z}_2$ global Ising symmetry: $P = \prod_i X_i$. 

The first model is the stabilizer circuit introduced in Ref.~\cite{hsieh_sang_2004_protected} (see Fig.~\ref{fig:Z2_app}(a)).
The circuit is composed of randomly chosen two-site operations organized in a brick-wall pattern in spacetime. Each operation can be either a $\mb{Z}_2$-symmetric random Clifford unitary gate 
with probability $p$ or a measurement with probability $1-p$.
Given that the operation is a measurement, it can be done in either $XI$ or $ZZ$ basis, with probabilities $q$ and $1-q$, respectively.
The phase diagram of this model is shown in Fig.~\ref{fig:Z2_app}(b). The diagram is similar to the CPLC's (Fig.~\ref{fig:phase_diagram}) as both contain a topological and a trivial insulating phase, as well as a critical phase. The difference is $\mathbb{Z}_2$ circuit can have a volume-law phase when $p$ is large, because it contains interacting fermion gates. 


We compute MN and MI in this model at the volume-to-trivial and volume-to-topological transition points, and plot the results in Fig.~\ref{fig:Z2_app}(c) and (d). At the volume-to-trivial transition we have $h_{\rm MI}\approx 0.7$ and $h_{\rm MN}\approx 3.0$, while on the volume-to-topological one we have $h_{\rm MI}\approx 1.4$ and $h_{\rm MN}\approx 2.5$.

The second model is similar to the hybrid random Haar circuit in Sec.~\ref{sec:rand_Clifford_Haar}, except the unitaries are now required to be $\mb{Z}_2$ symmetric, but otherwise random, and measurements are all in single-site $X$ basis.  
We plot the results of MI and MN at the critical point of this model in Fig.~\ref{fig:Z2_app}(e), whose fitting gives $h_{\rm MI}\approx 0.9$ and $h_{\rm MN}\approx 1.9$. We note that these exponents are smaller than their counterparts in the Haar circuits without symmetry shown in Fig.~\ref{fig:hybrid_numerics}(b), indicating a stronger correlation with the presence of $\mathbb{Z}_2$ symmetry. However, due to extremely small system size limits, the values of the exponent may not be taken for a serious quantitative analysis, when comparing with the data for Clifford circuits.

\begin{figure}
    \centering
    \subfigure[]{
        \includegraphics[width=.45\textwidth]{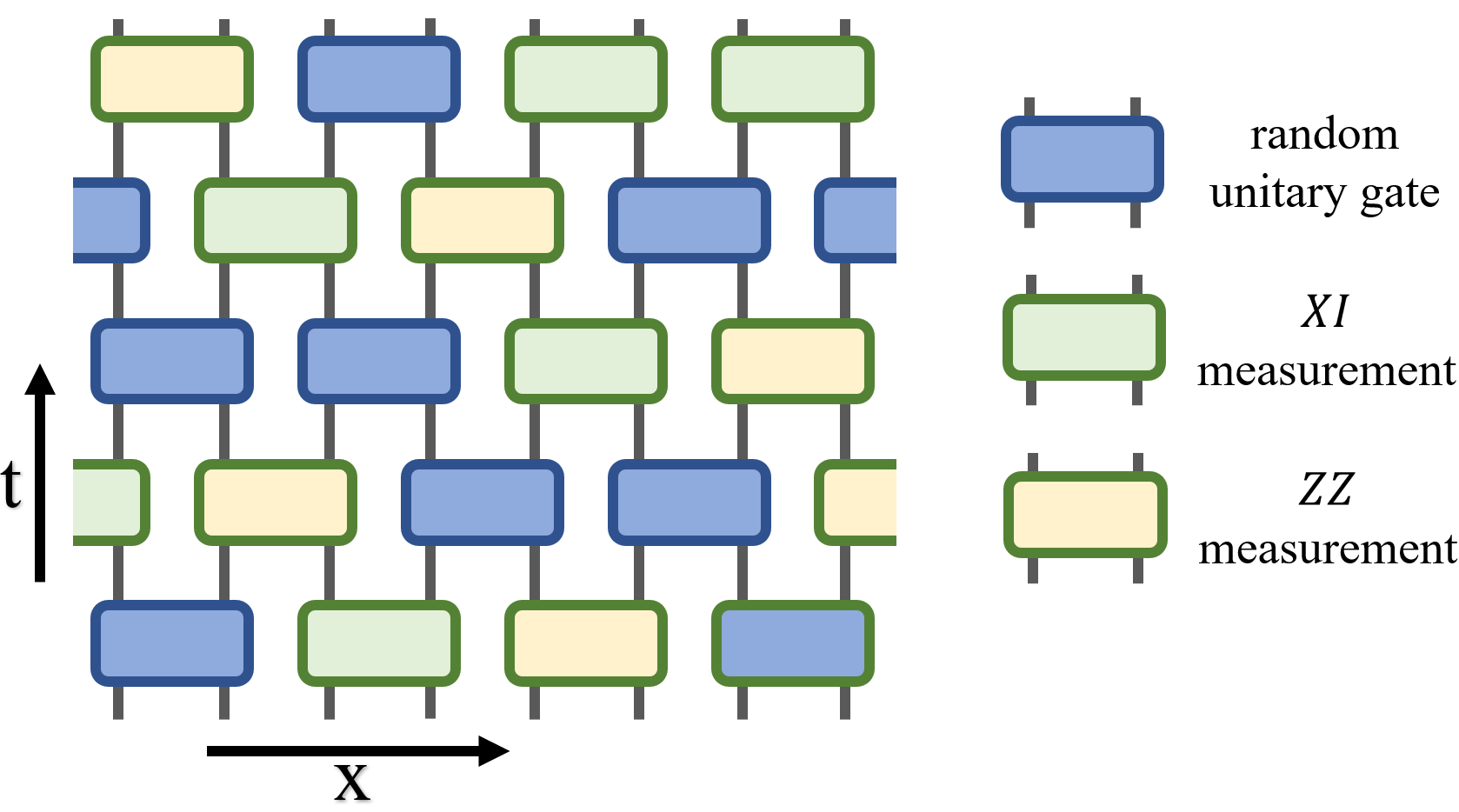}
    }
    \subfigure[]{
        \includegraphics[width=.45\textwidth]{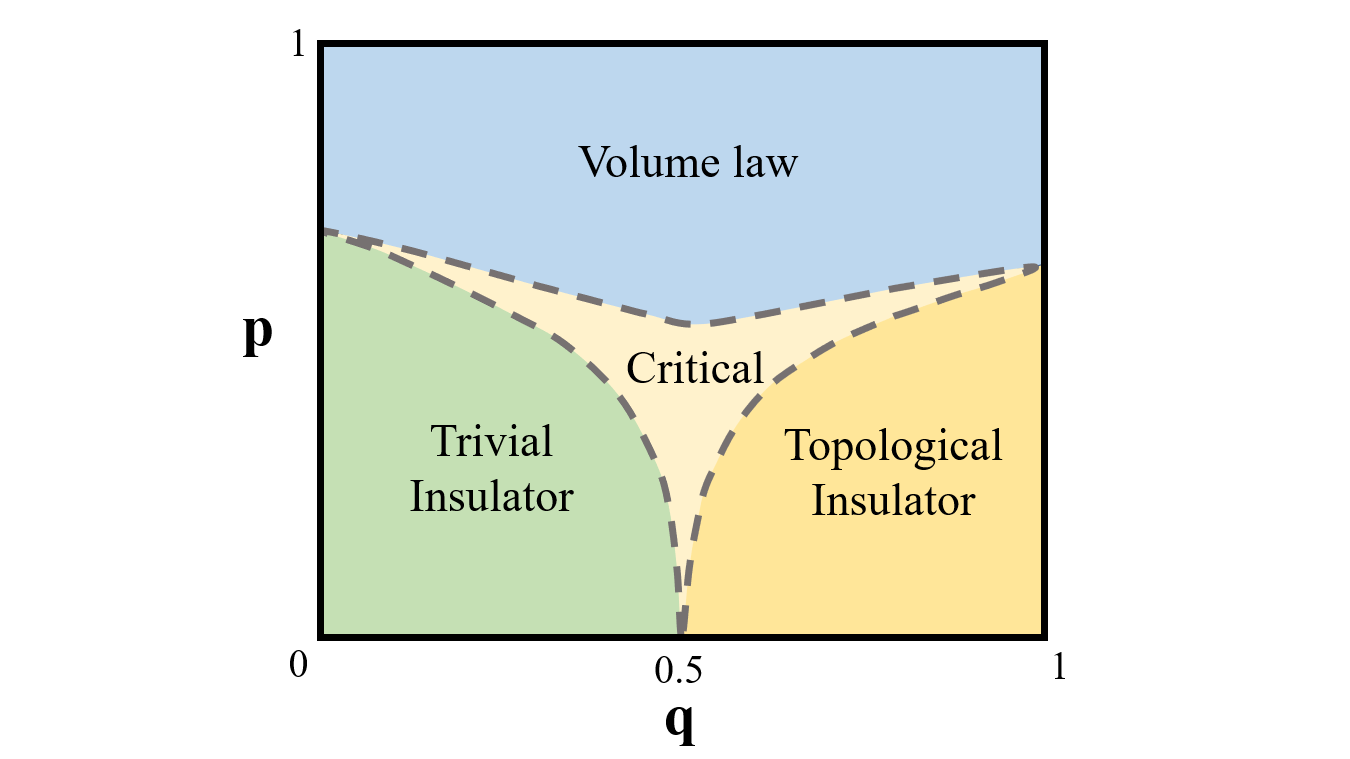}
    }
    \\
    \subfigure[]{
        \includegraphics[width=.45\textwidth]{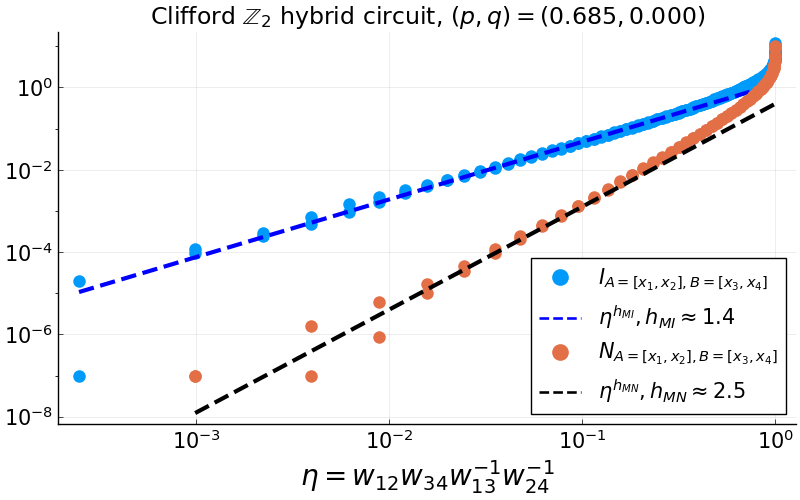}
    }
    \subfigure[]{
        \includegraphics[width=.45\textwidth]{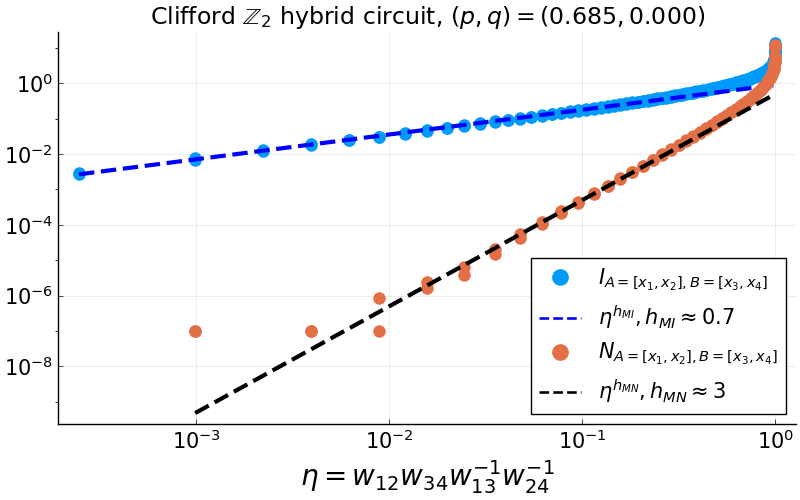}
    }\\
     \subfigure[]{
        \includegraphics[width=.45\textwidth]{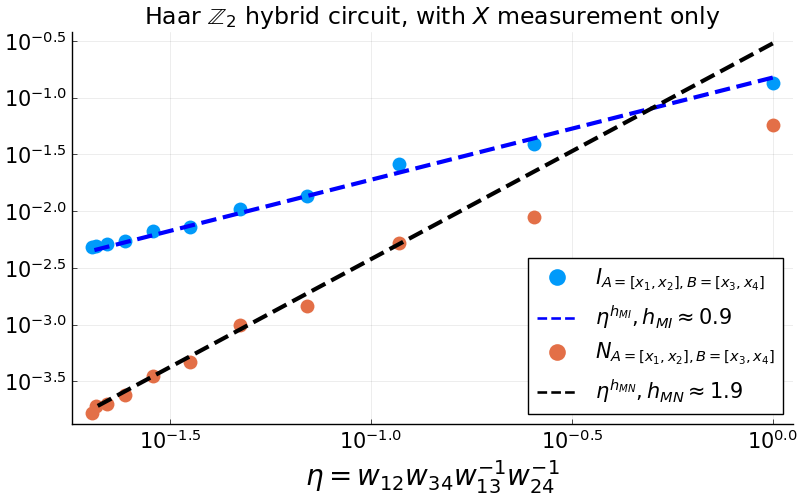}
    }
    \caption{Graphical illustration, phase diagram and numerical results for the $\mathbb{Z}_2$ hybrid circuit model.}
    \label{fig:Z2_app}
\end{figure}

By using the Jordan-Wigner transformation, both models can be represented in the Majorana picture, composed of Majorana parity measurements and four-Majorana unitary gates. Here the $\mathbb{Z}_2$ symmetry guarantees that unitary gates are spatially local in both qubit and Majorana picture.

\end{widetext}

\bibliography{refs}

\end{document}